\documentclass[twocolumn,aps,epsfig,nofootinbib,preprintnumbers]{revtex4}
\usepackage{graphicx}
\usepackage{epstopdf}
\usepackage{latexsym}
\usepackage{amssymb}
\usepackage{amsmath}
\usepackage{color}
\usepackage{float}
\usepackage{mathrsfs}
\usepackage{multirow}
\usepackage{tabularx,booktabs}
\newcolumntype{Y}{>{\centering\arraybackslash}X}

\usepackage[center]{subfigure}
\usepackage{makecell}
\setcellgapes{4pt}
\begin{document}

  \renewcommand\arraystretch{2}
 \newcommand{\bq}{\begin{equation}}
 \newcommand{\eq}{\end{equation}}
 \newcommand{\bqn}{\begin{eqnarray}}
 \newcommand{\eqn}{\end{eqnarray}}
 \newcommand{\nb}{\nonumber}
 \newcommand{\lb}{\label}
 \newcommand{\cb}{\color{blue}}
    \newcommand{\cc}{\color{cyan}}
        \newcommand{\cm}{\color{magenta}}
\newcommand{\rc}{\rho^{\scriptscriptstyle{\mathrm{I}}}_c}
\newcommand{\rd}{\rho^{\scriptscriptstyle{\mathrm{II}}}_c} 
\newcommand{\PRL}{Phys. Rev. Lett.}
\newcommand{\PL}{Phys. Lett.}
\newcommand{\PR}{Phys. Rev.}
\newcommand{\CQG}{Class. Quantum Grav.}

\title{Gravitational wave forms, polarizations, response functions and energy losses  of  triple systems  in Einstein-Aether theory}

\author{Kai Lin$^{1, 2}$}
%
\author{Xiang Zhao$^{3,4}$}
%
\author{Chao Zhang$^{3,4}$}
 %
\author{Tan Liu$^{5,6}$}
%
\author{Bin Wang$^{7, 8}$}
%
\author{Shaojun Zhang$^4$}
%
\author{Xing Zhang$^{5,6}$} 
%
\author{Wen Zhao$^{5,6}$}
%
\author{Tao Zhu$^4$}
%
\author{Anzhong Wang$^{3, 4}$\footnote{ Corresponding Author}}
\email{Anzhong_Wang@baylor.edu}

\affiliation{$^1$ Hubei Subsurface Multi-scale Imaging Key Laboratory, Institute of Geophysics and Geomatics, China University of Geosciences, Wuhan, Hubei, 430074, China}
\affiliation{$^2$ Escola de Engenharia de Lorena, Universidade de S\~ao Paulo, 12602-810, Lorena, SP, Brazil}
\affiliation{$^3$ GCAP-CASPER, Physics Department, Baylor University, Waco, TX 76798-7316, USA}
\affiliation{$^4$ Institute for Advanced Physics $\&$ Mathematics, Zhejiang University of Technology, Hangzhou 310032, China}
 \affiliation{$^5$ CAS Key Laboratory for Researches in Galaxies and Cosmology, Department of Astronomy, \\
University of Science and Technology of China, Chinese Academy of Sciences, Hefei, Anhui 230026, China}
\affiliation{$^6$ School of Astronomy and Space Science, University of Science and Technology of China, Hefei 230026, China}
\affiliation{$^7$ Center for Gravitation and Cosmology, Yangzhou University, Yangzhou 225009, China}
\affiliation{$^{8}$ School of Physics and Astronomy, Shanghai Jiao Tong University, Shanghai 200240, China}

\date{\today}

\begin{abstract}

Gravitationally bound hierarchies containing three or more components are very common in our Universe. In this paper we study {\em periodic} gravitational wave 
(GW) form, their  polarizations, response function, its Fourier transform, and energy loss rate of a triple system  through three different channels of radiation, 
the scalar, vector and tensor modes,  in Einstein-aether theory of gravity. The theory violates locally the Lorentz symmetry, and yet satisfies all the theoretical 
and observational constraints by properly choosing its four coupling constants $c_{i}$'s. In particular, in the weak-field approximations and with the recently 
obtained constraints of the theory, we first analyze  the energy loss rate of a binary system, and find that the dipole contributions from the scalar and vector modes 
could be of the order of ${\cal{O}}\left(c_{14}\right){\cal{O}}\left(G_Nm/d\right)^2$, where $c_{14} \; (\equiv c_{1} + c_{4})$ is constrained  to $c_{14}
\lesssim {\cal{O}}\left(10^{-5}\right)$ by current observations, and $G_N,\; m$ and $d$ are, respectively, the Newtonian constant, mass and size of the source.   
On the other hand, the ``strong-field" effects for a binary system of neutron stars are about six orders lower than that of GR. So, in this paper we ignore these 
``strong-field" effects and first develop the general formulas to  the lowest post-Newtonian order, by taking the coupling of the aether field with matter into account. 
Within this approximation, we find that the scalar breather mode and the scalar  longitudinal mode are  all suppressed by a factor of ${\cal{O}}\left(c_{14}\right)$ 
with respect to the transverse-traceless modes ($h_{+}$ and $h_{\times}$), while the vectorial modes $(h_{X}$ and $h_{Y}$) are suppressed by a factor of 
$c_{13} \lesssim {\cal{O}}\left(10^{-15}\right)$. Applying the general formulas to a triple system with periodic orbits, we find that the corresponding 
GW form, response function, and its Fourier transform depend sensitively on the configuration of the triple system, their orientation with respect to the 
detectors, and the binding energies of the three compact bodies.

 \end{abstract}


\maketitle

\section{Introduction}
 \renewcommand{\theequation}{1.\arabic{equation}} \setcounter{equation}{0}

The detection of the gravitational wave  from the coalescing of two massive black holes (one with  mass $36^{+5}_{-4}\; M_{\bigodot}$ and the other with mass 
$29^{+4}_{-4}\; M_{\bigodot}$)  by the advanced  Laser Interferometer Gravitational-Wave Observatory (aLIGO) marked the beginning of the era of the
gravitational wave (GW) astronomy \cite{GW150914}. Following it, five more GWs were detected  \cite{GW151226,GW170104,GW170608,GW170814,GW170817}, and one 
candidate was identified \cite{LVT151012}. The Advanced Virgo detector  (aVirgo) \cite{aVIRGO} joined the second observation run of aLIGO on August 1, 2017, and 
jointly  detected the last two GWs, GW170814 and GW170817 \cite{GW170814,GW170817}. Except GW170817, which was produced by the merger of binary
neutron stars (BNSs) \cite{GW170817},  all the rest were produced during the mergers of binary black holes (BBHs). The detection of GW170817 is important, not only 
because it confirmed that BNSs are indeed one of the most promising sources of GWs,  but also because it was companied by a short-duration gamma-ray burst (SGRB) 
\cite{GRB170817}, which enables a wealth of science unavailable from either messenger alone. In particular, it  allows us  simultaneously  to measure both  distance and
redshift of the source, with which we can study, for example, cosmology. 
  
 With the promise of increasing duration, observational sensitivity and the number of detectors, many more events are expected to be detected.
In particular, the Laser Interferometer Space Antenna (LISA)  \cite{LISA} is expected to observe tens of thousands of compact galactic binaries during its nominal four year 
mission lifetime  \cite{CR17}.  As a matter of fact, because of its high mass,   GW150914 would have been visible to LISA  for several  years prior to their coalescence \cite{Sesana16}. 
  
 Despite significant  investigations, the origins of these binary systems, particularly the  {heavier} BBHs, remain an open question (see, for example, \cite{Wang18} and references 
 therein). Not only from the point of view of theoretical simulations but also from the observational estimates, it is found very difficult to have such heavy black holes (BHs). {On the one hand},  most of BHs obtained from
 numerical simulations have masses lower than $10M_{\bigodot}$, unless the stellar metallicity is very low \cite{Fryer12}.  On the other hand, 
 \"Ozel {\em et al}   examined 16 low-mass X-ray binary systems containing BHs, and found that the masses of BHs hardly exceeded $20 M_{\bigodot}$, 
  and that there was a strongly peaked distribution at $7.8\pm 1.2 M_{\bigodot}$ \cite{Ozel10}.  Similar results were obtained by Farr {\em et al} \cite{Farr11}. 
  
To reconcile the above mentioned problem,   one of the mechanisms \cite{Wang18} is to consider a series mergers of such binary systems in a dense star cluster \cite{SH93,Banerjee17}. 
Assuming that such a merger initially occurred with stellar mass 
 ($\simeq 10 M_{\bigodot}$) BHs, due to the presence of many massive stars/BHs in the dense cluster, the merger product could easily combine with a third massive companion to 
 trigger another merger. One of the interesting properties of this scenario is that such formed BHs usually have very high spins, and can be easily identified with future aLIGO/aVirgo detections
 \cite{ASTA14,LLY15,FHF17,GB17}. Recently, Rodriguez {\em et al}  \cite{RACR18} considered  realistic models of globular clusters with fully post-Newtonian (PN) stellar dynamics for three- and four-body encounters, 
 and found  that nearly half of all binary BH mergers occur inside the cluster, and with about 10 $\%$ of those mergers entering the aLIGO/aVirgo band with eccentricities greater than $0.1$. 
 In particular, in-cluster mergers lead to the birth of a second generation (2G) of BHs with larger masses and high spins. These 2G BHs can reconcile   the upper BH mass limit 
 $\left(\lesssim 50 M_{\bigodot}\right)$ created by   the pair-instability  supernovae \cite{Woosley16}.  
 
 Gravitationally bound hierarchies containing three or more components are very common in our Universe \cite{Naoz16}. 
 Roughly speaking, about $13\%$ of low-mass stelar systems contains three or 
 more stars \cite{FC17}, and $96\%$ of low-mass binaries with periods shorter than 3 days are part of a larger hierarchy \cite{Tok06}. The simplest example is the 3-body system of our Sun, Earth and Moon.
  In fact, any star in the vicinity of a supermassive BH binary  naturally 
 forms a triple system. 
 
 Recently, a realistic triple system   was observed, named as PRS J0337 + 1715 \cite{Ransom14}, which consists of an inner  binary and a third companion.  The inner binary consists 
 of a pulsar  with mass  $m_1 = 1.44 M_{\bigodot}$ and a white dwarf with mass  $m_2 = 0.20 M_{\bigodot}$ in a 1.6 day orbit.  The outer  binary consists of the inner binary and a second dwarf with mass    
 $m_3 = 0.41 M_{\bigodot}$ in a  327 day orbit.  The two orbits are very circular with its eccentricities $e_I \simeq 6.9\times 10^{-4}$ for the inner binary and $e_O \simeq 3.5  \times 10^{-2}$ for the outer orbit.  
 The two orbital planes  are remarkably  coplanar with an inclination  $\lesssim 0.01^{o}$.  A triple system is an ideal place to test the strong equivalence principle \cite{Shao16}. 
 Remarkably, after 6-year observations, lately it  was found that the accelerations of the pulsar and its nearby white-dwarf companion 
  differ fractionally by no more than $2.6 \times 10^{-6}$ \cite{Archibald18}, which provides the most severe constraint on the violation of the strong equivalence principle. 
 
 In a  triple system, the existence of the third companion  can undertake  the Lidov-Kozai oscillations \cite{Lidov61,Kozai62}, and  {cause} the orbit of the inner binary to become nearly radial, whereby
 a rapid merger due to GW emissions can be resulted \cite{ST17,HN18}.   
 Such a  system can emit GWs in the  10 Hz frequency band \cite{Wen03,MKL17,Samsing18},  which are potential sources for the current ground-based detectors, such as  aLIGO, aVirgo and KAGRA \cite{KAGRA}. 
 It   can also emit GWs in the  frequency bands to be detectable  by LISA \cite{AS12}, and pulsar timing arrays \cite{Kocsis12}. In particular, 
  with such a high detectable event rate, it is expected that LISA  will detect many  triple or higher multiple systems \cite{RCTT18}. 
 
 In this paper, we shall study the {\em periodic} gravitational  wave forms, response functions, and energy losses of triple systems in Einstein-aether theory \cite{JM01}. This problem is  interesting particularly for the orbits 
 in which two of the bodies pass each other very closely and yet avoid their collisions, so they can produce periodic gravitational waves with intension, which are the natural sources for the future detections of GWs. 
 Certainly, this problem is also very challenging, as even in Newtonian theory, the systems allow
 chaotic and singular solutions, and only few periodic solutions are known  \cite{MQ14,THZ16} \footnote{The three-body problem can be traced back to Newton in 1680's. In the last 300 years, only three families 
 of periodic solutions were found  \cite{MQ14,THZ16}. In 2013 a breakthrough was made, and 11 new families of Newtonian planner 3-body problem with equal mass and totally zero-angular momentum were
  found numerically in \cite{SD13}.  In 2017,  695 families of such solutions (with equal mass and totally zero-angular momentum) were numerically found in \cite{LL17}.}.  When one of the 3-bodies is a test mass, 
  it reduces to the restricted 3-body problem, and a collinear solution was found by Euler \cite{Euler1767}. In 1772 Lagrange found a second class of periodic orbits for an equilateral triangle configuration 
  \cite{Lagrange1772} (A historic review of the subject can be found in \cite{MQ14}).  
  
  Gravitational wave forms of 3-body systems
  in general relativity (GR) were calculated up to the 1PN approximations with the  {orbits} of the 3-bodies are still Newtonian \cite{THA09,DSH14}.   In GR, neither analytical nor numerical
  solutions of 3-body problem of the full theory have been found, and most of the studies were restricted to PN approximations, see, for example, Refs.\cite{ICA07,LN08,Brum03,Naoz16,BED17,RX18} and references 
  therein. In particular, the 1PN collinear solution was found in \cite{YA10} and proved that it is unique in \cite{YA11}. The 1PN triangular solution and stability were studied,  respectively, in \cite{IYA11,YA12} and \cite{YTA15,YT17}. 
 Lately, the existence and uniqueness of the 1PN collinear solution in the scalar-tensor theory were studied in \cite{ZCX16,CZX17}. 
  
  In the framework of Einstein-aether theory, Foster \cite{Foster07} and Yagi {\em et al} \cite{Yagi14} derived the metric and equations of motion to the 1PN order for a N-body system. Recently, Will applied them to study
  the 3-body problem and obtained the accelerations of a 2-body system in the presence of the third body at the quasi-Newtonian order \cite{Will18}. For nearly circular coplanar orbits, he also calculated the 
  ``strong-field" Nordtvedt parameter $\hat{\eta}_{N}$. For the  PRS J0337 + 1715 system, ignoring the sensitivities of the 
 two white-dwarf companions,  Will found that  $\hat{\eta}_{N}$   is given by $\hat{\eta}_{N} = s_{1}/(1-s_1)$, where $s_1$ denotes the sensitivity of the pulsar. 
 
 In this paper, we shall focus ourselves on   periodic GWs  
  in the framework of Einstein-aether theory. The theory breaks locally the Lorentz symmetry by the presence of a globally time-like unit vector field - the aether,
  and  allows three different  types of gravitational modes, the scalar, vector and tensor  \cite{JM04}, and
 all the modes in principle move at different speeds \cite{Jacobson}. Recently, it was found \cite{OMW18} that the four independent coupling constants of the theory must satisfy the constraints of Eq.(\ref{2.8a})  given below, after
 several conditions are imposed. In the vacuum, GWs were also studied in \cite{JM04,GHLP18}, while GW forms and angular momentum loss were studied for binary systems in  \cite{HYY15,SY18},  respectively. 
 
The rest of the paper  is organized as follows:  in Sec. II we give a brief introduction to the Einstein-aether theory, and in Sec. III we first study the 
effects of the gravitational radiations from the scalar and vector modes to the energy loss for a binary system to the lowest PN order \cite{Foster06}, 
 and find that for a neutron star binary system  their contributions to the quadrupole, monopole and dipole are, respectively, the orders of ${\cal{O}}\left(10^{-5}\right),\; {\cal{O}}\left(10^{-5}\right)$ and ${\cal{O}}\left( {10^{-2}}\right)$ 
 lower than   the quadrupole contributions of GR (in which the scalar and vector modes are absent)  [cf. Eq.(\ref{4.1kb})], while the strong field
 effects are, respectively, the orders of ${\cal{O}}\left( {10^{-6}}\right),\; {\cal{O}}\left( {10^{-6}}\right)$ and ${\cal{O}}\left( {10^{-7}}\right)$ lower.  Additionally, the order for the cross term is of ${\cal{O}}\left( {10^{-6}}\right)$ lower  
  [cf. Eq.(\ref{4.1y})]. Similar conclusions can  {also be} obtained by analyzing the amplitudes of polarization modes of a binary system with non-vanishing sensitivities given in \cite{HYY15}.   Therefore, to the current (second) generation of GW detectors  
 \cite{Schutz18}, we can safely ignore these strong field effects. Then,  in Sec. IV we  consider the lowest PN approximations by taking the 
 coupling of the aether with matter fields into account. When such couplings  are turned off, our formulas reduce to the ones presented in \cite{Foster06}, subjected to some corrections of typos.
 From the general expressions for the polarization modes $h_{N}$ given by Eq.(\ref{polarizationsB}), we can see that  the scalar breather  and the scalar  longitudinal modes are always proportional to each other, so
  only five of the six polarization modes are independent. In addition, the two scalar modes are  all suppressed by a factor of ${\cal{O}}\left(c_{14}\right) \lesssim {\cal{O}}\left(10^{-5}\right)$ 
with respect to the transverse-traceless modes ($h_{+}$ and $h_{\times}$), while the vectorial modes $(h_{X}$ and $h_{Y}$) are suppressed by a factor of 
${\cal{O}}\left(c_{13}\right) \lesssim {\cal{O}}\left(10^{-15}\right)$.  In Sec. V, we apply these formulas to
 triple systems and obtain the GW forms, response functions, their  {Fourier transforms},  and energy losses for three representative cases. Our results show that the GW forms sensitively depend on not only the configurations of the 3-body
 orbits, but also their relative positions to the detectors, sharply in contrast to the 2-body problem \cite{Mag08,PW14}.  
 Our paper is ended with Sec. VI, in which we present our main  conclusions.

\section{Einstein-Aether Theory}
 \renewcommand{\theequation}{2.\arabic{equation}} \setcounter{equation}{0}

In  Einstein-aether ($\ae$-) theory, the fundamental variables of the gravitational  sector are \cite{JM01},
\bq
\lb{2.0a}
\left(g_{\mu\nu}, u^{\mu}, \lambda\right),
\eq
with the Greek indices $\mu,\nu = 0, 1, 2, 3$, and $g_{\mu\nu}$ is  the four-dimensional metric  of the space-time
with the  {signature} $(-, +,+,+)$ \cite{Foster06,GEJ07},  $u^{\mu}$  {is} the aether four-velocity, and $\lambda$ is a Lagrangian multiplier, which guarantees that the aether  four-velocity  is always timelike.
{In this paper, we will adopt the following conventions: all the repeated indices  $i, j, k, l \; (i, j, k, l = 1, 2, 3)$ will be summed over regardless they are up or down, but, repeated indices $a, b, c \; (a, b, c = 1, 2, 3)$ 
will not be summed over, unless  the summation is explicitly indicated.}  {In this paper, we also adopt units so that  the   speed of light is one ($c=1$).}
The general action of the theory  is given  by  \cite{Jacobson},
\bq
\lb{2.0}
S = S_{\ae} + S_{m},
\eq
where  $S_{m}$ denotes the action of matter,  and $S_{\ae}$  the gravitational action of the $\ae$-theory, given,  respectively, by
\bqn
\lb{2.1}
 S_{\ae} &=& \frac{1}{16\pi G_{\ae} }\int{\sqrt{- g} \; d^4x \Big[R(g_{\mu\nu}) + {\cal{L}}_{\ae}\left(g_{\mu\nu}, u^{\alpha}, {\lambda}\right)\Big]},\nb\\
S_{m} &=& \int{\sqrt{- g} \; d^4x \Big[{\cal{L}}_{m}\left(g_{\mu\nu}, u^{\alpha}; \psi\right)\Big]}.
\eqn
Here $\psi$ collectively denotes the matter fields, $R$    and $g$ are, respectively, the  Ricci scalar and determinant of $g_{\mu\nu}$,
 and
\bq
\lb{2.2}
 {\cal{L}}_{\ae}  \equiv - M^{\alpha\beta}_{~~~~\mu\nu}\left(D_{\alpha}u^{\mu}\right) \left(D_{\beta}u^{\nu}\right) + \lambda \left(g_{\alpha\beta} u^{\alpha}u^{\beta} + 1\right),
\eq
 where $D_{\mu}$ denotes the covariant derivative with respect to $g_{\mu\nu}$, and  $M^{\alpha\beta}_{~~~~\mu\nu}$ is defined as
\bqn
\lb{2.3}
M^{\alpha\beta}_{~~~~\mu\nu} \equiv c_1 g^{\alpha\beta} g_{\mu\nu} + c_2 \delta^{\alpha}_{\mu}\delta^{\beta}_{\nu} +  c_3 \delta^{\alpha}_{\nu}\delta^{\beta}_{\mu} - c_4 u^{\alpha}u^{\beta} g_{\mu\nu}.\nb\\
\eqn
Note that here we assume that matter fields couple not only to $g_{\mu\nu}$ but also to the aether field $u^{\mu}$,  in order to model effectively the radiation of a compact object, such as a neutron star \cite{Ed75}.
The four coupling constants $c_i$'s are all dimensionless, and $G_{\ae} $ is related to  the Newtonian constant $G_{N}$ via the relation \cite{CL04},
\bq
\lb{2.3a}
G_{N} = \frac{G_{\ae}}{1 - \frac{1}{2}c_{14}},
\eq
where $c_{ij}\equiv c_i +c_j$.

The variations of the total action with respect to $g_{\mu\nu},\; u^{\mu}$ and $\lambda$ yield, respectively, the field equations,
 \bqn
 \lb{2.4a}
 R^{\mu\nu} - \frac{1}{2} g^{\mu\nu}R - S^{\mu\nu} &=& 8\pi G_{\ae}  T^{\mu\nu},\\
 \lb{2.4b}
  \AE_{\mu} &=& 8\pi G_{\ae}  T_{\mu}, \\
   \lb{2.4c}
  g_{\alpha\beta} u^{\alpha}u^{\beta} &=& -1, 
 \eqn
where
 \bqn
 \lb{2.5}
  S_{\alpha\beta} &\equiv&
  D_{\mu}\Big[J^{\mu}_{\;\;\;(\alpha}u_{\beta)} + J_{(\alpha\beta)}u^{\mu}-u_{(\beta}J_{\alpha)}^{\;\;\;\mu}\Big]\nb\\
&& + c_1\Big[\left(D_{\alpha}u_{\mu}\right)\left(D_{\beta}u^{\mu}\right) - \left(D_{\mu}u_{\alpha}\right)\left(D^{\mu}u_{\beta}\right)\Big]\nb\\
&& + c_4 a_{\alpha}a_{\beta}    + \lambda  u_{\alpha}u_{\beta} - \frac{1}{2}  g_{\alpha\beta} J^{\delta}_{\;\;\sigma} D_{\delta}u^{\sigma},\nb\\
 \AE_{\mu} & \equiv &
 D_{\alpha} J^{\alpha}_{\;\;\;\mu} + c_4 a_{\alpha} D_{\mu}u^{\alpha} + \lambda u_{\mu},\nb\\
  T^{\mu\nu} &\equiv&  \frac{2}{\sqrt{-g}}\frac{\delta \left(\sqrt{-g} {\cal{L}}_{m}\right)}{\delta g_{\mu\nu}},\nb\\
T_{\mu} &\equiv& - \frac{1}{\sqrt{-g}}\frac{\delta \left(\sqrt{-g} {\cal{L}}_{m}\right)}{\delta u^{\mu}},
 \eqn
 with
\begin{equation}
 \lb{2.6}
J^{\alpha}_{\;\;\;\mu} \equiv M^{\alpha\beta}_{~~~~\mu\nu}D_{\beta}u^{\nu}\,,\quad
a^{\mu} \equiv u^{\alpha}D_{\alpha}u^{\mu}.
\end{equation}
From Eq.(\ref{2.4b}),  we find that
\bq
\lb{2.7}
\lambda = u_{\beta}D_{\alpha}J^{\alpha\beta} + c_4 a^2 - 8\pi G_{\ae}  T_{\alpha}u^{\alpha},
\eq
where $a^{2}\equiv a_{\lambda}a^{\lambda}$.

 It is easy to show that the Minkowski spacetime  is a solution of the Einstein-aether theory, in which the aether is aligned along the time direction, $\bar{u}_{\mu} = \delta^{0}_{\mu}$. 
 Then, the linear perturbations around the Minkowski background show that the theory in general possess three types of excitations, scalar, vector and tensor modes  \cite{JM04}, with their squared  {speeds given,  respectively, by}
\begin{eqnarray}
 \label{CSs}
 c_S^2 & = & \frac{c_{123}(2-c_{14})}{c_{14}(1-c_{13}) (2+c_{13} + 3c_2)}\,,\nonumber\\
 c_V^2 & = & \frac{2c_1 -c_{13} (2c_1-c_{13})}{2c_{14}(1-c_{13})}\,,\nonumber\\
 c_T^2 & = & \frac{1}{1-c_{13}},
\end{eqnarray}
where $c_{ijk} \equiv c_i + c_j + c_k$.  

In addition, among the 10 parameterized post-Newtonian (PPN) parameters \cite{Will06},  in the Einstein-aether theory the only two parameters 
that deviate from GR are $\alpha_1$ and $\alpha_2$, which measure the preferred frame effects. In terms of the four dimensionless coupling constants $c_i$'s, they are given by \cite{FJ06},
\bqn
\lb{2.3aa}
\alpha_1 &=& -  \frac{8\left(c_1c_{14} - c_{-}c_{13}\right)}{2c_1 - c_{-}c_{13}}, \nb\\
\alpha_2 &=&   \frac{1}{2}\alpha_1  + \frac{\left(c_{14}- 2c_{13}\right)\left(3c_2+c_{13}+c_{14}\right)}{c_{123}(2-c_{14})},~~~~~~~~
\eqn
where $c_{-} \equiv c_1 - c_3$.   In the weak-field regime, using lunar laser ranging and solar alignment with the ecliptic, 
Solar System observations constrain these parameters to very small values \cite{Will06},
\bq
\lb{CD5}
\left| \alpha_1\right| \le 10^{-4}, \quad 
 \left|\alpha_2\right| \le 10^{-7}.
 \eq

Recently,   the combination of the gravitational wave event GW170817 \cite{GW170817}, observed by the LIGO/Virgo collaboration, and the event of the gamma-ray burst
GRB 170817A \cite{GRB170817} provides  a remarkably stringent constraint on the speed of the spin-2 mode, 
\bq
\lb{CD6}
- 3\times 10^{-15} < c_T -1 < 7\times 10^{-16},
\eq
which, together with Eq.(\ref{CSs}), implies that 
\bq
\lb{2.8a}
\left |c_{13}\right| < 10^{-15}.
 \eq
 
 Imposing further the following conditions: (a) the theory is free of ghosts; (b) the squared speeds $c_I^2\; (I = S, V, T)$ must be non-negative; (c) $c_{I}^2-1$ must be greater than $-10^{-15}$ or so, 
 in order to avoid the existence of the vacuum gravi-\v{C}erenkov radiation by matter such as cosmic rays \cite{EMS05}; and (d) the theory must be consistent with the current observations on the primordial 
 helium abundance $\left|G_{cos}/G_{N} - 1\right| \lesssim 1/8$, where $G_{cos} \equiv G_{\ae}/(1+ (c_{13} + 3c_2)/2)$ \cite{CL04}, together with Eqs.(\ref{CD5}) and (\ref{2.8a}), 
 it was found that   the parameter space of the theory is restricted to \cite{OMW18}, 
 \bq
\lb{2.8b}
0 \lesssim c_{14} \lesssim c_{1}, \quad  c_{14} \lesssim c_2 \lesssim 0.095, \quad  c_{14} \lesssim 2.5\times 10^{-5}.
 \eq
 
Note that the above limits do not include the strong-field constraints \cite{SW13},
\bq
\lb{CD7}
\left|\hat\alpha_1\right| \le 10^{-5}, \quad 
 \left|\hat\alpha_2\right| \le 10^{-9},\; (\mbox{at $95\%$ confidence}),
 \eq
obtained from  the isolated millisecond pulsars PSR B1937 + 21 \cite{SPulsarA} and PSR J17441134  \cite{SPulsarB}, 
where ($\hat\alpha_1, \hat\alpha_2$) denotes the strong-field generalization of  ($\alpha_1, \alpha_2$)  \cite{DEF92}, because they depend on the sensitivity    $\sigma_{\ae}$, which  is not known for the new constraints of Eq.(\ref{2.8b}).
In fact, in the Einstein-$\ae$ther theory, they are given by \cite{Yagi14}, 
\bqn
\lb{2.3ac}
\hat\alpha_1 &=& \alpha_1 + \frac{c_-(8+\alpha_1)\sigma_{\ae}}{2c_1}, \nb\\
\hat\alpha_2&=& \alpha_2 + \frac{\hat\alpha_1 - \alpha_1}{2}   
- \frac{(c_{14} -  2)(\alpha_1 -2\alpha_2)\sigma_{\ae}}{2(c_{14} - 2c_{13})}.~~~~~
\eqn
 For details, we refer interested readers to \cite{OMW18}.

\section{Effects of the Aether field on Energy Loss Rate}  
 \renewcommand{\theequation}{3.\arabic{equation}} \setcounter{equation}{0}

  Before proceeding further, let us pause here for a while to consider the effects of the aether field on energy loss of a given system,
   in order to have a better understanding of the approximations to be taken in this paper. Although in this section we shall restrict ourselves only to  binary
  systems, we believe that such estimations are also valid for other systems, as long as they are weak enough, so the PN approximations are applicable. 
  In particular, our studies of triple systems   are consistent   with such estimations as to be shown below.   
  
  To the lowest PN order (by setting the sensitivity $\sigma_{\ae} = 0$), Foster found that the energy loss rate in Einstein-{\AE}ther 
  gravity is given by \cite{Foster06} (See also \cite{Yagi14} for the correction of a typo in the expression of $Z$ given below.), 
  \bqn
 \lb{4.1a}
\dot{\cal E}=-G_N\left\langle \frac{\cal A}{5}\dddot{Q}_{ij}\dddot{Q}_{ij}+{\cal B}\dddot{I}\dddot{I}+{\cal C}\dot{\Sigma}_i\dot{\Sigma}_i\right\rangle,
 \eqn
where
\bqn
\lb{4.1b}
Q_{ij} &\equiv& I_{ij} - \frac{1}{3}I \delta_{ij}, \quad I_{ij} \equiv \int{d^3x \; \rho x_i x_j}, \nb\\
I &\equiv& I_{ii}, \quad \Sigma_i \equiv  \int{d^3x \; t_{i}},
\eqn
and
\bqn
\lb{4.1c}
\cal A&=&\left(1-\frac{c_{14}}{2}\right)\left(\frac{1}{c_T}+\frac{2c_{14}c_{13}^2}{(2c_1-c_{13} c_-)^2c_V}\right.\nb\\
&&\left.+\frac{3(Z-1)^2 c_{14}}{2(2-c_{14})c_S}\right),\nb\\
\cal B&=&\frac{Z^2 c_{14}}{8c_S}, \quad Z \equiv \frac{(\alpha_1 - 2\alpha_2)(1-c_{13})}{3(2c_{13} - c_{14})}, \nb\\
\cal C&=&\frac{2}{3 c_{14}}\left(\frac{2-c_{14}}{c_V^3}+\frac{1}{c_S^3}\right).
\eqn
Here $\rho = T_{00}$ to the lowest order, and $t_{i}$ denotes the quadratic terms given in Eq.(\ref{3.6b}). 

It should be noted that the scalar (monopole) perturbations have contributions to all the three parts, quadrupole ($\dddot{Q}_{ij}$), dipole ($\dot{\Sigma}_{i}$) and monopole
($\dddot{I}$). The vector (dipole) perturbations  have contributions to both quadrupole   and dipole   terms, while the tensor perturbations have only
contributions to the quadrupole term. This can be seen clearly from the expressions for $\int{d\Omega \dot{\phi}_{ij}\dot{\phi}_{ij}}$, $\int{d\Omega \dot{\nu}_{i}\dot{\nu}_{i}}$
and $\int{d\Omega \dot{F}\dot{F}}$, given by Eqs.(102)-(104) in \cite{Foster06}, where   $F \equiv \Delta f$ and $\phi_{ij},\; \nu_i$ and $f$ are all defined in Eq.(\ref{3.2}) below. 

 In the case of GR ($c_i = 0$), we have,
\bq
\lb{4.1d}
 {\cal{A}}^{GR} = 1, \quad {\cal{B}}^{GR} = {\cal{C}}^{GR} = 0. 
 \eq
To see  clearly the effects of the aether field, in the rest of this section let us restrict ourselves  only to a binary system, for which Eq.(\ref{4.1a}) takes the form,
 \bqn
 \lb{4.1e}
\dot{\cal E}&=&-G_N\left\langle \left(\frac{G_N \mu m}{r^2}\right)^2\Bigg[\frac{8}{15}{\cal A}\left(12 v^2 -11 \dot{r}^2\right) \right.\nb\\
&& ~~~~~~~~~ \left. + 4 {\cal{B}}\dot{r}^2 +  {\cal C} \Sigma^2\Bigg]\right\rangle,
 \eqn
 where $m \equiv m_1 + m_2$, $\mu \equiv m_1m_2/m$,  ${\bf r} \equiv {\bf r}_1 - {\bf r}_2$, ${\bf v} \equiv \dot{\bf r}$, $ r \equiv |{\bf r}|$, and
 \bqn
 \lb{4.1f}
  {\Sigma \equiv  \left(\alpha_1 - \frac{2}{3}\alpha_2\right) \left(\frac{\Omega_1}{m_1} - \frac{\Omega_2}{m_2}\right)},
 \eqn
 with $\Omega_a$ denoting the binding energy of the a-th compact body, and
  \bqn
 \lb{4.1g}
 \frac{\Omega_a}{m_a}  \simeq {\cal{O}}\left(\frac{G_N m_a}{d_a}\right) \simeq   
 \begin{cases}
 10^{-3}, & {\mbox{white dwarfs}},\cr
 0.1 \sim 0.3, &  {\mbox{pulsars}},\cr
 \end{cases}
 \eqn
 where $d_a$ denotes the size of the a-th body. 
 
 For double  {pulsars}, we have $v^2 \simeq 10^{-6} \sim 10^{-5}$ \cite{Stairs03}. In addition, without loss of the generality, 
 we assume that $v^2$ and $\dot{r}^2$ are of the same order, ${\cal{O}}(v^2) \simeq {\cal{O}}(\dot{r}^2)$. Then, from the constraint $|c_{13}| < 10^{-15}$ [cf.(\ref{2.8a})], 
 we find that 
 \bq
 \lb{4.1h}
 c_T \simeq 1, \quad c_{V} \simeq {\cal{O}}\left(\frac{c_1}{c_{14}}\right)^{1/2}, \quad c_{S} \simeq {\cal{O}}\left(\frac{c_2}{c_{14}}\right)^{1/2}.
 \eq
 Hence, we obtian
 \bqn
 \lb{4.1i}
 {\cal{A}} &=& 1 +  {\cal{O}}\left(c_{14}\right)\Big[ {1} + {\cal{O}}\left(c_{S}^{-5}\right)\Big], \nb\\
  {\cal{B}} &=&   {\cal{O}}\left(c_{14}\right) {\cal{O}}\left(c_{S}^{-1}\right), \nb\\
   {\cal{C}} &=&   {\cal{O}}\left(c_{14}^{-1}\right)\left[{\cal{O}}\Big(c_{V}^{-3}\right) + {\cal{O}}\left(c_{S}^{-3}\right)\Big],
 \eqn
 where $c_{I} \gtrsim 1 \; (I = S, V, T)$ because of the  vacuum gravi-\v{C}erenkov effects \cite{EMS05}. Then,  {comparing Eq.(\ref{4.1d}) and 
 Eq.(\ref{4.1i}),} we can see that the contributions of the aether field to both of the quadrupole and monopole radiations  are
 at most of the order of $ {\cal{O}}\left(c_{14}\right) \lesssim 10^{-5}$. To see its contributions to the dipole radiation, we first note that
 \bqn
 \lb{4.1j}
&& \alpha_1 - \frac{2}{3}\alpha_2 \lesssim  {\cal{O}}\left(c_{14}\right),\nb\\
&& \Sigma \lesssim  {\cal{O}}\left( {c_{14}}\right) {\cal{O}}\left(\frac{G_N m}{d}\right).  
 \eqn
 Thus, we find that 
 \bqn
 \lb{4.1k}
{\cal{W}}_{\cal{C}} \equiv {\cal{C}}\Sigma^2   &\lesssim&  {\cal{O}}\left( {c_{14}}\right)  {\cal{O}}\left(\frac{G_N m}{d}\right)^2\nb\\
& \lesssim& 10^{ {-5}} \; {\cal{O}}\left(\frac{G_N m}{d}\right)^2.
 \eqn
 On the other hand,  the quadrupole and monopole radiations are given, respectively, by
  \bqn
 \lb{4.1ka}
{\cal{W}}_{\cal{A}} &\equiv&  \frac{8}{15}{\cal{A}}\left(12v^2 - 11\dot{r}^2\right)  {\lesssim}  \Big[1 + {\cal{O}}\left(c_{14}\right)\Big] {\cal{O}}\left(v^2\right),\nb\\
{\cal{W}}_{\cal{B}} &\equiv& 4{\cal{B}} \dot{r}^2 \lesssim {\cal{O}}\left(c_{14}\right) \;  {\cal{O}}\left(v^2\right).
 \eqn

 In particular, for a binary system of neutron stars,  taking $ {\cal{O}}\left(v^2\right) \simeq 10^{-5}$ and $ {\cal{O}}\left(\frac{G_N m}{d}\right) \simeq 10^{-1}$, we find that 
   \bqn
 \lb{4.1kb}
{\cal{W}}_{\cal{A}}^{\mathrm{NS}} &\simeq&  \Big[1 + {\cal{O}}\left(10^{-5}\right)\Big]{\cal{O}}\left(10^{-5}\right),\nb\\
{\cal{W}}_{\cal{B}}^{\mathrm{NS}} &\lesssim& {\cal{O}}\left(10^{-10}\right),\quad
{\cal{W}}_{\cal{C}}^{\mathrm{NS}} \lesssim  {\cal{O}}\left(10^{ {-7}}\right).
 \eqn

 \subsection{Strong Field Effects }

Strong field effects can be important  in the vicinity of the compact bodies, such as neutron stars, as the fields inside such bodies  can be  very strong.
Following Eardley \cite{Ed75},  these effects  can be included by considering the action of one-particle \cite{Foster07},
\bqn
\lb{4.1l}
S_A &=&  -\int{d\tau_A  \tilde{m}_A[ {\gamma_A}]} \nb\\
&=&  - \tilde{m}_A\int{d\tau_A \Bigg[1 + \sigma_A (1- {\gamma_A}) }\nb\\
&& + \frac{1}{2}(\sigma_A + \sigma_A^2 + \bar{\sigma}_A)(1- {\gamma_A})^2 + ...\Big].
\eqn
Here  $ {\gamma_A} = -u^{\mu} v^A_{\mu}$, where $v^A_{\mu}$ is the four-velocity of the body, and $A$ labels the body, $\tau_A$ is the proper time along the body's curve. 
When the  {body} is at rest with respect to the aether, we have $ {\gamma_A}|_{v^A_{\mu} = u_{\mu}} = 1$ and $ \tilde{m}_A =  \tilde{m}_A[ {\gamma_A}]|_{ {\gamma_A}=1}$. The sensitivities $\sigma_A$ and $\bar\sigma_A$ are defined as,
\bqn
\lb{4.1m}
\sigma_A  &\equiv& - \left.\frac{d\ln\tilde{m}_A[ {\gamma_A}]}{d\ln {\gamma_A}}\right|_{ {\gamma_A} = 1},\nb\\
\bar\sigma_A  &\equiv&  \left.\frac{d^2\ln\tilde{m}_A[ {\gamma_A}]}{d\ln {\gamma_A}^2}\right|_{ {\gamma_A} = 1},
\eqn
which can be determined by considering asymptotic properties of perturbations of static stellar configurations. Indeed, $\sigma_A$ was calculated for neutron stars in \cite{Yagi14}. But, unfortunately, the calculations were
done by setting $\alpha_1 = \alpha_2 = 0$, which are no longer valid when the new constraint (\ref{2.8a}) is taken into account \footnote{Note that from Eq.(\ref{2.3aa}) 
we can write $c_2$ and $c_4$ in terms of $\alpha_1$ and $\alpha_2$. Then, with the new constraint (\ref{2.8a}), to have a self-consistent expansion of any given function
$F(c_1, c_2, c_3, c_{4}) \equiv {\cal{F}}(c_1, \alpha_1, \alpha_2, c_{13})$ in terms of the small quantities $\alpha_1, \; \alpha_2$ and $c_{13}$, one must expand $ {\cal{F}}(c_1, \alpha_1, \alpha_2, c_{13})$  at least to the third-order of $\alpha_1$,  
the second-order of $\alpha_2$ (plus their mixed terms, such as $\alpha_1^2\alpha_2$), and to the first-order of $c_{13}$, considering the fact that the constraints (\ref{CD5}) and (\ref{2.8a}) are in different orders.    
If one naively sets $\alpha_1,\; \alpha_2,\; c_{13}$ all to zero at the same time, then it will be ended up with $c_2 = 0$, $c_3 = c_4 = - c_1$ and $\left|c_1\right| \le 10^{-15}$, 
which are not only too strict, but also inconsistent, as this is equivalent to assume that these three quantities  were
constrained all to the same order.}. However, to our current purpose, the exact values of $\sigma_A$ is not important, and we can
simply use the expression of the small $\sigma_A$ limit \cite{Foster07,Yagi14},
\bq
\lb{4.1n}
s_A =  {\left(\alpha_1 - \frac{2}{3}\alpha_2\right)\frac{\Omega_A}{m_A}} + {\cal{O}}\left(\frac{G_N m}{d}\right)^2,
\eq
where $s_A \equiv \sigma_A/(1+\sigma_A)$. 

After taking the strong field effects into account, Eq.(\ref{4.1e}) got four different kinds of corrections, one to each of the three terms presented in Eq.(\ref{4.1e}), plus a crossing term. It is given explicitly by 
Eq.(116) in \cite{Yagi14} \footnote{In Ref.\cite{Foster07}, it is given by Eq.(89). However, some typos appeared in this equation,  and were corrected in  \cite{Yagi14}.}. In the following, let us consider these corrections 
term by term. 
First, to the first term of Eq.(\ref{4.1e}), the coefficient ${\cal{A}}$ is replaced by $({\cal{A}} + {\cal{S}}{\cal{A}}_2 +  {\cal{S}}^2{\cal{A}}_3)$, where
\bqn
\lb{4.1o}
 {\cal{S}} \equiv s_1\mu_2 +  {s_2\mu_1}, 
 \eqn
with $\mu_A = m_A/m \simeq {\cal{O}}(1)$. Thus, we have 
\bqn
\lb{4.1p}
 {\cal{S}}  \simeq {\cal{O}}\left(s_A\right) \lesssim   {\cal{O}}\left( {c_{14}}\right) \; {\cal{O}}\left(\frac{G_N m}{d}\right).
 \eqn
In addition, we also have, 
\bqn
\lb{4.1q}
Z &\simeq&  {\cal{O}}(1),\nb\\
{\cal{A}}_2 &\equiv& \frac{2(Z-1)}{(c_{14} -2)c_S^3} + \frac{2c_{13}}{(2c_1 - c_{-}c_{13}) c_V^3}\nb\\
  &\simeq& {\cal{O}}\left(c_S^{ {-5}}\right),\nb\\
{\cal{A}}_3 &\equiv& \frac{1}{2c_{14}c_V^5} + \frac{2}{3c_{14}(2 - c_{14}) c_S^5}\nb\\
&\simeq&  {\cal{O}}(c_{14}^{-1})\Big[{\cal{O}}\left(c_V^{-5}\right)+{\cal{O}}\left(c_S^{-5}\right)\Big].
\eqn
Thus, we find that the correction to the quadrupole term is given by
\bqn
\lb{4.1r}
\delta{\cal{W}}_{\cal{A}}   &\equiv& \left({\cal{S}}{\cal{A}}_2 +  {\cal{S}}^2{\cal{A}}_3\right)\left(12v^2 - 11\dot{r}^2\right) \nb\\
&\simeq&  {\cal{O}}( {c_{14}})\Big[{\cal{O}}\left(c_V^{-5}\right)+{\cal{O}}\left(c_S^{-5}\right)\Big] {\cal{O}}\left(\frac{G_N m}{d}\right)^2  {\cal{O}}\left(v^{2}\right)  \nb\\
&& + {\cal{O}}( {c_{14}})  {\cal{O}}\left(c_S^{ {-5}}\right)  {\cal{O}}\left(\frac{G_N m}{d}\right)\; {\cal{O}}\left(v^{2}\right)\nb\\
&\lesssim& {\cal{O}}( {c_{14}})    {\cal{O}}\left(\frac{G_N m}{d}\right) \; {\cal{O}}\left(v^{2}\right)\nb\\
& \lesssim& 10^{ {-5}}\;    {\cal{O}}\left(\frac{G_N m}{d}\right) \; {\cal{O}}\left(v^{2}\right).
\eqn

Second, the correction to the coefficient  ${\cal{B}}$ is $({\cal{S}}{\cal{B}}_2 +  {\cal{S}}^2{\cal{B}}_3)$, where
\bqn
\lb{4.1s}
{\cal{B}}_2 &\equiv& \frac{Z}{3(c_{14}-2)c_S^3} \simeq  {\cal{O}}\left(c_S^{-3}\right),\nb\\
{\cal{B}}_3 &\equiv& \frac{1}{9c_{14}(2-c_{14})c_S^5} \simeq  {\cal{O}}(c_{14}^{-1}) \; {\cal{O}}\left(c_S^{-5}\right).
\eqn
Therefore, we have
\bqn
\lb{4.1t}
\delta{\cal{W}}_{\cal{B}}   &\equiv& 4\left({\cal{S}}{\cal{B}}_2 +  {\cal{S}}^2{\cal{B}}_3\right)\dot{r}^2\nb\\
& {\lesssim}&  \Bigg[{\cal{O}}( {c_{14}}) {\cal{O}}\left(c_S^{-3}\right)  {\cal{O}}\left(\frac{G_N m}{d}\right)  \nb\\
&& +{\cal{O}}( {c_{14}}) {\cal{O}}\left(c_S^{-5}\right)  {\cal{O}}\left(\frac{G_N m}{d}\right)^2\Bigg]  {\cal{O}}\left(v^{2}\right)\nb\\
&\lesssim& {\cal{O}}( {c_{14}})   \; {\cal{O}}\left(\frac{G_N m}{d}\right)\;  {\cal{O}}\left(v^{2}\right)\nb\\
& \lesssim& 10^{ {-5}}\;    {\cal{O}}\left(\frac{G_N m}{d}\right)\;  {\cal{O}}\left(v^{2}\right),
\eqn
which has the same order as that of $\delta {{\cal{W}}_{\cal{A}}}$, as shown by Eq.(\ref{4.1r}). 

For the third   kind of corrections, they are involved with the velocity $V^i$ of the center-of-mass and $V^in^i$, where 
$n^i$ is the unity vector, defined by $n^i \equiv r^i/r$ {, where $r$ is the module of the vector $r^i$}. As pointed out by Foster in \cite{Foster07}, $V^i$ is not directly measurable, but the validity of leading PN order for 
the double pulsar \cite{Kramer06} requires $V^2 \lesssim 3 v^2$, while for other systems, they require  $V^2 \lesssim {\cal{O}}(100) v^2$. In any case, without loss of the generality,
we assume that ${\cal{O}}\left(V^i n^i\right)^2 \simeq {\cal{O}}\left(V^2\right)$. In addition,
\bqn
\lb{4.1u}
{\cal{C}}_2 &\equiv&  \frac{1}{6c_{14}c_V^5} \lesssim {\cal{O}}(c_{14}^{-1}) \; {\cal{O}}\left(c_V^{-5}\right).
\eqn
Thus, we find that 
\bqn
\lb{4.1v}
\delta{\cal{W}}_{\cal{C}}   &\equiv& \left(s_1-s_2\right)^2 \Bigg[\left(\frac{18}{5}{\cal{A}}_3 + 2{\cal{C}}_2\right)V^2 \nb\\
&& + \left(\frac{6}{5}{\cal{A}}_3 + 36{\cal{B}}_3 - 2{\cal{C}}_2\right)\left(V^in^i\right)^2\Bigg]\nb\\
& {\lesssim}& {{\cal{O}}(c_{14}) \Bigg[{\cal{O}}\left(c_S^{-3}\right)+{\cal{O}}\left(c_V^{-3}\right)\Bigg]}\nb\\
&& \times  {\cal{O}}\left(\frac{G_N m}{d}\right)^2  {\cal{O}}\left(V^2\right)\nb\\
&\lesssim&   {{\cal{O}}(c_{14}) \;  {\cal{O}}\left(\frac{G_N m}{d}\right)^2 \;  {\cal{O}}\left(V^2\right)}\nb\\
&\lesssim& 10^{ {-5}}  \;  {\cal{O}}\left(\frac{G_N m}{d}\right)^2 \;  {\cal{O}}\left(V^2\right).
\eqn

The fourth kind of corrections is involved with the crossing terms, $(V^i n^i v^j n^j)$ and $(v^iV^i)$. Again, without loss of the generality,
we assume that $ {\cal{O}}(V^i n^i v^j n^j) \simeq {\cal{O}}\left(V^i v^i\right)$. Then, the fourth kind of
corrections is given by
\bqn
\lb{4.1x}
\delta{\cal{W}}_{\cal{D}}   &\equiv& \left(s_1-s_2\right) \Bigg[12\left({\cal{B}}_2 + 2{\cal{S}} {\cal{B}}_3\right)(V^i n^i v^j n^j) \nb\\
&& + \frac{8}{5}\left({\cal{A}}_2 +  2{\cal{S}} {\cal{A}}_3\right)V^i\left(3v^i -2 n^i v^jn^j\right)\Bigg]\nb\\
& {\lesssim}& {\cal{O}}\left( {c_{14}}\right) {\cal{O}}\left(\frac{G_N m}{d}\right)  {\cal{O}}\left(v^iV^i\right)\Bigg\{{\cal{O}}\left(c_S^{-3}\right)\nb\\
&& + {\cal{O}}\left(c_{14}\right){\cal{O}}\left(\frac{G_N m}{d}\right)\Big[{\cal{O}}\left(c_S^{-5}\right)+{\cal{O}}\left(c_V^{-5}\right)\Big]\Bigg\}\nb\\
&\lesssim& {\cal{O}}\left( {c_{14}}\right) {\cal{O}}\left(\frac{G_N m}{d}\right)  {\cal{O}}\left(v^iV^i\right)\nb\\
&\lesssim&10^{ {-5}} \; {\cal{O}}\left(\frac{G_N m}{d}\right)  {\cal{O}}\left(v^iV^i\right).
\eqn
 
 For the binary system of neutron stars, taking
${\cal{O}}\left(v^iV^i\right) \simeq  {\cal{O}}\left(v^2\right)  \simeq  {\cal{O}}\left(V^2\right) \simeq  {\cal{O}}\left(10^{-5}\right)$ and $ {\cal{O}}\left({G_N m}/{d}\right) \simeq 10^{-1}$, we find that 
\bqn
\lb{4.1y}
\delta{\cal{W}}^{\mathrm{NS}}_{\cal{A}}    &\lesssim& {\cal{O}}\left(10^{ {-11}}\right), \quad
\delta{\cal{W}}^{\mathrm{NS}}_{\cal{B}}    \lesssim {\cal{O}}\left(10^{ {-11}}\right),\nb\\
\delta{\cal{W}}^{\mathrm{NS}}_{\cal{C}}    &\lesssim& {\cal{O}}\left(10^{ {-12}}\right), \quad
\delta{\cal{W}}^{\mathrm{NS}}_{\cal{D}}    \lesssim {\cal{O}}\left(10^{ {-11}}\right),
\eqn
which are all {smaller} than the quadrupole, monopole and dipole terms ${\cal{W}}^{\mathrm{NS}}_{\cal{A}}$, ${\cal{W}}^{\mathrm{NS}}_{\cal{B}}$ and  ${\cal{W}}^{\mathrm{NS}}_{\cal{C}}$. 
With the current accuracy of GW observations, it is hard to detect these effects \cite{Schutz18,PW14}. 
This justifies our current studies of triple systems only up to the lowest PN order. 

 {It should be noted that similar conclusions  {can also be} obtained by analyzing the  the polarization modes of a binary system with non-vanishing sensitivities  given in \cite{HYY15}.}

\section{Gravitational Radiation  in Einstein-aether Theory}
 \renewcommand{\theequation}{4.\arabic{equation}} \setcounter{equation}{0}

As shown in the last section, the strong field effects are very small with the current constraints of Eq.(\ref{2.8a}) \cite{OMW18}, and 
can be safely ignored for the current generation of detectors. So, in the rest of this paper, we shall consider gravitational radiation  in Einstein-aether theory to the lowest PN order, similar to what was already done in 
\cite{Foster06}, but without setting ${T}_{\mu} = 0$ for our future studies. Recall that ${T}_{\mu}$ originates from the direct coupling between matter and aether. Therefore, in the following we shall try to 
provide some detailed derivations of the formulas with the risk of repeating some materials already presented previously in \cite{Foster06}, although we shall try to limit these to their minimum
 \footnote{This also allows us to correct some  typos. Ref. \cite{Foster06} has been published in \cite{Foster06b}, but in Ref. \cite{Foster06} some  corrections of typos were made and also 
 with the same signature as used in the current paper, while  in \cite{Foster06b} the signature ($+1, -1, -1, -1$) was used.}.

As mentioned in the previous section,  the Minkowski spacetime is a vacuum solution of the Einstein-aether theory with the aether being along the time direction, i.e.,
\bqn
 \lb{3.0}
\left(\bar{g}_{\mu\nu}, \bar{u}^{\alpha}, \bar\lambda, \bar{T}_{\mu\nu}, \bar{T}_{\mu}\right) = \left(\eta_{\mu\nu}, \delta^{\alpha}_{0}, 0, 0, 0\right),
\eqn
 where $\eta_{\mu\nu} \equiv {\mbox{diag.}} \left(-1, 1, 1, 1\right)$ represents the Minkowski spacetime metric in the Cartesian  coordinates $x^{\mu} = (t, x^i)$.  Then, let us consider the linear perturbations,
 \bqn
 \lb{3.1}
h_{\mu\nu}=g_{\mu\nu}-\eta_{\mu\nu},~~~w^0=u^0-1,~~~w^i=u^i,
 \eqn
where   the perturbations $h_{\mu\nu}$, $w^0$ and $w^i$ are decomposed to the forms  \cite{Foster06},
 \bqn
 \lb{3.2}
h_{0i}&=&\gamma_i+\gamma_{,i},~~~w_i=\nu_i+\nu_{,i},\nb\\
h_{ij}&=&\phi_{,ij}+\frac{1}{2}P_{ij}[f]  +2\phi_{(i,j)}+ \phi_{ij},
 \eqn
with $i, j = 1, 2, 3$,  $(a, b) \equiv (ab + ba)/2$, and
 \bqn
 \lb{3.3}
&& \partial^{i}\gamma_{i}= \partial^{i}\nu_{i}= \partial^{i}\phi_{i} = 0,  \quad  \partial^{j}\phi_{ij}= 0, \quad \phi_{i}^{\;\; i}=0,\nb\\
&& P_{ij}[f] \equiv \delta_{ij}\Delta f - f_{,ij}, \quad \Delta \equiv \delta^{ij}\partial_i \partial_j.
 \eqn
All  spatial indices are raised or lowered by $\delta^{ij}$ and $\delta_{ij}$, respectively. For example, $\partial^ i \equiv \delta^{ij}\partial_{j}$, and so on.
Therefore, we have six scalars, $h_{00}$,  $w^0$, $\gamma$, $\nu$, $f$ and $\phi$; three transverse vectors, $\gamma
_i$, $\nu_i$ and $\phi_i$; and one transverse-traceless tensor, $\phi_{ij}$.

Note that, to the zeroth-order, $T_{\mu\nu}$ and $T_{\mu}$ all vanish identically, while to the first-order, we can  decompose them as,
\bqn
\lb{3.5d}
&& T_{0 i} = \Psi_{,i} +  \Psi_i, \quad T_i =\Theta_{,i} + \Theta_i, \nb\\
&&  T_{ij} = \Pi_{,ij} + \frac{1}{2}P_{ij}[ \Upsilon] + 2\Pi_{(i, j)} + \Pi_{ij},
\eqn
where
\bqn
\lb{3.5e}
&& \partial^i \Psi_i =  \partial^i \Theta_i = \partial^i \Pi_i = 0, \nb\\
&&     \partial^{j}\Pi_{ij}= 0, \quad \Pi_{i}^{\;\; i}=0.
\eqn
Therefore, in the matter sector, in general we have six scalars, $T_{00}$, $T_0$, $\Psi$, $\Theta$, $\Pi$ and $\Upsilon$; three vectors, $\Psi_i$, $\Theta_i$
 and $\Pi_i$; and one tensor $\Pi_{ij}$.

Under the following coordinate transformations,
\bq
\lb{3.4}
\tilde{t} = t + \xi^0\,, \quad \tilde{x}^{i} = x^i +  \xi^{i} + \partial^i\xi\,,
\eq
where $ \partial_i\xi^{i} = 0$,  we find
 \bqn
 \lb{3.5a}
&&\tilde{h}_{00}=h_{00}+2\dot{\xi^0}, \quad \tilde{w}^0=w^0+\dot{\xi^0},\nb\\
&&\tilde{\gamma}=\gamma+\xi^0-\dot{\xi}, \quad \tilde{\nu}=\nu+\dot{\xi},\nb\\
&& \tilde{f}=f, \quad \tilde{\phi}=\phi-2\xi,\\
\lb{3.5b}
&&\tilde{\gamma}_i=\gamma_i-\dot{\xi_i},\quad \tilde{\nu}_i=\nu_i+ \dot{{\xi_i}},\nb\\
&&\tilde{\phi}_i=\phi_i-\xi_i,\\
\lb{3.5c}
&&\tilde{\phi}_{ij}=\phi_{ij}.
 \eqn
Thus, out of the six scalar fields, we can construct four gauge-invariant quantities, while out of the three vector fields, two gauge-invariant quantities can be constructed, which can be 
chosen, respectively, by \footnote{Since they are all gauge-invariant, any  function of them is also gauge-invariant.}
\bqn
\lb{3.5ca}
&& \Phi^{\mathrm{I}} \equiv h_{00} - 2w^{0}, \quad \Phi^{\mathrm{II}} \equiv \frac{1}{2}\Delta{f},\nb\\
&& \Phi^{\mathrm{III}} \equiv \nu + \frac{1}{2}\dot\phi - \frac{1}{4}\dot{f},\nb\\  
&& \Phi^{\mathrm{IV}} \equiv \dot\gamma - w^{0} -\frac{1}{2}\ddot\phi + \frac{1}{4}\ddot{f}, 
\eqn
and
\bqn
\lb{3.5cb}
\Psi^{\mathrm{I}}_i &\equiv& \gamma_i + \nu_i, \quad 
\Psi^{\mathrm{II}}_i \equiv  \gamma_i - \dot{\phi}_i. 
\eqn
Clearly, the tensor mode $\phi_{ij}$ is already gauge-invariant.

On the other hand, since $T_{\mu\nu} = {\cal{O}}\left(\epsilon\right)$ and $T_{\mu} = {\cal{O}}\left(\epsilon\right)$, where   $\epsilon = {\cal{O}}(\xi_{\mu}) \ll 1$,  we find that
\bqn
\lb{3.5f}
&&\tilde{T}_{\mu\nu}\left(\tilde{x}\right) =T_{\alpha\beta}\left(x\right) \frac{\partial x^{\alpha}}{\partial \tilde{x}^{\mu}} \frac{\partial x^{\beta}}{\partial \tilde{x}^{\nu}}
= T_{\mu\nu}\left(x\right) + {\cal{O}}\left(\epsilon^2\right),\nb\\
&&\tilde{T}_{\mu}\left(\tilde{x}\right) =T_{\alpha}\left(x\right) \frac{\partial x^{\alpha}}{\partial \tilde{x}^{\mu}}
= T_{\mu}\left(x\right) + {\cal{O}}\left(\epsilon^2\right),
\eqn
that is,   to the first-order of $\epsilon$, all the quantities of the matter sector remain the same, and are gauge-invariant.

With the above analysis in mind, to the leading order, we find that Eqs.(\ref{2.4a}) - (\ref{2.4c}) reduce to,
 \bqn
 \lb{3.6a}
&& {G}_{\mu\nu}- S_{\mu\nu}=8\pi G_{\ae} \left(T_{\mu\nu}+t_{\mu\nu}\right),\\
\lb{3.6b}
&& {\AE}_{\mu}=8\pi G_{\ae}  (T_{\mu}+t_{\mu}),\\
\lb{3.6c}
&& h_{00}=2w^0,
 \eqn
where, to simplify the notations and without causing any confusions, except $t_{\mu\nu}$ and $t_{\mu}$, we use the same notations of Eqs.(\ref{2.4a}) - (\ref{2.4c}) to denote the linearized ones,  as
they all vanish for the Minkowski background.  The quantities  $t_{\mu\nu}$ and $t_{\mu}$ represent the nonlinear source terms \cite{Foster06} \footnote{Note that here we use $t_{\mu}$ to  {denote} the nonlinear
source term of the aether field, instead of $\sigma_{\mu}$ \cite{Foster06}, as we shall reserve the latter for other uses.}.
In particular, to the linear order, we have
\bqn
\lb{3.7}
\lambda &=&  \partial_{\alpha} J^{\alpha}_{\;\;\; 0} - 8\pi  {G_{\ae}} T_0,\nb\\
\AE_{\mu} &=& \partial_{\alpha} J^{\alpha}_{\;\;\; \mu} - \left(\partial_{\alpha} J^{\alpha}_{\;\;\; 0} - 8\pi  {G_{\ae}} T_0\right)\delta^0_{\mu},
\eqn
and
 \bqn
 \lb{3.8}
{G}_{00}&=&-\frac12\Delta^2f,\nb\\
{G}_{0i}&=&-\frac12\Delta(\gamma_i-\dot{\phi}_i)-\frac12\Delta\dot{f}_{,i},\nb\\
{G}_{ij}&=&-\frac12(\Delta\phi_{ij}-\ddot{\phi}_{ij})+(\ddot{\phi}_{(i,j)}-\dot{\gamma}_{(i,j)})-\frac12\ddot{f}_{,ij}\nb\\
&&  +\frac14P_{ij}[\Delta f-4w^0+4\dot{\gamma}-\ddot{f}-2\ddot{\phi}], \nb\\
S_{00}&=&c_{14}\Delta(\dot{\nu}+\dot{\gamma}-w^0) - {\AE}_0, \nb\\
S_{0i}&=&c_{14}(\ddot{\nu}_i+\ddot{\gamma}_i)-\frac{c_-}{2}\Delta(\nu_i+\gamma_i)\nb\\
&& +c_{14}\partial_{i}\left(\ddot{\nu}+\ddot{\gamma}-\dot{w}^0\right) \nb\\
&=&\frac12\Delta\left[c_{13}(\nu_i+\dot{\phi}_i)+c_{123}(2\nu+\dot{\phi})_{,i}+c_2\dot{f}_{,i}\right] \nb\\
&& - {\AE}_i, \nb\\
S_{ij}&=&\frac{c_{13}}{2}\ddot{\phi}_{ij}+c_{13}\left(\dot{\nu}_{(i,j)}+\ddot{\phi}_{(i,j)}\right)\nb\\
&& +\frac12P_{ij}\left[c_2(2\dot{\nu}+\ddot{\phi}+\ddot{f})\right.\nb\\
&& +\left. \frac{c_{13}}{2}\ddot{f}\right]+\frac12\left[c_{123}(\ddot{\phi}+2\dot{\nu})+c_2\ddot{f}\right]_{,ij}.
 \eqn
Setting
\bq
\lb{3.9}
\tau_{\mu\nu} \equiv T_{\mu\nu} - T_{\mu}\delta^0_{\nu} + \left(t_{\mu\nu}- t_{\mu}\delta^0_{\nu}\right),
\eq
we find that Eq.(\ref{3.6a}) can be cast in the form,
\bqn
 \lb{3.10}
{G}_{\mu\nu}- S_{\mu\nu} -  \AE_{\mu}  \delta^0_{\nu} =8\pi  {G_{\ae}} \tau_{\mu\nu}.
 \eqn
It should be noted that  $\tau_{\mu\nu}$ defined by Eq.(\ref{3.9})  in general is not symmetric.  In particular, we have
$\tau_{0i} \not= \tau_{i0}$.  Then, it can be shown that the left-hand side of the above equation satisfies,
 \bqn
 \lb{3.11}
\partial^{\nu}\left({G}_{\mu\nu}- S_{\mu\nu} -  \AE_{\mu}  \delta^0_{\nu}\right)  = 0,
 \eqn
 which leads to the following conservation laws,
  \bqn
 \lb{3.12}
\partial^{\nu}\tau_{\mu\nu} = \partial_{i}\tau_{\mu i} - \dot{\tau}_{\mu 0}  = 0.
 \eqn
When $T_{\mu} = 0$, it can be shown that the above equations reduce to the ones given in \cite{Foster06}
\footnote{A term $\partial_{\nu} J^{\nu 0} \delta^{0}_{\mu}$ in the left-hand side of Eq.(23) in \cite{Foster06} is missing.}.

Similar to the linearized terms $T_{\mu\nu}$ and $T_{\mu}$, we can also decompose the nonlinear terms $t_{\mu\nu}$ and $t_{\mu}$ in the forms of Eqs.(\ref{3.5d}) and (\ref{3.5e}),
\bqn
\lb{3.13}
&& t_{0 i} = \psi_{,i} +  \psi_i, \quad t_i =\theta_{,i} + \theta_i, \nb\\
&&  t_{ij} = \pi_{,ij} + \frac{1}{2}P_{ij}[ \upsilon] + 2\pi_{(i, j)} + \pi_{ij},
\eqn
where
\bqn
\lb{3.14}
&& \partial^i \psi_i =  \partial^i \theta_i = \partial^i \pi_i = 0, \nb\\
&&     \partial^{j}\pi_{ij}= 0, \quad \pi^{i}_{\;\; i}=0.
\eqn
 Then,  we find that $\tau_{\mu\nu}$ has the following non-vanishing components,
\bqn
\lb{3.15}
 \tau_{00} &=& T_{00} - T_{0} + \left(t_{00} - t_0\right) \nb\\
 \tau_{0 i} &=& T_{0i} + t_{0i} \nb\\
                &=& \Psi_{,i} + \Psi_{i} + \left(\psi_{,i} + \psi_{i}\right)  \nb\\
                &\equiv & \tau_{,i} + \tau_{i},\nb\\
\tau_{i 0} &=& T_{0i} - T_i + \left(t_{0i} - t_{i}\right) \nb\\
               &=& \Psi_{,i} - \Theta_{,i} + \Psi_i - \Theta_{i} + \left[ \left(\psi - \theta\right)_{,i} + \left(\psi_i - \theta_{i}\right)\right]\nb\\
               &\equiv & \chi_{,i} + \chi_{i},\nb\\
  \tau_{ij} &=& T_{ij} + t_{ij} \nb\\
  &=& \digamma_{,ij} + \frac{1}{2}P_{ij}[\varrho] + 2\digamma_{(i, j)} + \digamma_{ij},
\eqn
where
\bqn
\lb{3.16}
&& \tau \equiv \Psi + \psi, \quad \tau_i \equiv \Psi_i + \psi_i,\nb\\
&&  \chi \equiv \tau - \Theta - \theta, \quad  \chi_i \equiv \tau_i - \Theta_i - \theta_i,\nb\\
&& \digamma \equiv \Pi + \pi, \quad
\varrho \equiv \Upsilon + \upsilon, \nb\\
&& \digamma_i \equiv \Pi_i + \pi_i, \quad
\digamma_{ij} \equiv \Pi_{ij} + \pi_{ij}.
\eqn
Clearly, such defined three vectors $\tau_i$, $\chi_i$ and $\digamma_i$ are transverse, and the tensor $\digamma_{ij}$ is traceless and transverse, i.e., 
\bqn
\lb{3.16b}
&&  \partial^i \tau_i  =    \partial^i  \chi_i  =
  \partial^i \digamma_i  = 0, \nb\\
&&
 \partial^i\digamma_{ij}  = 0, \quad  \digamma^{i}_{\;\;\; i}=0.
\eqn

 With the above decompositions, it can be shown that the field equations can be divided into two groups, one represents the propagation equations for the scalar, vector and tensor modes, and the other
represents the type of Poisson equations. In terms of the gauge-invariant quantities defined in Eqs.(\ref{3.5ca}) and (\ref{3.5cb}), the first group is given by
 \bqn
 \lb{3.17a}
 &&\Box_S \Phi^{\mathrm{II}}  =-\frac{16\pi G_{\ae} c_{14}}{2-c_{14}}\Bigg[\frac{1}{2}\Delta\varrho -\frac{1 + c_2}{c_{123}}\Delta\digamma\nb\\
&& ~~~~~~~~~~~~~~ ~~~~~~~~~~~~ +\frac{1}{c_{14}}\left(\tau_0-\Gamma_0\right)\Bigg],\\
 \lb{3.17b}
&&\Box_V \Psi^{\mathrm{I}}_{i}
=-\frac{16\pi G_{\ae}}{2c_1-c_-c_{13}}\left(c_{13}\tau_i-\Gamma_i\right),\\
 \lb{3.17c}
&&\Box_T   \phi_{ij} =-16\pi G_{\ae}\digamma_{ij},
 \eqn
where  
$\Box_I \equiv \Delta - c_I^{-2}\partial_t^2$,  and
 \bqn
 \lb{3.18a}
\tau_0 &\equiv& T_{00} + t_{00}, \quad \Gamma_0 \equiv T_0 + t_0,\nb\\
\Gamma &\equiv& \Theta + \theta, \quad \Gamma_i \equiv \Theta_i + \theta_i,
\eqn
with $c_I$'s being given by Eq.(\ref{CSs}).

The second group is given by,
 \bqn
 \lb{3.19a}
&& \Delta\left[c_{13}(\nu_i+\dot{\phi}_i)+\gamma_i-\dot{\phi}_i\right]
=-16\pi G_{\ae}\left(\tau_i-\Gamma_i\right),\nb\\
\\
 \lb{3.19b}
&& \Delta\left[F-2c_{14}w^0+2c_{14}(\dot{\gamma}+\dot{\nu})\right]
=-16\pi G_{\ae}\left(\tau_0-\Gamma_0\right),\nb\\
\\
 \lb{3.19c}
&& \Delta\left[(1+c_2)\dot{f}+c_{123}(\dot{\phi}+2\nu)\right]_{,i}
=-16\pi G_{\ae}\left(\tau-\Gamma\right)_{,i},\nb\\
\\
 \lb{3.19d}
&& \Delta\left[(1+c_2)\ddot{f}+c_{123}(\ddot{\phi}+2\dot\nu)\right]
=-16\pi G_{\ae}\Delta\digamma, \nb\\
 \eqn
which all take the form,
\bq
\lb{3.20}
\Delta \psi = - 4\pi \tau.
\eq

When far away from the source, the above Poisson equations have  solutions of the form,
\bq
\lb{3.21}
 \psi(t, \vec{x}) = \frac{1}{R} \int_{V'}{d^3x' \tau(t, x')} + {\cal{O}}\left(\frac{1}{R^2}\right),
\eq
where $R \equiv |\vec{x}| \gg d$, with $d$ denoting the size of the source. As argued in \cite{Foster06}, the contributions of this part   to the wave forms are negligible, and without loss of the generality, we can safely set it  to zero,
\bq
\lb{3.21}
 \psi(t, \vec{x}) \simeq 0, \; (R \gg d),
\eq
in the wave zone.

\subsection{Polarizations of Gravitational Waves in Einstein-aether Theory}

To consider the polarizations of gravitational waves in Einstein-aether theory, let us consider the time-like geodesic deviations.  In the spacetime described by the metric, $g_{\mu\nu} = \eta_{\mu\nu} + h_{\mu\nu}$,
 the spatial part, $\zeta_i$,   takes the form \cite{GHLP18},
 \bqn
 \lb{3.22}
\ddot{\zeta}_i=-R_{0i0j}\zeta^j\equiv\frac{1}{2}\ddot{{\cal{P}}}_{ij}\zeta^j,
 \eqn
where $\zeta_{\mu}$ describes the  deviation vector   between two nearby trajectories  of test particles, and
  \bqn
 \lb{3.23}
R_{0i0j}&\simeq& \frac{1}{2}(h_{0j,0i}+h_{0i,0j}-h_{ij,00}-h_{00,ij})\nb\\
&=&\frac{1}{2}\Big[2\dot{\gamma}_{(i,j)}+2\dot{\gamma}_{,ij}-2w^0_{,ij} -\ddot{\phi}_{ij}\nb\\
&& -2\ddot{\phi}_{(i,j)}-\ddot{\phi}_{,ij}- \frac12\delta_{ij}\Delta\ddot{f} +\frac12\ddot{f}_{,ij}\Big]\nb\\
&=&-\frac{1}{2}\ddot{\phi}_{ij}+\dot{\Psi}^{\mathrm{II}}_{(i,j)}+\Phi^{\mathrm{IV}}_{,ij}-\frac{1}{2}\delta_{ij}\ddot{\Phi}^{\mathrm{II}}.
 \eqn
When deriving the last expression of the above equation,   we had used Eqs.(\ref{3.5ca}) and (\ref{3.5cb}).

 In the wave zone ($R \gg d$), Eqs.(\ref{3.21}) and (\ref{3.19a})-(\ref{3.19d})  imply that
  \bqn
 \lb{3.19aa}
&&  c_{13}(\nu_i+\dot{\phi}_i)+\gamma_i-\dot{\phi}_i
=0,\\
 \lb{3.19bb}
&&  F-2c_{14}w^0+2c_{14}(\dot{\gamma}+\dot{\nu})
=0,\\
 \lb{3.19cc}
&&  (1+c_2)\dot{f}+c_{123}(\dot{\phi}+2\nu) =0.
 \eqn
From Eq.(\ref{3.19aa}), we find that 
\bq
\lb{3.19ee}
\dot{\phi}_i = \frac{\gamma_i + c_{13}\nu_i} {1-c_{13}}.
\eq
Inserting it into Eq.(\ref{3.5cb}), we obtain
\bq
\lb{3.19ff}
\Psi^{\mathrm{II}}_i=-\frac{c_{13}}{1-c_{13}}\Psi^{\mathrm{I}}_i.
\eq
On the other hand, the combination of Eqs.(\ref{3.5ca}) and (\ref{3.19bb}) yields, 
\bqn
\lb{4.46a}
\Phi^{\mathrm{II}}= - c_{14}\left(\Phi^{\mathrm{IV}}+\dot{\Phi}^{\mathrm{III}}\right), 
\eqn
while from Eq.(\ref{3.19cc})  we obtain,   
\bqn
\lb{4.46b}
2c_{123}{\Phi}^{\mathrm{III}} +\left(1+c_2+\frac{c_{123}}{2}\right)\dot{f}=0.
\eqn
Then, combining it with Eq.(\ref{3.5ca}) we have
\bqn
\lb{4.46c}
2c_{123}\Delta\dot{\Phi}^{\mathrm{III}}&=&-\left(1+c_2+\frac{c_{123}}{2}\right)\Delta\ddot{f}\nb\\
&=&-2\left(1+c_2+\frac{c_{123}}{2}\right)\ddot{\Phi}^{\mathrm{II}}\nb\\
&=&-2\left(1+c_2+\frac{c_{123}}{2}\right)c_S^2\Delta\Phi^{\mathrm{II}}\nb\\
&=&-\frac{c_{123}\left(2-c_{14}\right)}{c_{14}\left(1-c_{13}\right)}\Delta\Phi^{\mathrm{II}}, \nb
\eqn
that is, 
\bqn
\lb{4.46d}
\dot{\Phi}^{\mathrm{III}}=-\frac{2-c_{14}}{2c_{14}\left(1-c_{13}\right)}\Phi^{\mathrm{II}}.
\eqn
Note that in writing the above expressions, we had used    ${\square}_S\Phi^{\mathrm{II}}=0$ in the wave zone.
The combination of   Eqs.(\ref{4.46a}) and (\ref{4.46d}) yields, 
 \bqn
\lb{3.24a}
\Phi^{\mathrm{IV}}&=&\frac{c_{14}-2c_{13}}{2-c_{14}}\dot{\Phi}^{\mathrm{III}}=\frac{c_{14}-2c_{13}}{2c_{14}(c_{13}-1)}\Phi^{\mathrm{II}}.\nb\\
 \eqn
 
 In the wave zone,    we also have \cite{Foster06},  
 \bqn
 \lb{3.24c}
\Psi^{\mathrm{I}}_{i,j}&=&-\frac{1}{c_V}\dot{\Psi}^{\mathrm{I}}_i N_j,
\nb\\
\Phi^{\mathrm{II}}_{,i}&=&-\frac{1}{c_S}\dot{\Phi}^{\mathrm{II}} N_i,
 \eqn
where $N_k$ denotes the unit vector along the direction between the source and the observer. Then, inserting the above expressions into Eq.(\ref{3.22})
we obtain
 \bqn
 \lb{3.25}
{{\cal{P}}}_{ij}&=&\phi_{ij}-\frac{2c_{13}}{(1-c_{13})c_{V}}\Psi^{\mathrm{I}}_{(i}N_{j)}\nb\\
&&-\frac{c_{14}-2c_{13}}{c_{14}(c_{13}-1)c^2_{S}}\Phi^{\mathrm{II}}N_iN_j+\delta_{ij}\Phi^{\mathrm{II}}.
 \eqn

 Assuming that  (${\bf e}_X, {\bf e}_Y, {\bf e}_Z$) are three unity vectors and form an orthogonal basis with  ${\bf e}_Z \equiv \bf{N}$, so that
 (${\bf e}_X, {\bf e}_Y$) lay on the plane orthogonal to the propagation direction $\bf{N}$ of the gravitational wave, we find that,  in the coordinates $x^{\mu} = (t, x^i)$,
 these three vectors
 can be specified by two angles, $\vartheta$ and $ \varphi$, via the relations \cite{PW14}, 
\bqn
\lb{rotationsA}
{\bf e}_{X} &=& \left(\cos\vartheta\cos\varphi, \cos\vartheta\sin\varphi, - \sin\vartheta\right), \nb\\
{\bf e}_{Y} &=& \left(-\sin\varphi, \cos\varphi, 0\right), \nb\\
{\bf e}_{Z} &=& \left(\sin\vartheta\cos\varphi, \sin\vartheta\sin\varphi, \cos\vartheta\right).
\eqn
Then, we can define the six  polarizations $h_{N}$'s by 
\bqn
\lb{polarizations}
h_+&\equiv&\frac{1}{2}\left({\cal{P}}_{XX}-{\cal{P}}_{YY}\right), \quad
h_\times\equiv\frac{1}{2}\left({\cal{P}}_{XY}+{\cal{P}}_{YX}\right), \nb\\
h_b&\equiv& \frac{1}{2}\left({\cal{P}}_{XX}+{\cal{P}}_{YY}\right),  \quad
h_L\equiv {\cal{P}}_{ZZ}, \nb\\
h_X&\equiv&\frac{1}{2}\left({\cal{P}}_{XZ}+{\cal{P}}_{ZX}\right), \quad  
h_Y\equiv \frac{1}{2}\left({\cal{P}}_{YZ}+{\cal{P}}_{ZY}\right),  \nb\\
\eqn
where ${\cal{P}}_{XY} \equiv {\cal{P}}_{ij}e^{i}_X e^{j}_Y$, and so on. However,  
 in Einstein-aether theory, only  five of them are independent:  two from each of the vector and tensor modes,   and one from  the scalar mode.
 
 In order to calculate the wave forms, let us first assume that the detector is located at ${\bf{x}}$ with $R \equiv |{\bf{x}}| \gg d$, where $d$ denotes the size
of the source. Then, under the gauge,
  \bq
\lb{4.13}
\phi_i=0, \; \nu=\gamma=0, 
  \eq
we find  
  \bqn
\lb{4.10}
\phi_{ij}&=&\frac{2 {G_{\ae}}}{R}(\ddot{Q}_{ij})^{TT},\\
\lb{4.11}
\nu_i&=&-\frac{2 {G_{\ae}}}{(2c_1-c_{13}c_-)R} \left(\frac{c_{13}\ddot{Q}_{ij}N_j}{(1-c_{13})c_V}-2\Sigma_i\right)^T,\nb\\
\gamma_i&=&-c_{13}\nu_i,\\
\lb{4.12}
{F}&=&\frac{ {G_{\ae}}}{R}\frac{c_{14}}{2-c_{14}}\left( 6(Z-1)\ddot{Q}_{ij}N_iN_j \right.\nb\\
&&\left.-\frac{8}{c_{14}c_S}\Sigma_i N_i+2Z\ddot{I}\right), \nb\\
h_{00}&=&2\omega^0=\frac{1}{c_{14}}F,~~~~\phi=-\frac{1+c_2}{c_{123}}f,\\
\lb{4.14}
\Psi^{\mathrm{I}}_{i}&=&(1-c_{13})\nu_i, \quad
\Phi^{\mathrm{II}}  = \frac{1}{2}F,
 \eqn
 {where, for any given symmetric tensor $S_{ij}$, we have $S_{ij}^{TT}=\Lambda_{ij,kl}S_{kl}$ and $S_i^{T}=P_{ij}S_j$, where $\Lambda_{ij,kl}$ and $P_{ij}$
are the projection operators   defined, respectively, by Eqs.(1.35) and (1.39) in  \cite{Mag08}.}

 Inserting  {Eq.}(\ref{3.25}) into  {Eq.}(\ref{polarizations}) and using the above equations, we find that
 \bqn
\lb{polarizationsB}
h_+& = &  
 \frac{ {G_{\ae}}}{R}\ddot{Q}_{kl}e_+^{kl},\quad 
h_\times = \frac{ {G_{\ae}}}{R}\ddot{Q}_{kl}e_\times^{kl}, \nb\\
h_b &=&\frac{ {c_{14}G_{\ae}}}{R(2-c_{14})} \Bigg[3(Z-1)\ddot{Q}_{ij}e_Z^ie_Z^j \nb\\
&&~~~~~~~~~~~~~~~~ -\frac{4}{c_{14}c_S}\Sigma_i e_Z^i+Z\ddot{I}\Bigg], \nb\\
h_L &=& \left[1-\frac{c_{14}-2c_{13}}{c_{14}(c_{13}-1)c_S^2}\right]h_b, \nb\\
h_X &=& {\frac{2c_{13} {G_{\ae}}}{(2c_1-c_{13}c_-)c_V R}}\nb\\
 &&\times \left[\frac{c_{13}\ddot{Q}_{jk}e_Z^k}{(1-c_{13})c_V}-2\Sigma_j\right]e_X^j ,\nb\\
h_Y &=&  \frac{2c_{13}G_{\ae}}{(2c_1-c_{13}c_-)c_V R}\nb\\
 &&\times \left[\frac{c_{13}\ddot{Q}_{jk}e_Z^k}{(1-c_{13})c_V}-2\Sigma_j\right]e_Y^j ,
 \eqn
where $e_+^{kl} \equiv e_X^k e_X^l-e_Y^k e_Y^l$ and $e_\times^{kl} \equiv e_X^k e_Y^l+e_Y^k e_X^l$. From the above expressions we can see that
the scalar longitudinal  mode $h_L$ is proportional to the scalar  breather mode $h_b$. So, out of these six components, only five of them are independent.

It is remarkable to note that both of the
scalar breather  and the scalar longitudinal   modes are suppressed by a factor $c_{14} \lesssim {\cal{O}}\left(10^{-5}\right)$
with respect to the transverse-traceless modes $h_{+}$ and $h_{\times}$, while the vectorial modes $h_{X}$ and $h_{Y}$ are suppressed by a factor
 $c_{13}\lesssim {\cal{O}}\left(10^{-15}\right)$.

\subsection{Response Function}

To study the GW forms, an important quantity is the response function $h(t)$, with respect to a specific detector, which, for the sake of simplicity, is 
assumed to have two orthogonal arms, such as aLIGO, aVIRGO or KAGRA. 
Assume that  the two arms of a detector are along, respectively,  ${\bf e}_1$- and  ${\bf e}_2$-directions. Then, from them we can 
construct another unity (space-like) vector ${\bf e}_3$ that forms an orthogonal basis together with $(\bf{e}_1, \; \bf{e}_2$), that is, ${\bf e}_{i}\cdot {\bf e}_j = \delta_{ij}$. 
The choice of this frame is independent of the one  (${\bf e}_X, {\bf e}_Y, {\bf e}_Z$), which 
we just introduced in the last subsection. But, we can always rotate properly from one frame to get the other with three independent angles, $\theta, \phi$ and $\psi$ [cf. Fig. 11.5 and 
 Eqs.(11.319a)-(11.319c) of \cite{PW14}], given by,
\bqn
\lb{rotationsB}
{\bf e}_{1} &=&\left(\cos\theta\cos\phi\cos\psi - \sin\phi\sin\psi\right){\bf e}_{X}  \nb\\
&& + \left(\cos\theta\cos\phi\sin\psi + \sin\phi\cos\psi\right){\bf e}_{Y} \nb\\
&& - \sin\theta\cos\phi  \; {\bf e}_{Z},\nb\\
{\bf e}_{2} &=&\left(\cos\theta\sin\phi\cos\psi + \cos\phi\sin\psi\right){\bf e}_{X}  \nb\\
&& + \left(\cos\theta\sin\phi\sin\psi - \cos\phi\cos\psi\right){\bf e}_{Y} \nb\\
&& - \sin\theta\sin\phi  \; {\bf e}_{Z},\nb\\
{\bf e}_{3} &=& - \sin\theta\cos\psi \; {\bf e}_{X}  -  \sin\theta\sin\psi \; {\bf e}_{Y} \nb\\
&& - \cos\theta  \; {\bf e}_{Z}.
\eqn
Then,  the response function    $h(t)$ is defined as,  
 \bqn
 \lb{4.4}
h(t) &\equiv& \frac{1}{2}\left(e^i_1e^j_1 - e^i_2e^j_2\right) {\cal{P}}^{ij} \nb\\
&=& F_{+}h_{+} + F_{\times} h_{\times}  + F_b h_b + F_L h_L \nb\\
&& + F_{X} h_X + F_{Y} h_Y, 
 \eqn
 where
 \bqn
\lb{RFs}
F_{+} &\equiv&\frac{1}{2} \left(1+\cos^2\theta\right) \cos 2\phi\cos 2\psi - \cos\theta\sin2\phi\sin2\psi,  \nb\\
F_{\times} &\equiv&\frac{1}{2} \left(1+\cos^2\theta\right) \cos 2\phi\sin 2\psi + \cos\theta\sin2\phi\cos2\psi,  \nb\\
F_{b} &\equiv&- \frac{1}{2} \sin^2\theta \cos 2\phi,  \nb\\
F_{L} &\equiv& \frac{1}{2} \sin^2\theta \cos 2\phi,  \nb\\
F_{X} &\equiv&-\sin\theta \left(\cos\theta\cos 2\phi\cos \psi - \sin2\phi\sin\psi\right),  \nb\\
F_{Y} &\equiv&-\sin\theta \left(\cos\theta\cos 2\phi\sin \psi + \sin2\phi\cos\psi\right).
\eqn
Hence, its Fourier transformation takes the form,  
 \bq
 \lb{4.4}
\tilde{h}(f) = { \frac{1}{2\pi}\int{h(t) e^{-i2\pi f t} dt}}.
 \eq

\section{Gravitational Wave Forms and Radiations of  Triple Systems}
 \renewcommand{\theequation}{5.\arabic{equation}} \setcounter{equation}{0}

Consider a triple system with masses $m_a$ and positions ${\bf x}_a(t)$, where $a$ specifies the three bodies, $a = 1, 2, 3$.  
Defining 
${\bf r}_1 \equiv {\bf x}_2 - {\bf x}_3$, ${\bf r}_2 \equiv {\bf x}_3 - {\bf x}_1$, ${\bf r}_3 \equiv {\bf x}_1 - {\bf x}_2$, and $r_a \equiv |{\bf r}_a|$, we find that 
  \bqn
 \lb{4.2}
\dddot{Q}_{ij}&=&\frac{2}{3}G_N\sum\limits_aP^{a}_{ ij}, \nb\\
\dddot{I}&=&-2 G_N \Bigg(m_2 m_3\frac{\dot{r}_1}{(r_1)^2}+m_1 m_3\frac{\dot{r}_2}{(r_2)^2}\nb\\
&& ~~~~~~~~~~~~ + m_1 m_2\frac{\dot{r}_3}{(r_3)^2}\Bigg),      \nb\\
{\Sigma}_i&=& {\left(\alpha_1-\frac{2}{3}\alpha_2\right)\sum\limits_{a}\left(v^i_a \Omega_a\right)},\nb\\
\dot{\Sigma}_i&=&-\left(\alpha_1-\frac{2}{3}\alpha_2\right)G_N\sum\limits_{a}\left({\cal D}_a \frac{r^i_a}{(r_a)^3}\right), 
 \eqn
where $\Omega_a$ is the binding energy of the a-th body, as mentioned previously, and   
  \bqn
 \lb{4.3}
{\cal D}_c&\equiv&{\sum\limits_{a,b}}\epsilon^{abc}\Omega_a m_b,\nb\\
P^{c}_{ij}&\equiv&{\sum\limits_{a,b}}\frac{\left|\epsilon^{abc}\right|m_a m_b}{2 (r_c)^4}\Bigg[(r_c)^2\dot{r}_c\delta_{ij}-6r_c\frac{d\left(r^i_c  r^j_c \right)}{dt}\nb\\
&& ~~~~~~~~~~~~~~~~~~ +9\dot{r}_c r^i_c r^j_c \Bigg].
 \eqn
Here $\epsilon^{abc}$ is the Levi-Civita symbol. Setting  $r_a^i r_a^i \equiv (r_a)^2$,  ${r_a^i\dot{r}_a^i \equiv r_a \dot{r}_a,}$ and $\dot{r}_a^i\dot{r}_a^i \equiv (v_{a})^2$,
we obtain
 \bqn
 \lb{4.5}
\left(\dddot{Q}_{ij}\right)^2 &=&\frac{8}{3}G_N^2 \Bigg\{\left[\frac{m_2^2m_3^2}{(r_1)^4}\left(12v_1^2-11(\dot{r}_1)^2\right)+\frac{1}{3}P^{2}_{ ij}P^{3}_{ ij}\right]\nb\\
&&+ \left[\frac{m_1^2m_2^2}{(r_3)^4}\left(12v_3^2-11(\dot{r}_3)^2\right)+\frac{1}{3}P^{1}_{ ij}P^{2}_{ ij}\right]\nb\\
&&+ \left[\frac{m_1^2m_3^2}{(r_2)^4}\left(12v_2^2-11(\dot{r}_2)^2\right)+\frac{1}{3}P^{1}_{ ij}P^{3}_{ ij}\right]\Bigg\},\nb\\
\eqn
 \bqn
 \lb{4.6}
\left(\dddot{I}\right)^2&=&4G_N^2 \Bigg\{\left[m_2^2 m_3^2\frac{(\dot{r}_1)^2}{(r_1)^4}+2m_2 m_3 m_1^2\frac{\dot{r}_2\dot{r}_3}{(r_2)^2 (r_3)^2}\right]\nb\\
&&+  \left[m_1^2 m_3^2\frac{(\dot{r}_2)^2}{(r_2)^4}+2m_1 m_3 m_2^2\frac{\dot{r}_1\dot{r}_3}{(r_1)^2 (r_3)^2}\right]\nb\\
&&+  \left[m_1^2 m_2^2\frac{(\dot{r}_3)^2}{(r_3)^4}+2m_1 m_2 m_3^2\frac{\dot{r}_1\dot{r}_2}{(r_1)^2 (r_2)^2}\right]\Bigg\},\nb\\
\eqn
 \bqn
 \lb{4.7}
\left(\dot{\Sigma}_i\right)^2  &=&\left(\alpha_1-\frac{2}{3}\alpha_2\right)^2G_N^2 \Bigg\{\left[\frac{{\cal D}_1^2 }{(r_1)^4}+\frac{2{\cal D}_2{\cal D}_3}{(r_2)^3 (r_3)^3} r_2^i r_3^i\right]\nb\\
&&+  \left[\frac{{\cal D}_2^2 }{(r_2)^4}+\frac{2{\cal D}_1{\cal D}_3}{(r_1)^3 (r_3)^3} r_1^i r_3^i\right]\nb\\
&&+  \left[\frac{{\cal D}_3^2 }{(r_3)^4}+\frac{2{\cal D}_1{\cal D}_2}{(r_1)^3 (r_2)^3} r_1^i r_2^i\right]\Bigg\},\nb\\
 \eqn
where
 \bqn
 \lb{4.8}
P^{1}_{ ij}P^{2}_{ ij}&=&m_1 m_2 m_3^2\left[\frac{36}{(r_1)^3 (r_2)^3}\frac{d\left(r_1^i r_1^j\right)}{dt}\frac{d\left(r_2^i r_2^j\right)}{dt}\right.\nb\\
&&-\frac{54 r_1^i r_1^j\dot{r}_1}{(r_1)^4 (r_2)^3}\frac{d\left(r_2^i r_2^j\right)}{dt}-\frac{54 r_2^i r_2^j \dot{r}_2}{(r_1)^3 (r_2)^4}\frac{d\left(r_1^i r_1^j\right)}{dt}\nb\\
&&\left.+\frac{81 r_1^i r_1^j r_2^i r_2^j\dot{r}_1\dot{r}_2}{(r_1)^4 (r_2)^4}-\frac{3\dot{r}_1\dot{r}_2}{(r_1)^2 (r_2)^2}\right],
 \eqn
and so on.

Then, from Eqs.(\ref{polarizationsB}) we can see that  $h_N({\bf x}; {\bf x}_{a}(t))$, where ${\bf x}_{a}(t)$'s are the trajectories of the 
three bodies. So, once ${\bf x}_{a}(t)$'s are known, from  {Eqs.(\ref{polarizationsB}) and (\ref{4.2})} we can study the polarizations of GWs emitted by this triple system. 
On the other hand, inserting  {Eqs.(\ref{4.5}) - (\ref{4.8})} into Eq.(\ref{4.1a}) we obtain the energy loss rate $\dot{\cal{E}}(t)$. 

However, in the framework of Einstein-aether theory  the trajectories of a triple system have not been studied, yet. So, in this paper we shall use the Newtonian trajectories of the triple systems
\footnote{Corrections  due to the aether effects are expected to be small, and should be consistent with the   lowest PN order approximations adopted in this paper.},
which have been intensively studied in the past three hundred years, and various periodic solutions have been found, see for example, \cite{LL17} and references therein. Some of them have been
also used to study the GW forms in the framework of GR. In particular, in \cite{THA09} it was shown that the quadrupole GW form of a figure eight trajectory
discovered by Moore in  {1993} \cite{Moore93} is indistinguishable from that of a binary system. In addition, Dmitrasinovic, Suvakov and Hodomal calculated the quadrupole   wave forms
and the corresponding luminosities for the $13 + 11$ periodic orbits of three-body problems in Newtonian gravity \cite{DSH14}, discovered, respectively, in \cite{SD13} and \cite{S14}.
Among other things, they found that all these $13+11$ orbits produce different waveforms and their luminosities vary by up to 13 order of magnitude in the mean, and up to 20 order for the 
peak values.


\begin{figure}[h!]
{
\includegraphics[width=\linewidth]{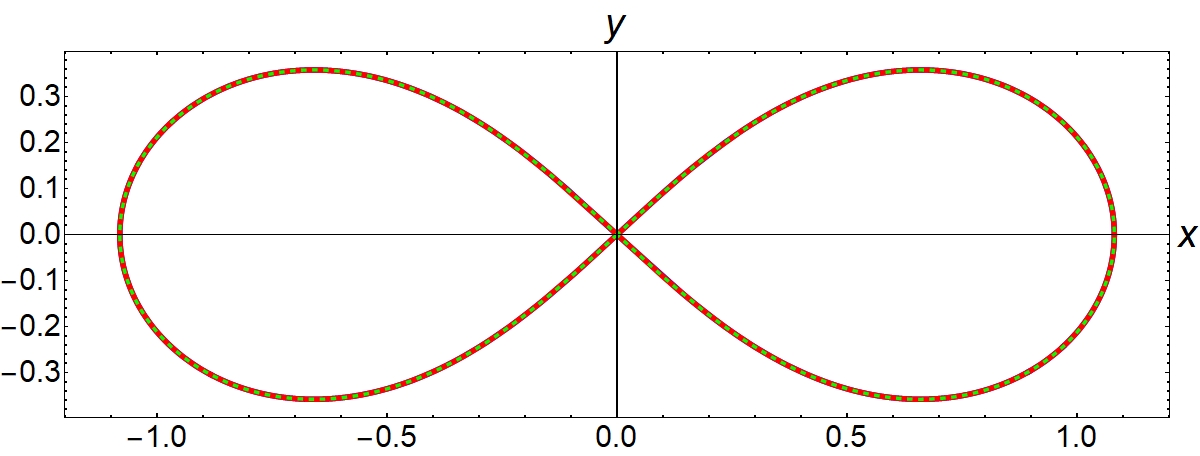} 
}
\caption{Trajectory of the Simo's figure-eight  3-body system \cite{Simo02}. In this plot, we set $m = 1$. }
\label{fig1ab}
\end{figure}

\begin{figure}[h!]
{
\includegraphics[width=\linewidth]{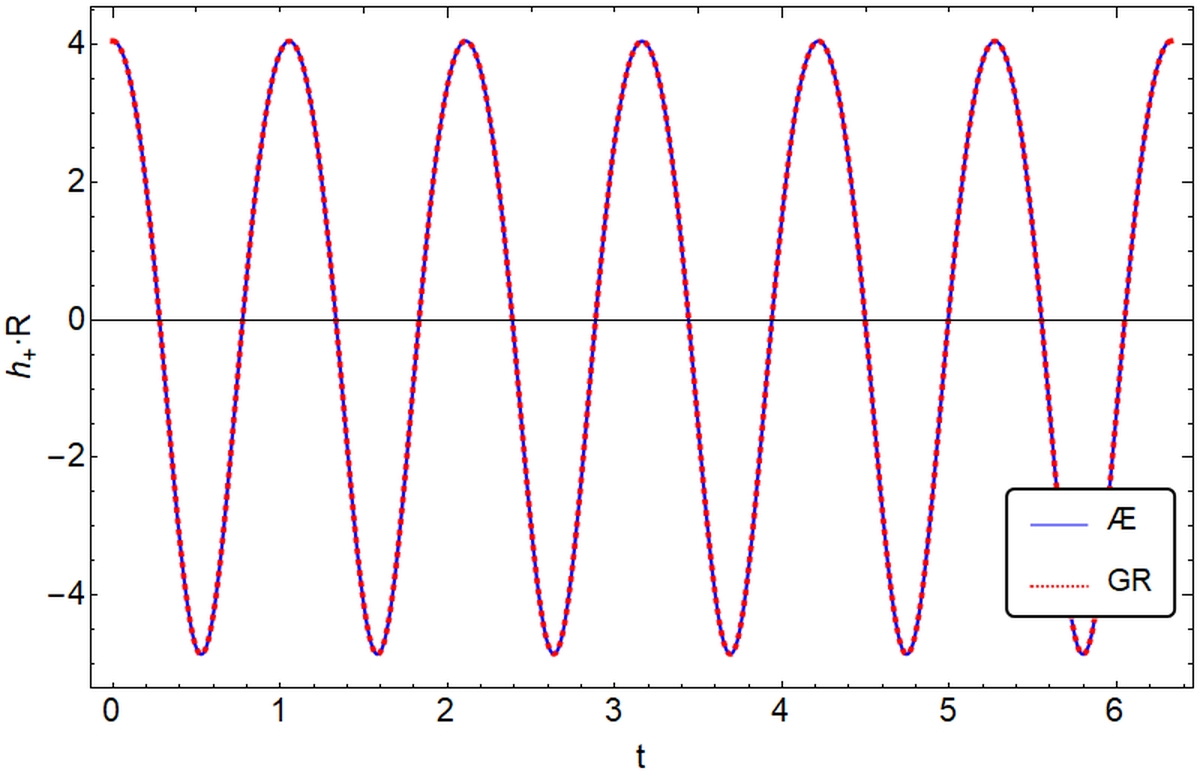}
\includegraphics[width=\linewidth]{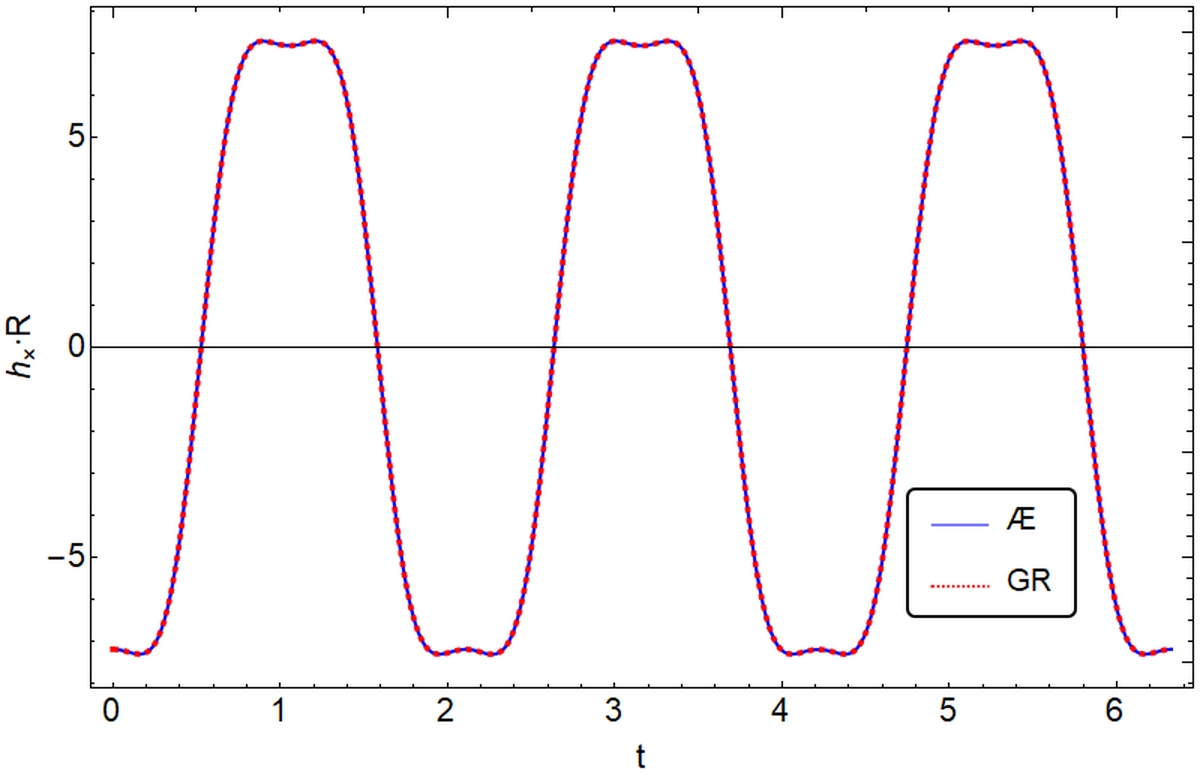}
}
\caption{The polarization modes $h_{+}$ and $h_{\times}$ defined in Eq.(\ref{polarizations}) for the Simo's figure-eight 3-body system  in both GR and $\ae$-theory, 
where the modes are propagating along the positive $z$-direction, and $(\Omega_1,\; \Omega_2,\; \Omega_3)  = (-0.1, \;  -2.76\times 10^{-6},\;  -2.9\times 10^{-5})$.}
\label{fig2ab1}
\end{figure}

\begin{figure}[h!]
{
\includegraphics[width=\linewidth]{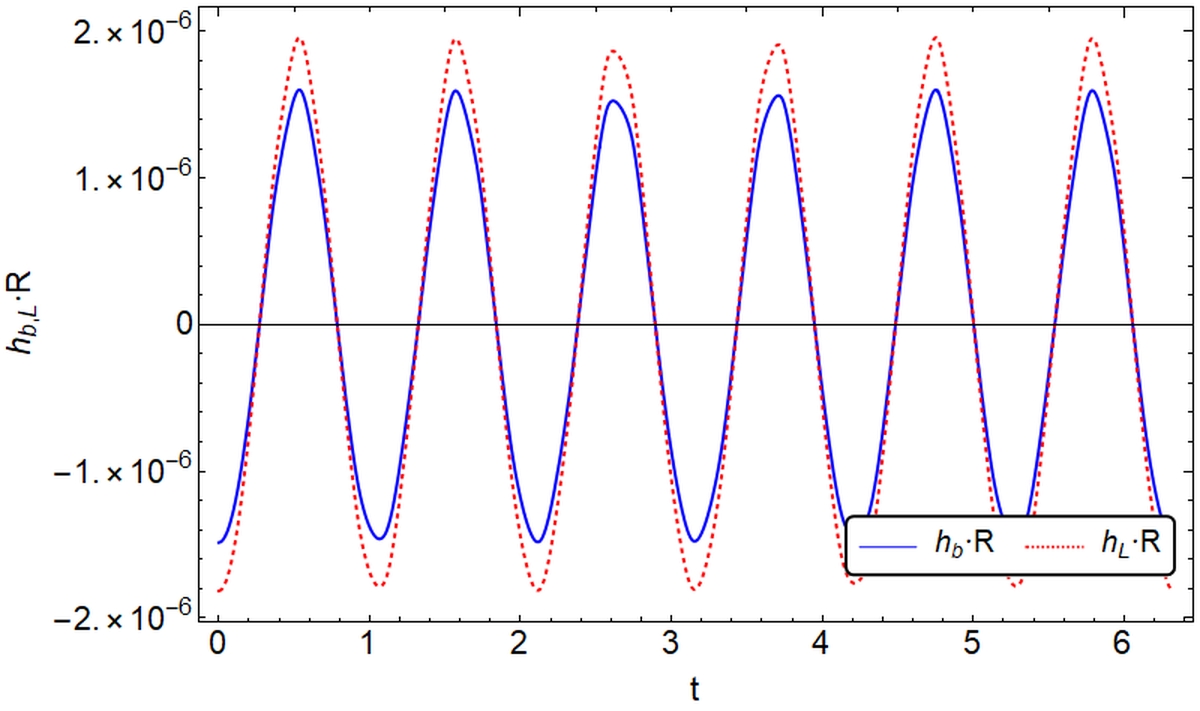}
}
\caption{The polarization modes $h_{b}$ and  $h_{L}$ defined in Eq.(\ref{polarizations}) for the Simo's figure-eight 3-body system  in $\ae$-theory, where the modes are propagating along the positive $z$-direction.,
and $(\Omega_1,\; \Omega_2,\; \Omega_3)  = (-0.1, \;  -2.76\times 10^{-6},\;  -2.9\times 10^{-5})$.}
\label{fig2ab2}
\end{figure}


In this paper, we shall consider some of the trajectories provided in \cite{Website} \footnote{In the configurations provided in this site, the three bodies are assumed all to have equal masses, $m_1 = m_2 = m_3 =m$, and also 
the unites were chosen so that $G_N = 1, \; m = 1$. Therefore, in our numerical simulations presented in this paper, we adopt the same units so that $ m = 1 = G_N$. However, restoring the physical units, this is equivalent to set 
$m_i  = 1/G_N \simeq 1.5 \times 10^{10} \; kg \ll M_{\bigodot}$.}.   

Before doing so, let us first  consider the  GW form of the Simo's  figure-eight trajectory \cite{Simo02}, 
studied in \cite{DSH14}. In Fg. \ref{fig1ab},  the trajectory of the 3-body problem is  {plotted} out in the ($x, y$)-plane for many periods, in order to make sure that our numerical codes converge well after a sufficiently long
run.  Assuming that the detector is along the $z$-axis, we plot out the polarization modes $h_{+}$ and $h_{\times}$ in Fig. \ref{fig2ab1}. In this figure, we plot these modes given in GR as well as in Einstein-aether theory. As 
we noted  previously, the contributions from the aether field is  of the order of ${\cal{O}}\left(10^{-5}\right)$ lower than that of GR. This can be seen clearly from this figure, in which the lines are almost identical 
in both of theories. Note that    when  plotting these figures, we assumed that the binding energies of the three bodies  are, respectively, $\Omega_1 = -0.1, \; 
\Omega_2 = -2.76\times 10^{-6}$ and $\Omega_1 = -2.9\times 10^{-5}$,  which are the same as these given for the PRS J0337+ 1715 \cite{Ransom14}, although here the three masses are assumed to be equal. 
 
In Fig. \ref{fig2ab2} we plot out the polarization modes $h_{b}$ and $h_{L}$ in $\ae$-theory, which all vanish in GR. Comparing it with Fig. \ref{fig2ab1} it can be seen that the amplitudes of  these modes are about five orders lower than 
$h_{+}$ and $h_{\times}$, which is again consistent with our analysis given in Sec. III.


\begin{figure}[h!]
{
\includegraphics[width=\linewidth]{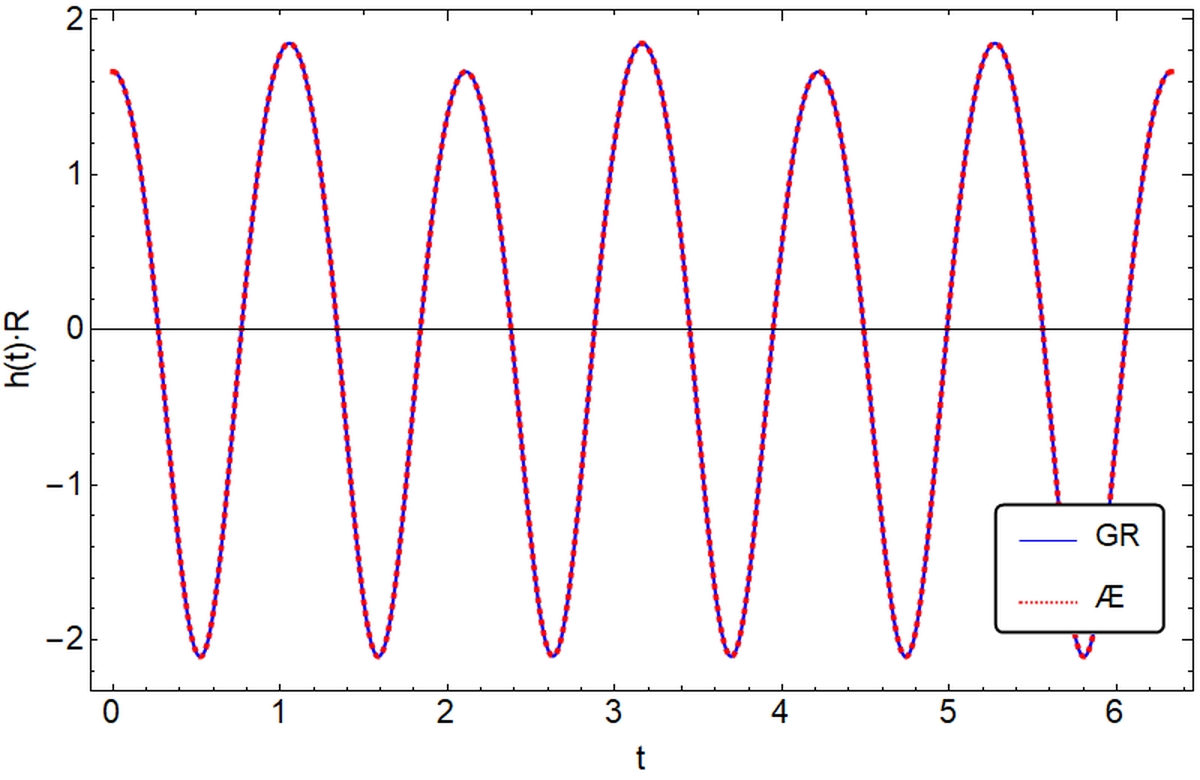}
}
\caption{ The response function $h(t)$ for the Simo's figure-eight 3-body system  in GR and $\ae$-theory, where the modes are propagating along the positive $z$-direction, 
and $(\Omega_1,\; \Omega_2,\; \Omega_3)  = (-0.1, \;  -2.76\times 10^{-6},\;  -2.9\times 10^{-5})$.}
\label{fig3ab}
\end{figure}

\begin{figure}[h!]
{
\includegraphics[width=\linewidth]{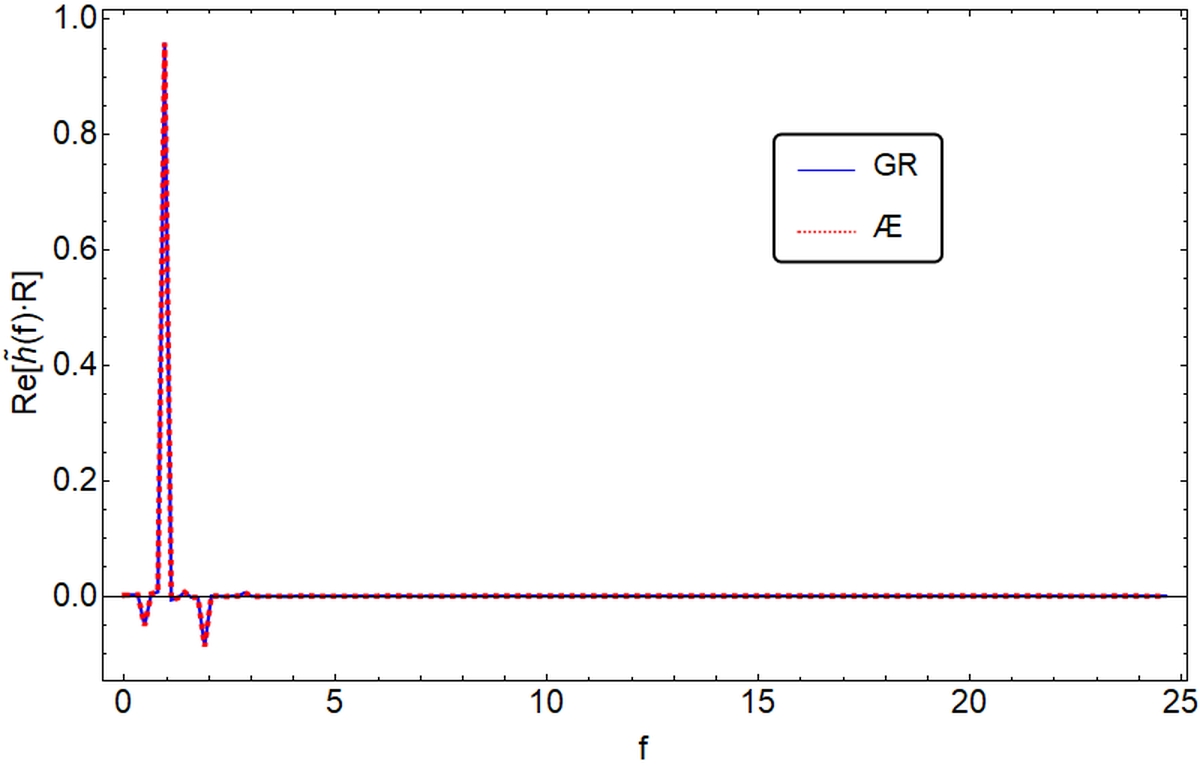}
\includegraphics[width=\linewidth]{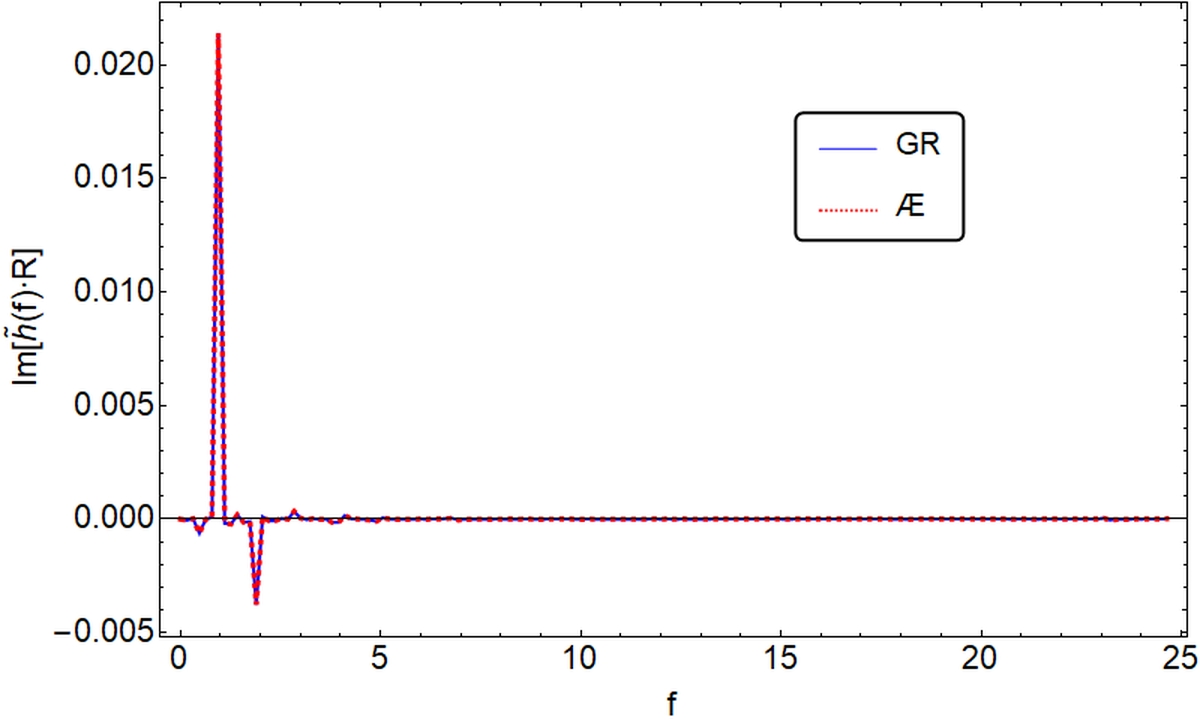}	
	}
	\caption{The Fourier transform $\tilde{h}(f)$ of the response function  $h(t)$  for the  Simo's figure-eight 3-body system  in GR and $\ae$-theory, where the modes are propagating along the positive $z$-direction,
	and $(\Omega_1,\; \Omega_2,\; \Omega_3)  = (-0.1, \;  -2.76\times 10^{-6},\;  -2.9\times 10^{-5})$. }
	\label{fig4ab}
\end{figure}

\begin{figure}[h!]
\includegraphics[width=\linewidth]{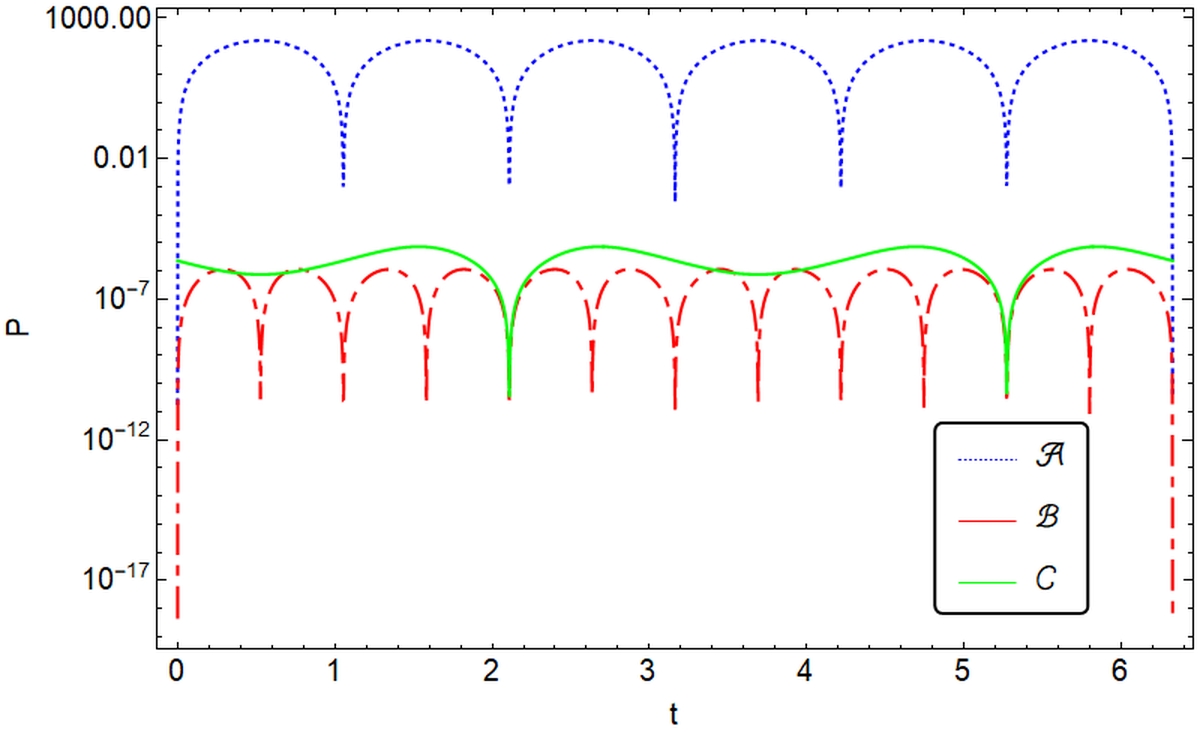}
\caption{ The radiation power $P (\equiv - \dot{\cal{E}})$ of the  Simo's figure-eight 3-body system  in   $\ae$-theory, where the modes are all propagating along the positive $z$-direction,
and $(\Omega_1,\; \Omega_2,\; \Omega_3)  = (-0.1, \;  -2.76\times 10^{-6},\;  -2.9\times 10^{-5})$. 
 The  dotted (blue), dash-dotted (red)  and  solid (green)  lines  denote, respectively, 
the parts of quadrupole, monopole and dipole radiations given in Eq.(\ref{4.1a}).}
\label{fig5ab}
\end{figure}


In Figs. \ref{fig3ab}, \ref{fig4ab} and \ref{fig5ab}, we plot out the corresponding response function $h(t)$, its Fourier transform $\tilde{h}(f)$ and the radiation power $P (\equiv - \dot{\cal{E}})$ for the Simo's figure-eight 3-body system. 
From Fig. \ref{fig5ab} we can see that both of the dipole and monopole  contributions are suppressed with the orders given in Section III. 

Note that in drawing the above figures, we had set    {$c_1=4\times 10^{-5}, c_2=9\times 10^{-5}, c_3=-c_1$, and $c_4=-2\times 10^{-5}$},   a condition that will be adopted for the
rest of this paper.  For such  choices, the coupling constants  $c_i$'s clearly satisfy the theoretical and observational constraints of the $\ae$-theory \cite{OMW18}, given by Eq.(\ref{2.8b}).  

In addition, with these choices, we have $c_{13} = 0$, and then from Eq.(\ref{polarizationsB}) we find that  the vectorial modes $h_{X}$ and $h_{Y}$ are identically zero,
\bq
\lb{hxhy}
h_{X }= h_{Y},\; (c_{13} =0).
\eq
So, in the rest of this paper we only  need to consider the $h_{+},\; h_{\times},\; h_ b$ and $h_{L}$ modes.

\begin{figure}[h!]
{
\includegraphics[width=\linewidth]{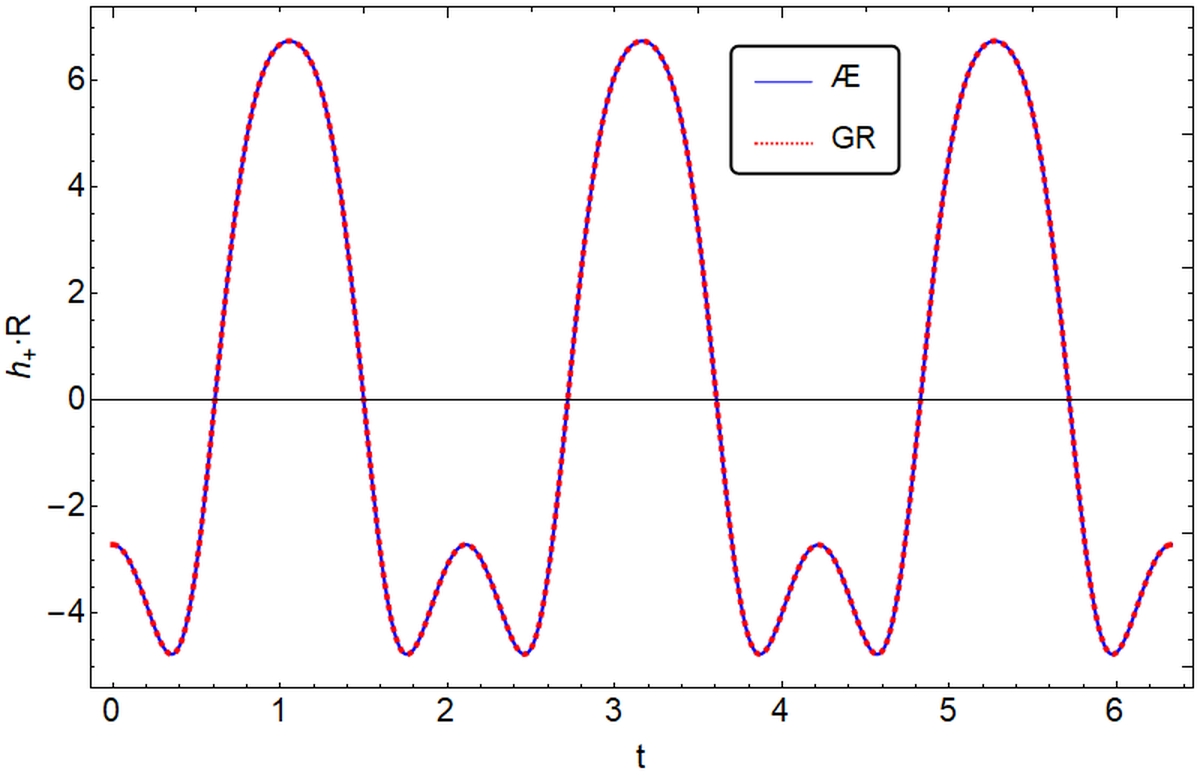}
\includegraphics[width=\linewidth]{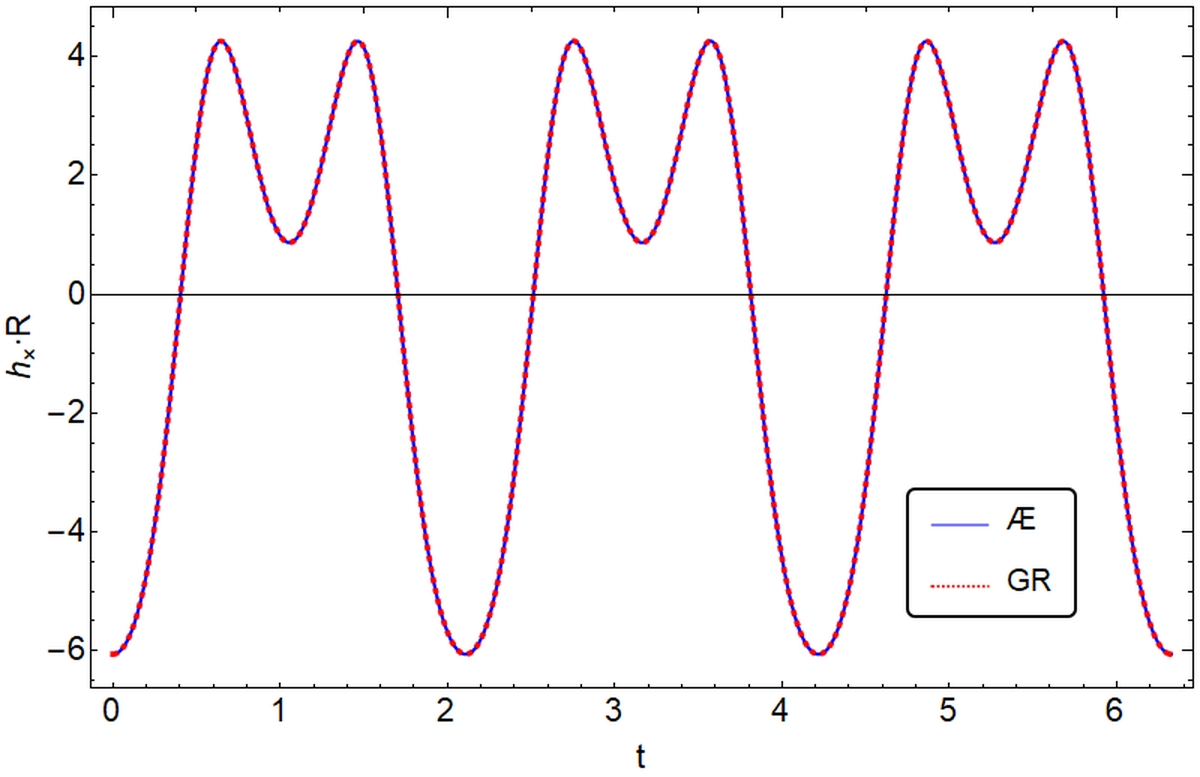}
}
\caption{The polarization modes $h_{+}$ and $h_{\times}$ defined in Eq.(\ref{polarizations}) for the Simo's figure-eight 3-body system  in both GR and $\ae$-theory, 
where the modes are propagating along the direction specified by  $(\vartheta, \varphi;   \theta, \phi, \psi) = (0.6, 5.2; 1.3, 1.2, 1.8)$ with
 $\left(\Omega_1,\; \Omega_2,\; \Omega_3)  = (-0.1, \;  -2.76\times 10^{-6},\;  -2.9\times 10^{-5}\right)$.}
\label{fig2aa1}
\end{figure}

\begin{figure}[h!]
{
\includegraphics[width=\linewidth]{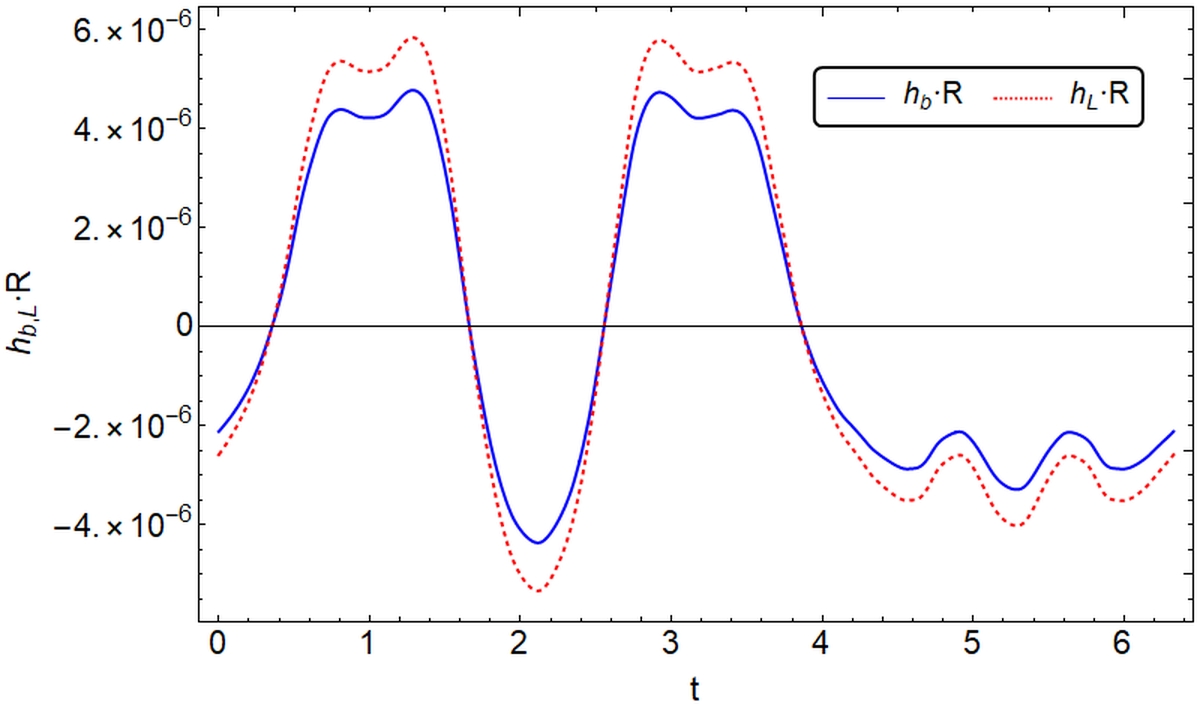}
}
\caption{The polarization modes $h_{b}$ and  $h_{L}$ defined in Eq.(\ref{polarizations}) for the Simo's figure-eight 3-body system  in $\ae$-theory, 
where the modes are propagating along the direction specified by  $(\vartheta, \varphi;   \theta, \phi, \psi) = (0.6, 5.2; 1.3, 1.2, 1.8)$ with
 $\left(\Omega_1,\; \Omega_2,\; \Omega_3)  = (-0.1, \;  -2.76\times 10^{-6},\;  -2.9\times 10^{-5}\right)$.}
\label{fig2aa2}
\end{figure}

\begin{figure}[h!]
{
\includegraphics[width=\linewidth]{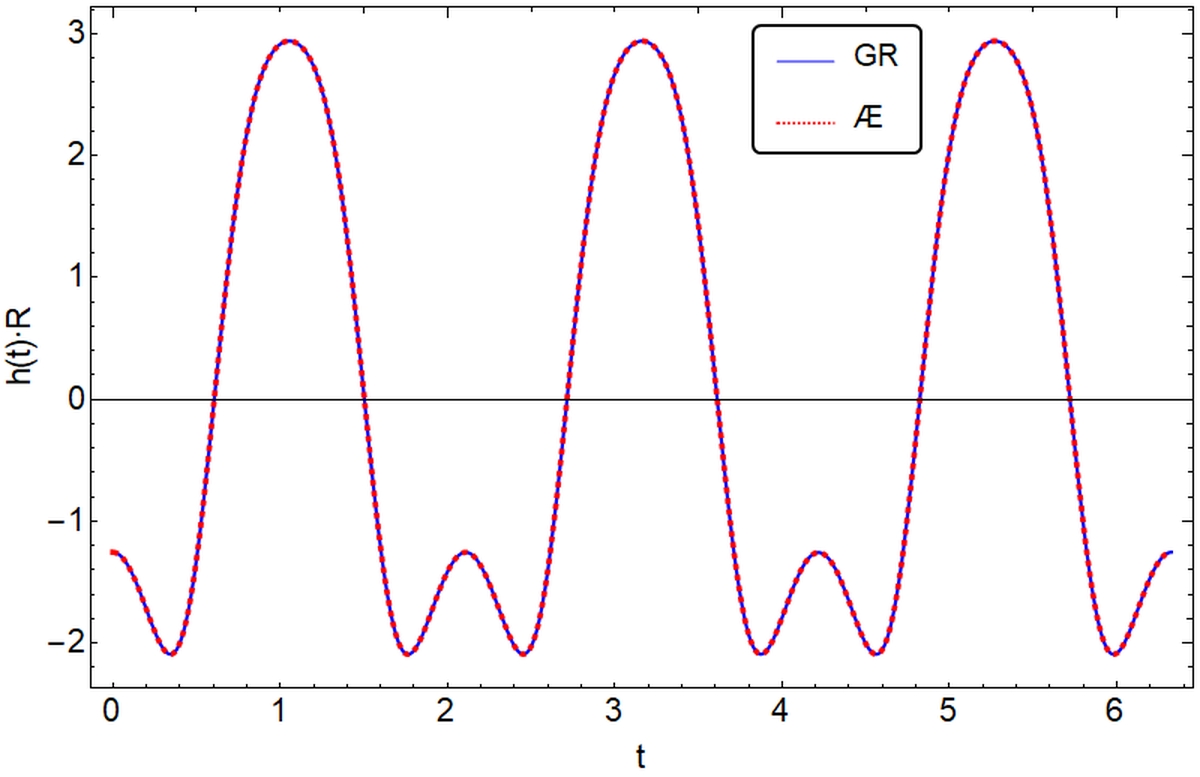}
}
\caption{ The response function $h(t)$ for the Simo's figure-eight 3-body system  in GR and $\ae$-theory, 
where the modes are propagating along the direction specified by  $(\vartheta, \varphi;   \theta, \phi, \psi) = (0.6, 5.2; 1.3, 1.2, 1.8)$ with
 $\left(\Omega_1,\; \Omega_2,\; \Omega_3)  = (-0.1, \;  -2.76\times 10^{-6},\;  -2.9\times 10^{-5}\right)$. }
\label{fig3aa}
\end{figure}

\begin{figure}[h!]
{
\includegraphics[width=\linewidth]{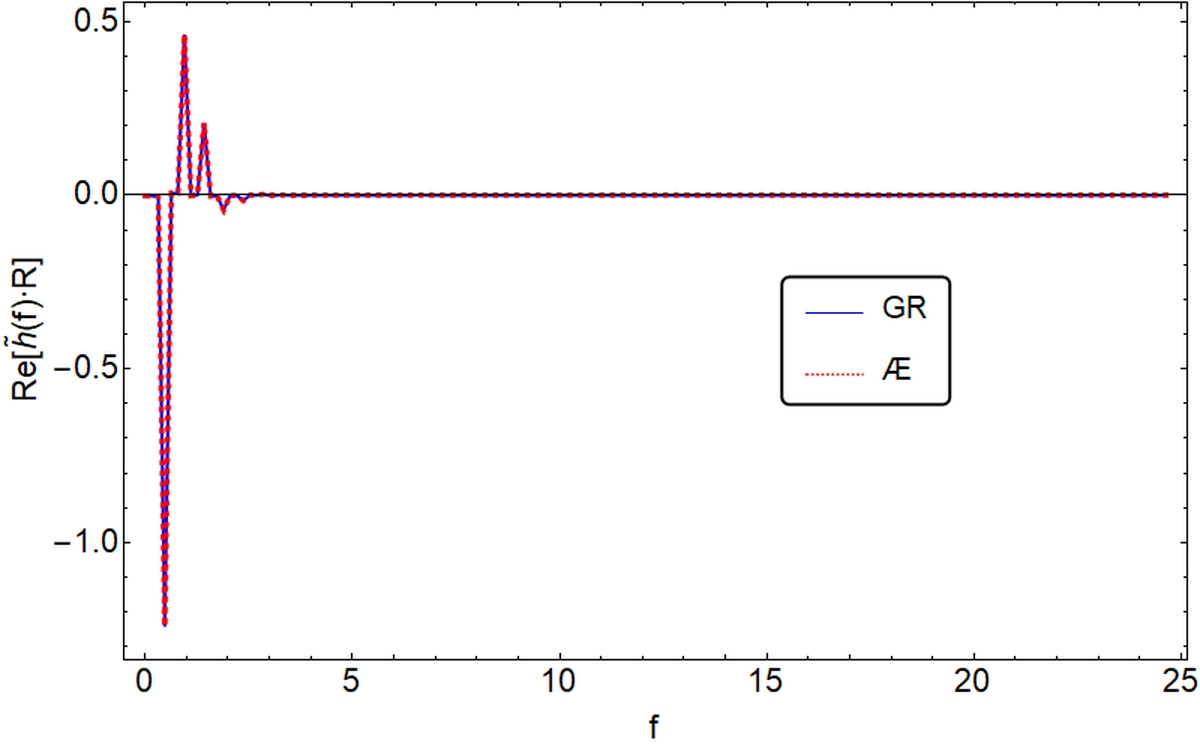}
\includegraphics[width=\linewidth]{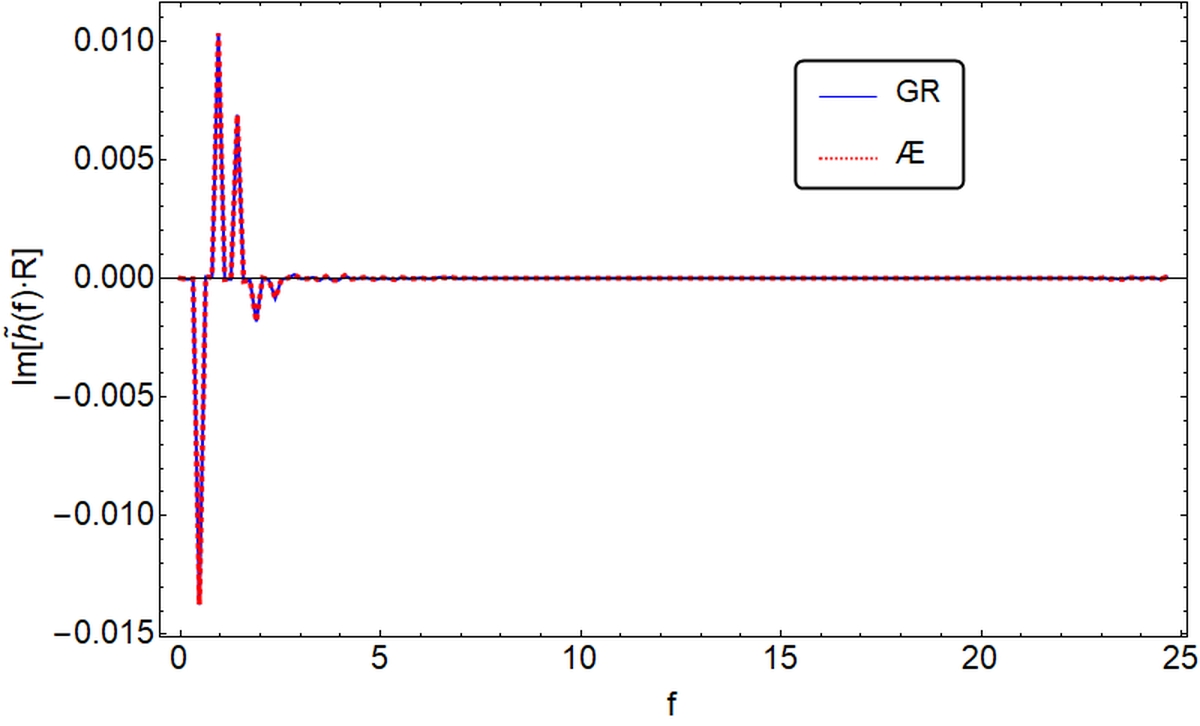}	
	}
	\caption{The Fourier transform $\tilde{h}(f)$ of the response function  $h(t)$  for the  Simo's figure-eight 3-body system  in GR and $\ae$-theory, 
	where the modes are propagating along the direction specified by  $(\vartheta, \varphi;   \theta, \phi, \psi) = (0.6, 5.2; 1.3, 1.2, 1.8)$ with
 $\left(\Omega_1,\; \Omega_2,\; \Omega_3)  = (-0.1, \;  -2.76\times 10^{-6},\;  -2.9\times 10^{-5}\right)$.}
	\label{fig4aa}
\end{figure}

\begin{figure}[h!]
\includegraphics[width=\linewidth]{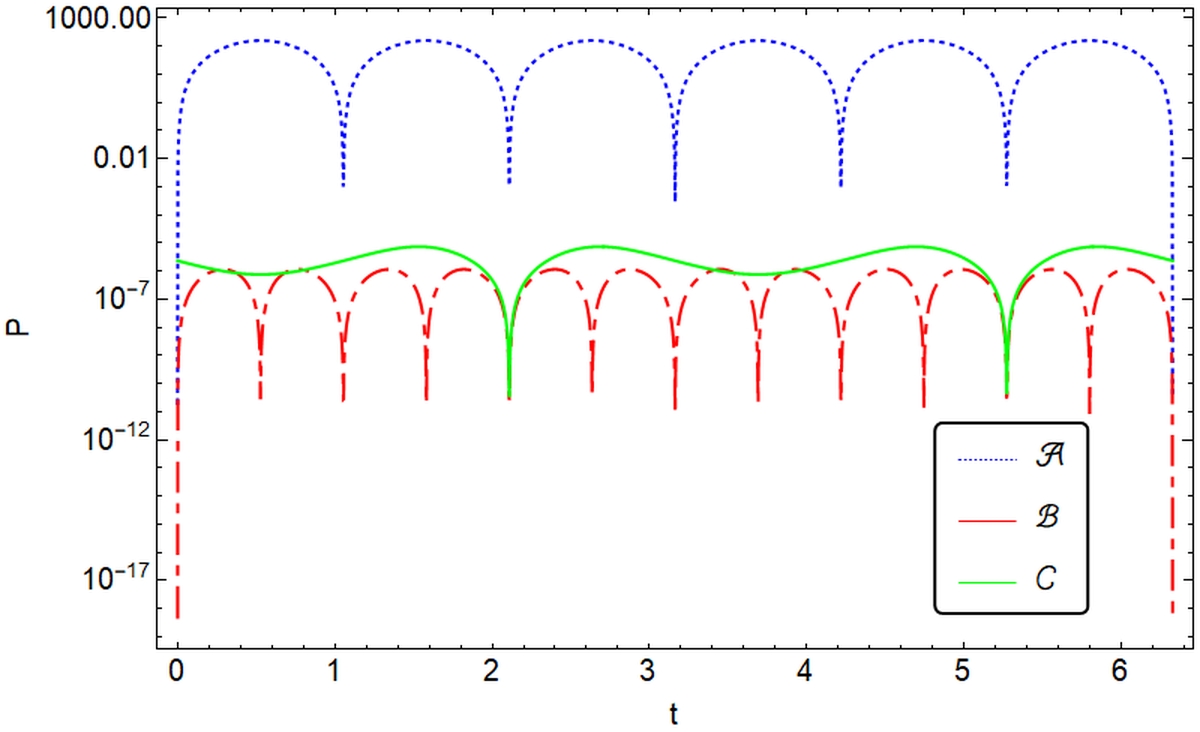}
\caption{ The radiation power $P (\equiv - \dot{\cal{E}})$ of the  Simo's figure-eight 3-body system  in   $\ae$-theory, 
where the modes are propagating along the direction specified by  $(\vartheta, \varphi;   \theta, \phi, \psi) = (0.6, 5.2; 1.3, 1.2, 1.8)$ with
 $\left(\Omega_1,\; \Omega_2,\; \Omega_3)  = (-0.1, \;  -2.76\times 10^{-6},\;  -2.9\times 10^{-5}\right)$. 
 The  dotted (blue), dash-dotted (red)  and  solid (green)  lines  denote, respectively, 
the parts of quadrupole, monopole and dipole radiations given in Eq.(\ref{4.1a}).}
\label{fig5aa}
\end{figure}

Yet, we also plot the above figures by assuming that the orbit is in the ($x, y$)-plane, while the detector is along the $z$-axis. The same conditions were also assumed in  \cite{DSH14}. 
However, as we mentioned in the last section, the locations and orientations of the sources as well as the detectors are all independent, which are specified by the five angles $(\vartheta, \varphi;   \theta, \phi, \psi)$, defined 
in Eqs.(\ref{rotationsA}) and (\ref{rotationsB}). For different choices of these parameters, the wave forms and response function   will be also different. In Figs. \ref{fig2aa1}-\ref{fig5aa} we plot the mode
functions $h_{N}$, response function $h(t)$, its Fourier transform $\tilde{h}(f)$ and  the radiation powers $P_{{\cal{A}}}$, $P_{{\cal{B}}}$ and $P_{{\cal{C}}}$, respectively, for $(\vartheta, \varphi;   \theta, \phi, \psi) = (0.6, 5.2; 1.3, 1.2, 1.8)$,
while still chose $(\Omega_1,\; \Omega_2,\; \Omega_3)  = (-0.1, \;  -2.76\times 10^{-6},\;  -2.9\times 10^{-5})$. 
Clearly, the corresponding mode functions, response function and its Fourier transform are all different from the case in which the 3-bodies are in the $(x, y$)-plane, while the detector is localized along the $z$-axis. However, 
 the radiation powers are the same and are independent of the choice of these five angular parameters, as can be seen from Figs. \ref{fig5ab} and \ref{fig5aa}.


\begin{figure}[h!]
{
\includegraphics[width=\linewidth]{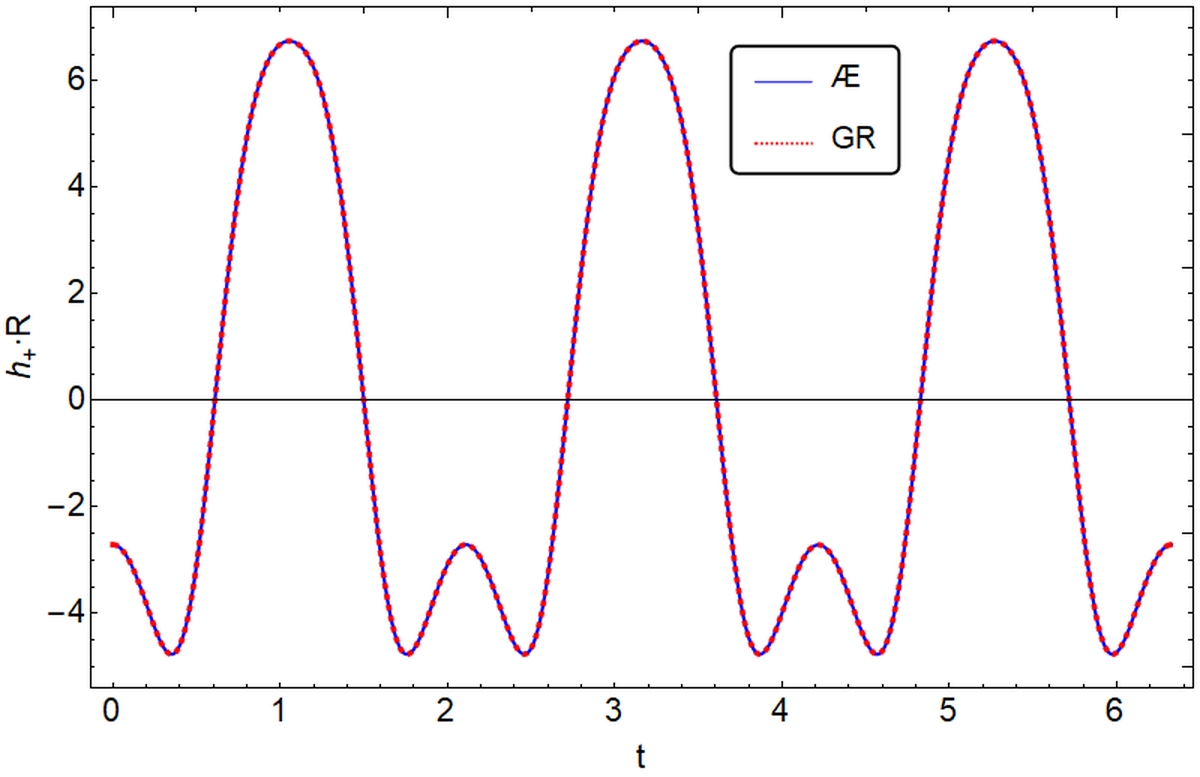}
\includegraphics[width=\linewidth]{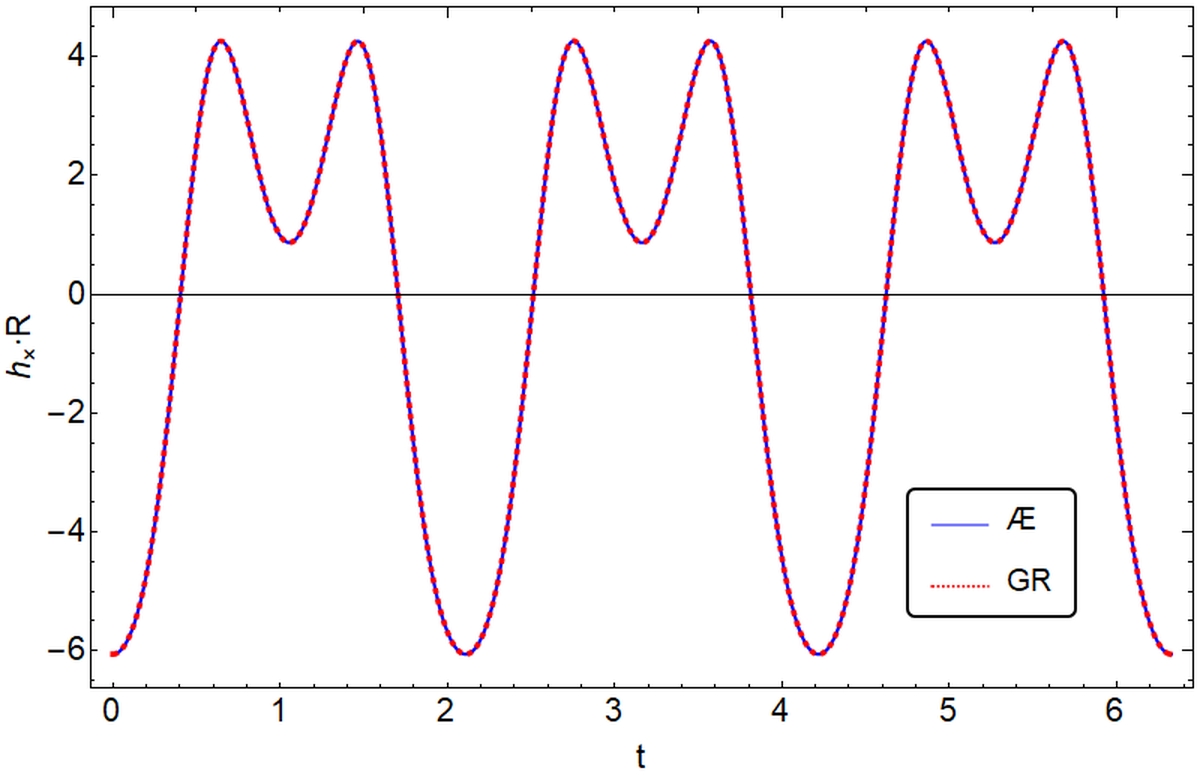}
}
\caption{The polarization modes $h_{+}$ and $h_{\times}$ defined in Eq.(\ref{polarizations}) for the Simo's figure-eight 3-body system  in both GR and $\ae$-theory, 
where the modes are propagating along the direction specified by  $(\vartheta, \varphi;   \theta, \phi, \psi) = (0.6, 5.2; 1.3, 1.2, 1.8)$,
with $\Omega_1 = \Omega_2 = \Omega_3 = -10^{-2}$.}
\label{fig2ac1}
\end{figure}

\begin{figure}[h!]
{
\includegraphics[width=\linewidth]{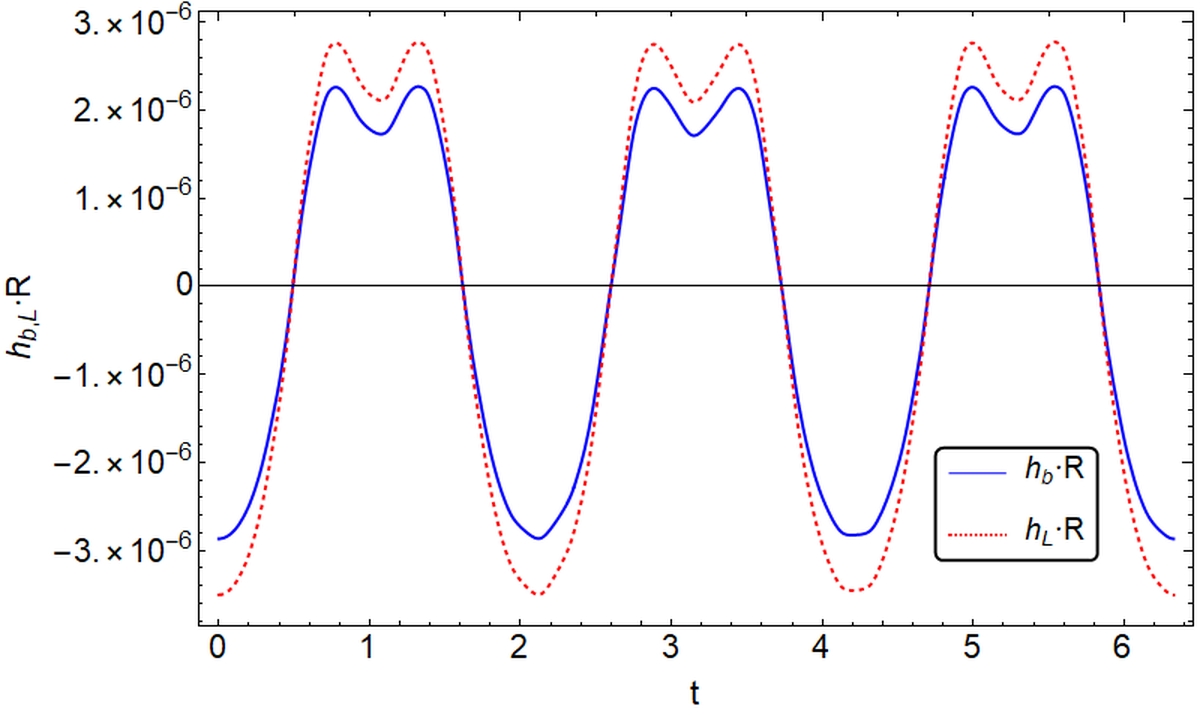}
}
\caption{The polarization modes $h_{b}$ and  $h_{L}$ defined in Eq.(\ref{polarizations}) for the Simo's figure-eight 3-body system  in $\ae$-theory, 
where the modes are propagating along the direction specified by  $(\vartheta, \varphi;   \theta, \phi, \psi) = (0.6, 5.2; 1.3, 1.2, 1.8)$,
with $\Omega_1 = \Omega_2 = \Omega_3 = -10^{-2}$.}
\label{fig2ac2}
\end{figure}

\begin{figure}[h!]
{
\includegraphics[width=\linewidth]{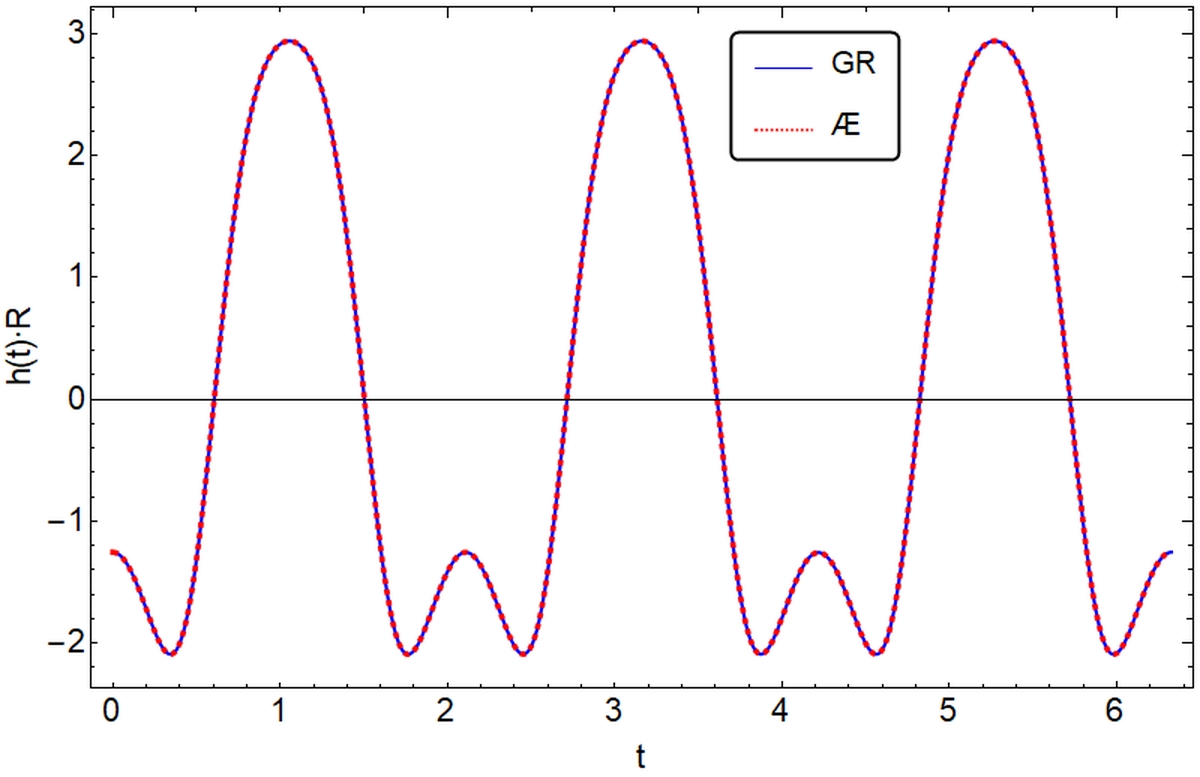}
}
\caption{ The response function $h(t)$ for the Simo's figure-eight 3-body system  in GR and $\ae$-theory, 
where the modes are propagating along the direction specified by  $(\vartheta, \varphi;   \theta, \phi, \psi) = (0.6, 5.2; 1.3, 1.2, 1.8)$,
with $\Omega_1 = \Omega_2 = \Omega_3 = -10^{-2}$. }
\label{fig3ac}
\end{figure}

\begin{figure}[h!]
{
\includegraphics[width=\linewidth]{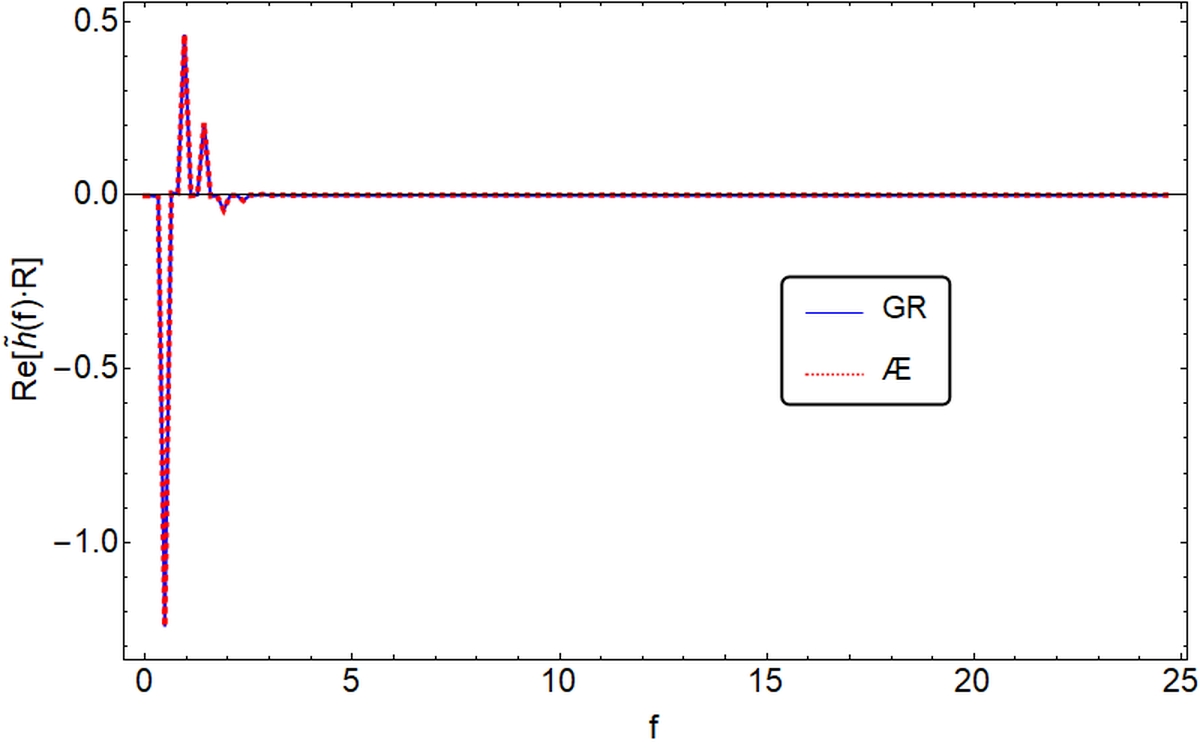}
\includegraphics[width=\linewidth]{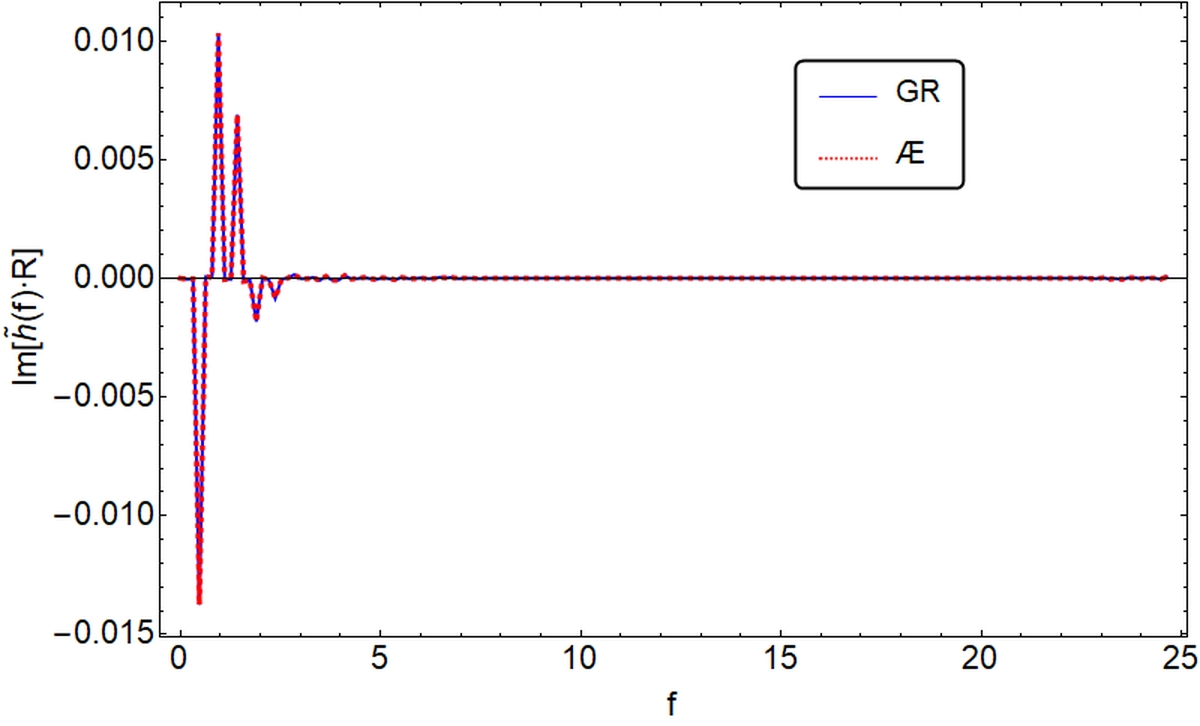}	
	}
	\caption{The Fourier transform $\tilde{h}(f)$ of the response function  $h(t)$  for the  Simo's figure-eight 3-body system  in GR and $\ae$-theory, 
	where the modes are propagating along the direction specified by  $(\vartheta, \varphi;   \theta, \phi, \psi) = (0.6, 5.2; 1.3, 1.2, 1.8)$,
with $\Omega_1 = \Omega_2 = \Omega_3 = -10^{-2}$.  }
	\label{fig4ac}
\end{figure}

\begin{figure}[h!]
\includegraphics[width=\linewidth]{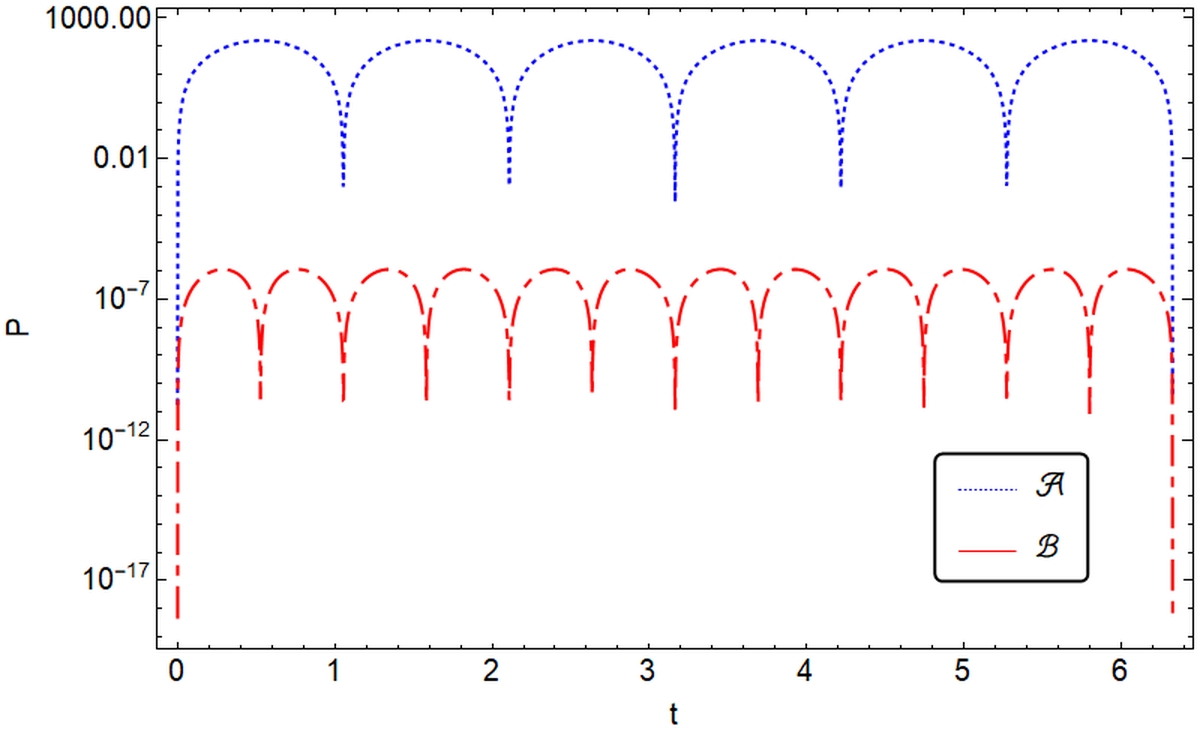}
\caption{The radiation power $P (\equiv - \dot{\cal{E}})$ of the  Simo's figure-eight trajectory of 3-body system in GR and $\ae$-theory, 
where the modes are propagating along the direction specified by  $(\vartheta, \varphi;   \theta, \phi, \psi) = (0.6, 5.2; 1.3, 1.2, 1.8)$,
with $\Omega_1 = \Omega_2 = \Omega_3 = -10^{-2}$.  The  dotted (blue)  and  solid (red)  lines denote, respectively,  the parts of quadrupole and 
monopole  radiations given in Eq.(\ref{4.1a}). Because of the choice of the binding energies
$\Omega_a$ and masses $m_a$ are all the same for the three compact objects, the dipole contributions, denoted by the ${\cal{C}}$ part in Eq.(\ref{4.1a}), are identically zero.}
\label{fig5ac}
\end{figure}

To study the effects of the binding energies of the three bodies on the wave forms and energy losses, let us consider the same case as shown by Figs. \ref{fig2aa1} - \ref{fig5aa} but now with the same binding energy,
$\Omega_a=-10^{-2}$. With this choice,  the dipole contributions are identically zero, as one can see from Eqs.(\ref{4.3}) and (\ref{4.7}), since now we have ${\cal D}_c = 0, \; (c =1, 2, 3)$ and
$\dot{\Sigma}_i^2 = 0$. This does not contradict with the results given by Eq.(\ref{4.1kb}), as there we assumed that $\Omega_1/m_1 - \Omega_2/m_2 \simeq {\cal{O}}\left(\Omega/m\right)$ [cf. Eqs.(\ref{4.1f}) and (\ref{4.1j})].
But, here due to our choice of $\Omega_a$ and $m_a$, we have $\Omega_1/m_1 - \Omega_2/m_2 = 0$ and so on. 
In Figs. \ref{fig2ac1} - \ref{fig5ac} we plot the mode
functions $h_{N}$, response function $h(t)$, its Fourier transform $\tilde{h}(f)$ and  the radiation powers $P_{{\cal{A}}}$ and $P_{{\cal{B}}}$, respectively, for $(\vartheta, \varphi;   \theta, \phi, \psi) = (0.6, 5.2; 1.3, 1.2, 1.8)$,
while setting  $(\Omega_1,\; \Omega_2,\; \Omega_3)  = (-10^{-2},   -10^{-2}, -10^{-2})$. 
Clearly, the corresponding mode functions, response function, its Fourier transform and  radiation powers are all different from the case in which the 3-bodies are in the $(x, y$)-plane, while the detector is localized along the $z$-axis. 
In particular, the dipole contributions vanish now, as explained above.


\begin{figure}[h!]
		\includegraphics[width=\linewidth]{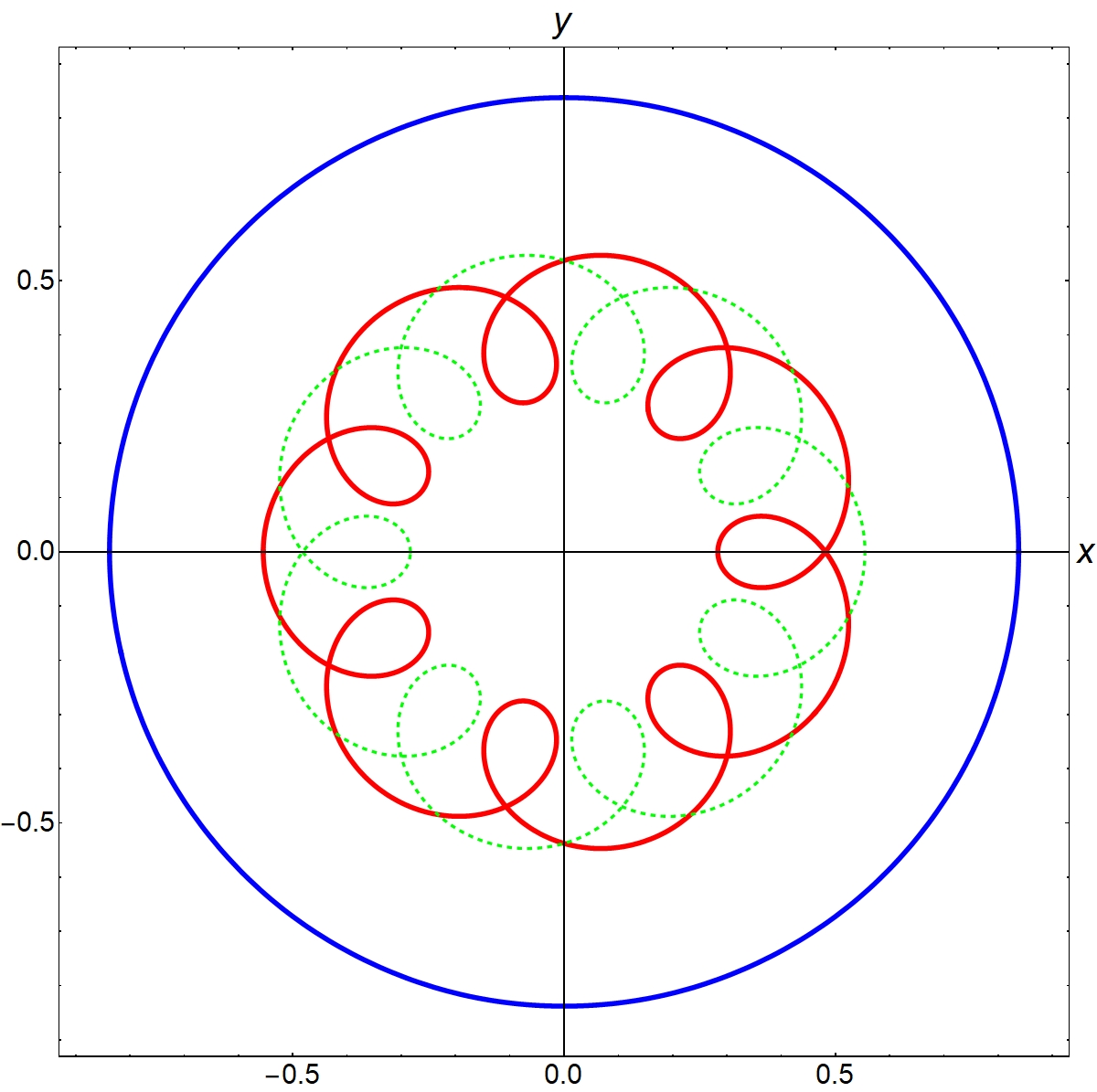}	
	\caption{Trajectory of the 3-body system for the Broucke R7 figure provided in \cite{Website}.}
	\label{fig1c}
\end{figure}

\begin{figure}[h!]
	{
		\includegraphics[width=\linewidth]{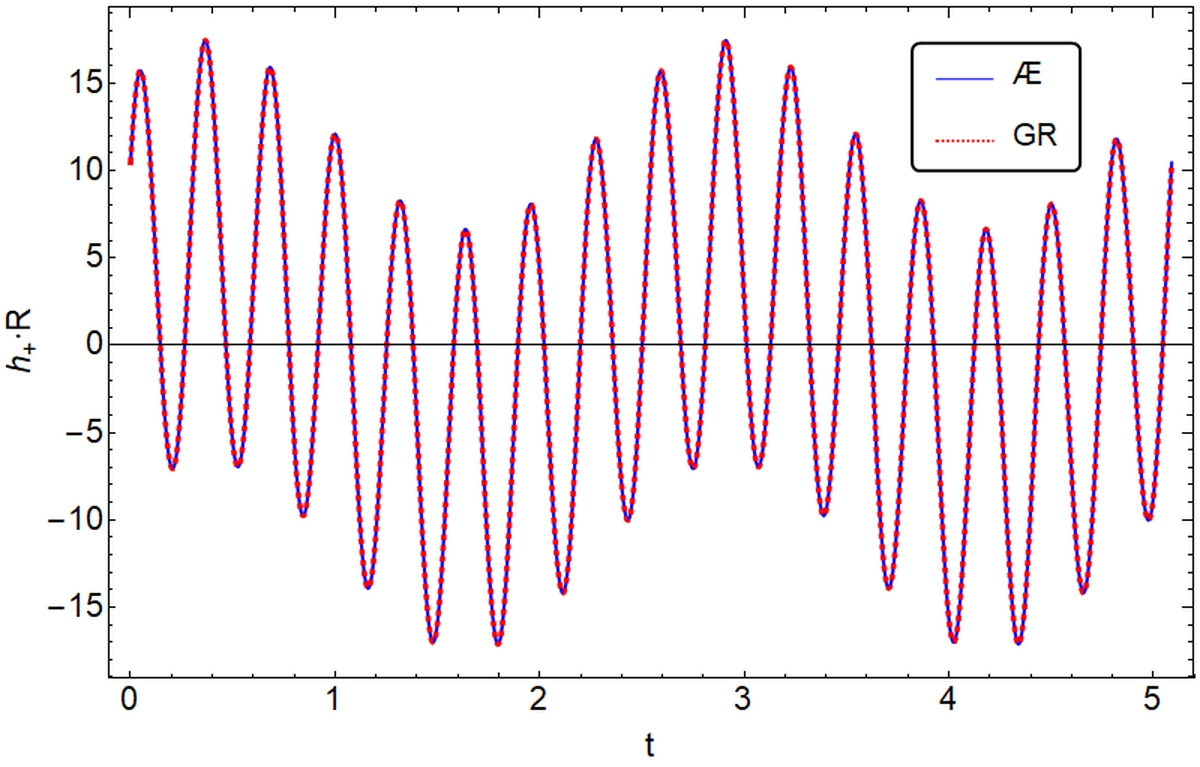}
\includegraphics[width=\linewidth]{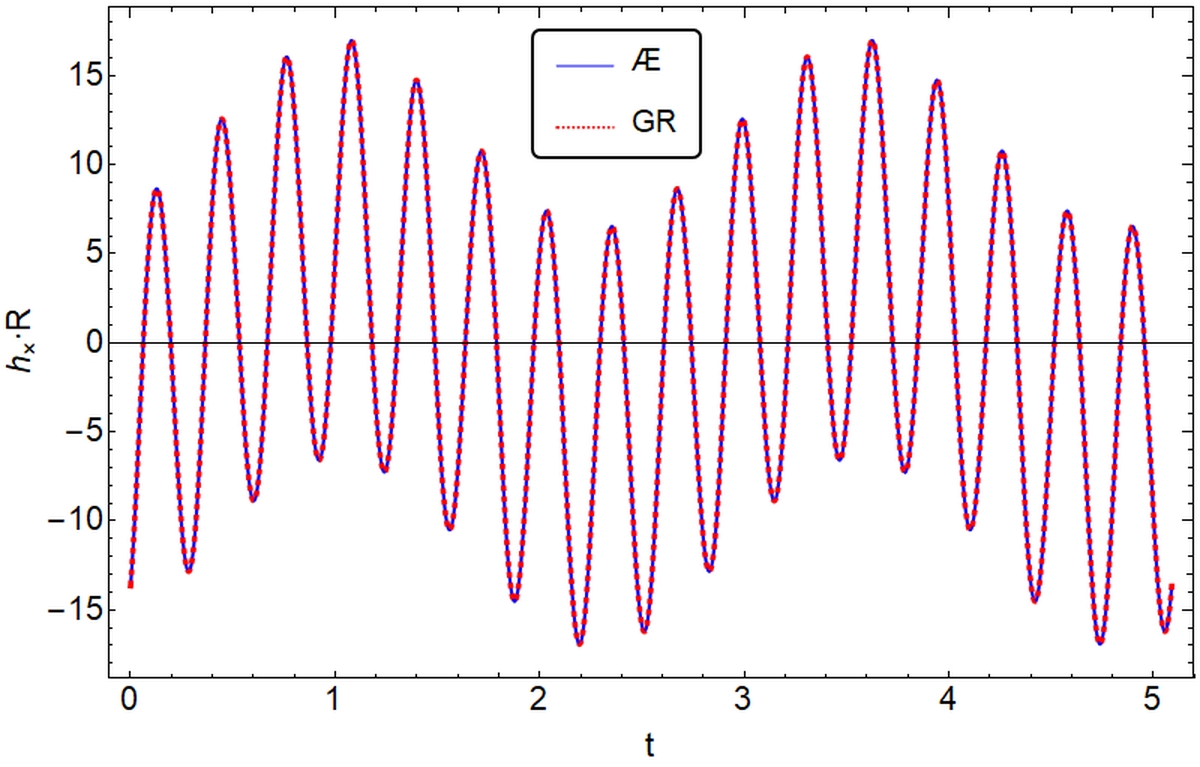}
\includegraphics[width=\linewidth]{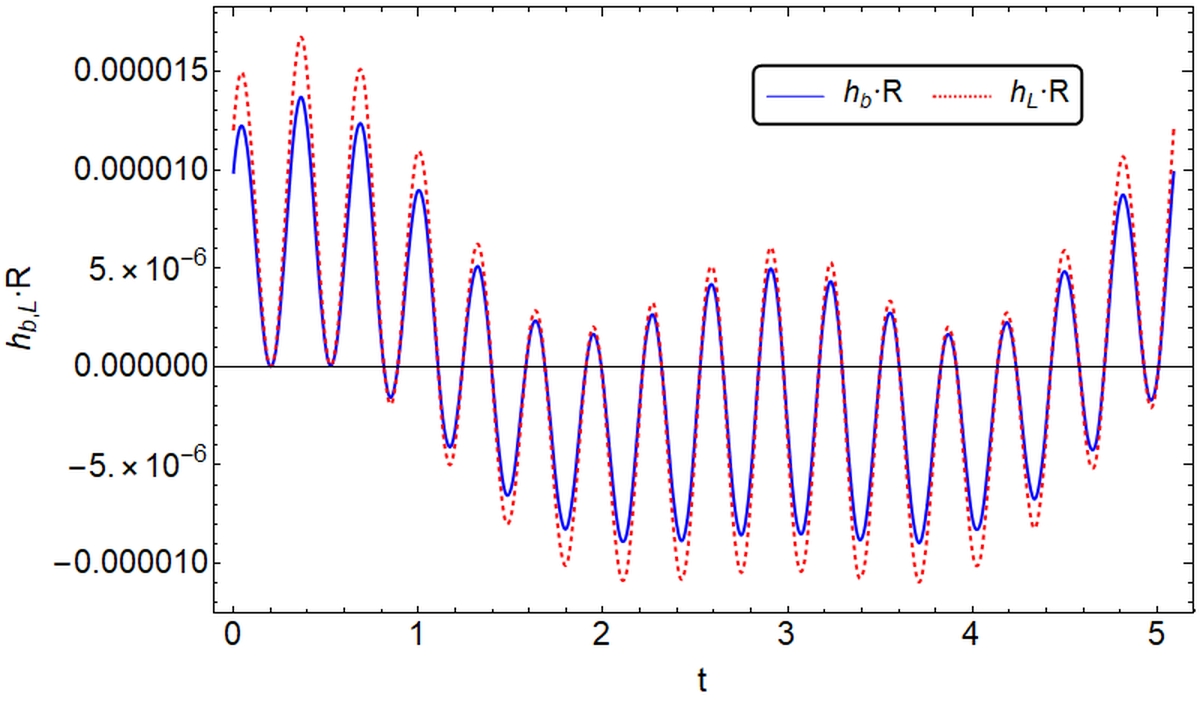}
		
	}
	\caption{The polarization modes $h_{N}$ defined in Eq.(\ref{polarizations}) for the Broucke R7 3-body system with the choice,
	$(\vartheta, \varphi;   \theta, \phi, \psi) = (0.6, 5.2; 1.3, 1.2, 1.8)$ and $(\Omega_1,\; \Omega_2,\; \Omega_3)  = (-0.1, \;  -2.76\times 10^{-6},\;  -2.9\times 10^{-5})$.}
	\label{fig2c}
\end{figure}

\begin{figure}[h!]
	{
		\includegraphics[width=\linewidth]{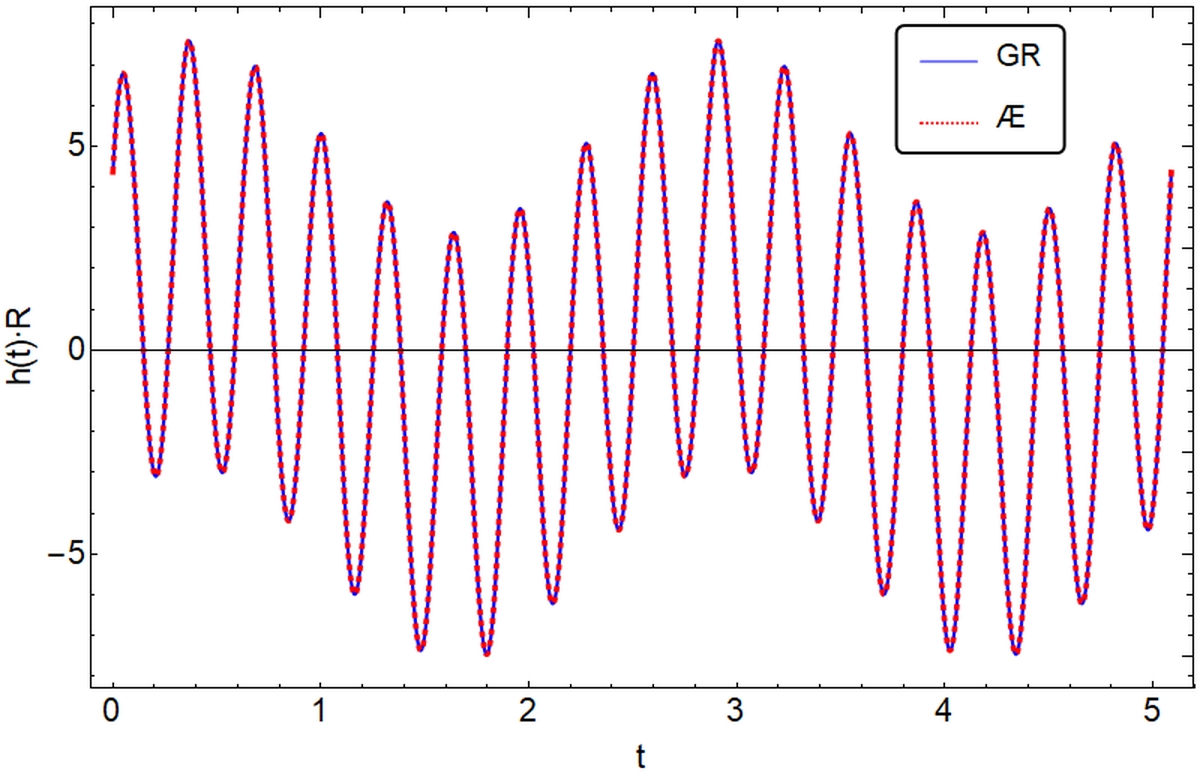}
		
	}
	\caption{ The response function $h(t)$ for the Broucke R7  3-body system with the choice,
	$(\vartheta, \varphi;   \theta, \phi, \psi) = (0.6, 5.2; 1.3, 1.2, 1.8)$ and $(\Omega_1,\; \Omega_2,\; \Omega_3)  = (-0.1, \;  -2.76\times 10^{-6},\;  -2.9\times 10^{-5})$. }
	\label{fig3c}
\end{figure}

\begin{figure}[h!]
	{
		\includegraphics[width=\linewidth]{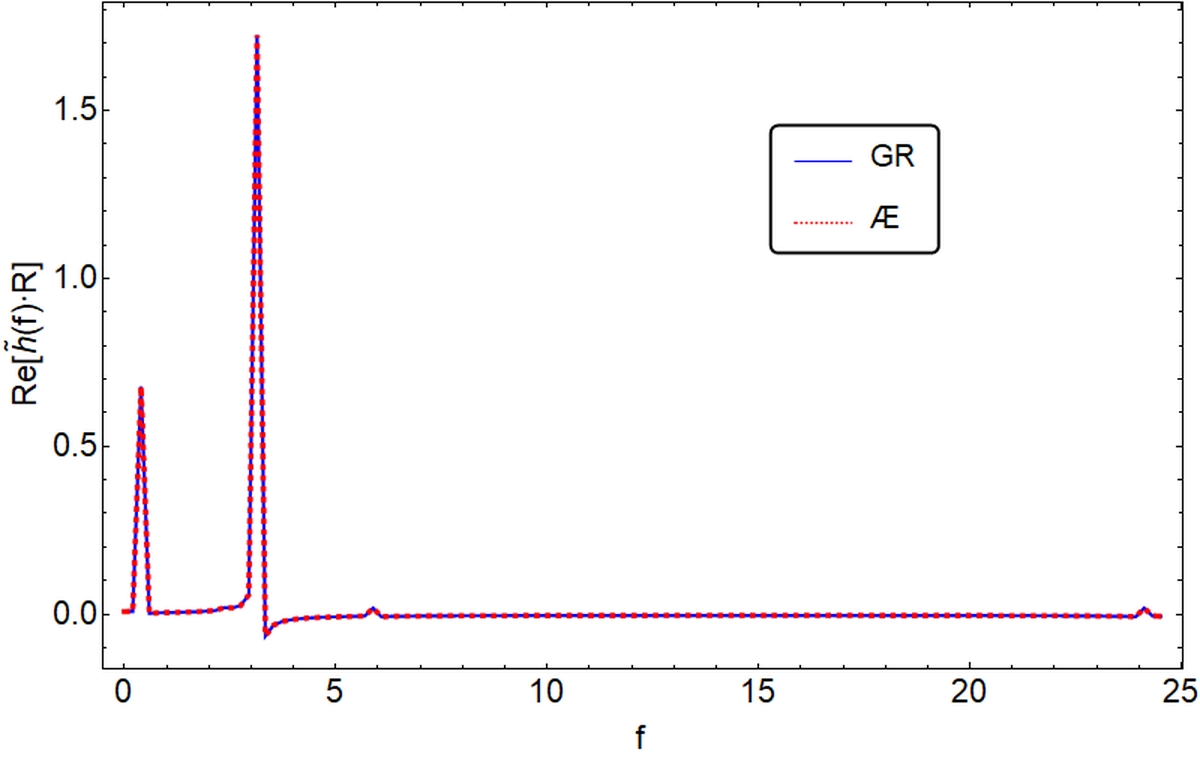}
\includegraphics[width=\linewidth]{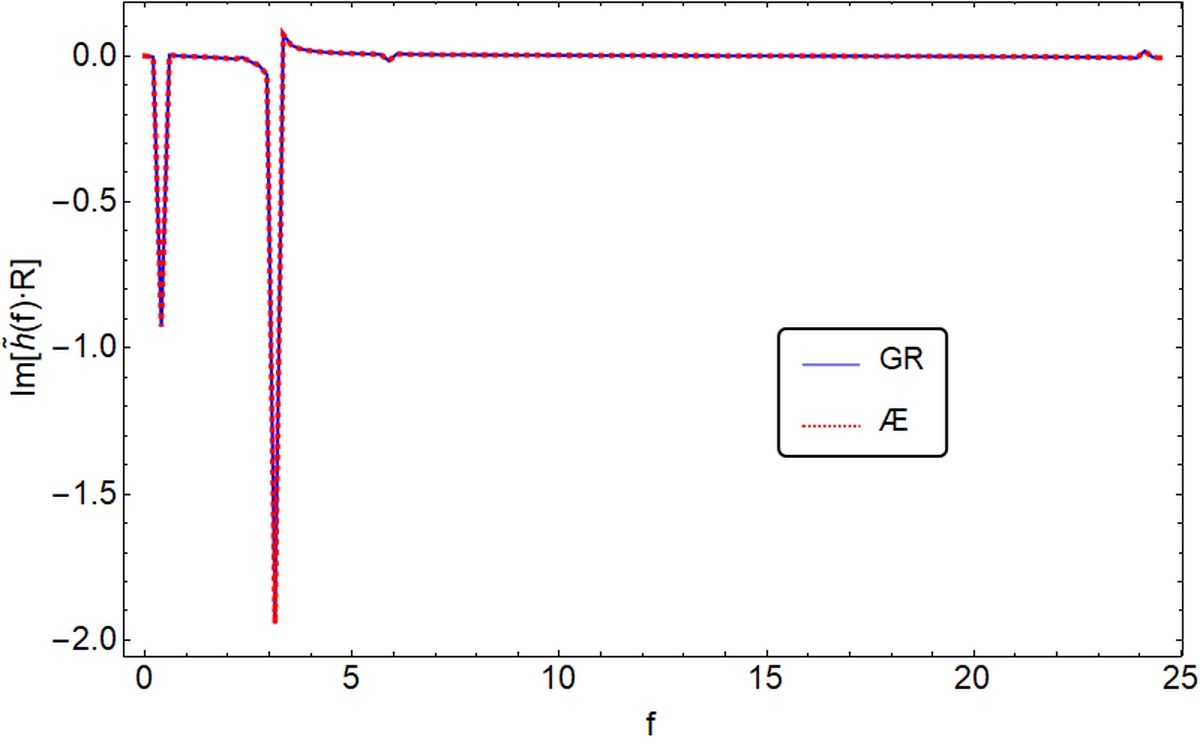}
		
	}
	\caption{The Fourier transform $\tilde{h}(f)$ of the response function  $h(t)$  for the Broucke R7 3-body system with the choice,
	$(\vartheta, \varphi;   \theta, \phi, \psi) = (0.6, 5.2; 1.3, 1.2, 1.8)$ and $(\Omega_1,\; \Omega_2,\; \Omega_3)  = (-0.1, \;  -2.76\times 10^{-6},\;  -2.9\times 10^{-5})$.  }
	\label{fig4c}
\end{figure}

\begin{figure}[h!]
\includegraphics[width=\linewidth]{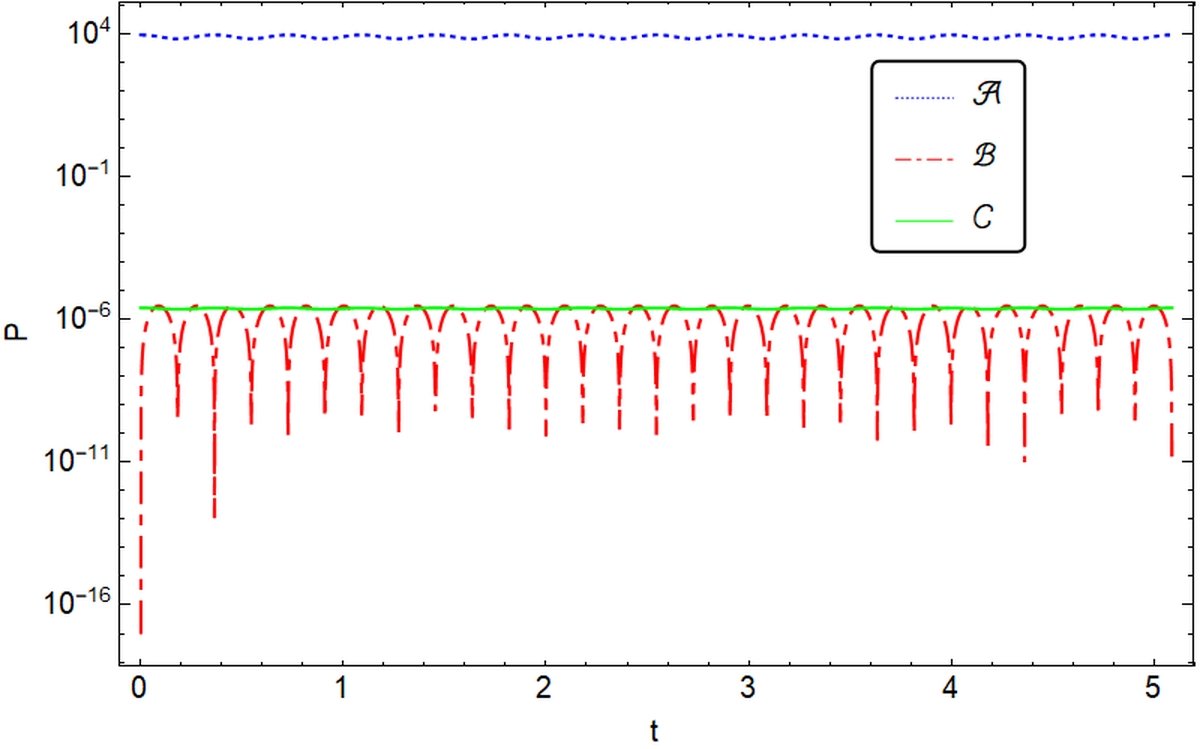}
\caption{The radiation power $P (\equiv - \dot{\cal{E}})$ of the 3-body system of the Broucke R7 figure with the choice,
	$(\vartheta, \varphi;   \theta, \phi, \psi) = (0.6, 5.2; 1.3, 1.2, 1.8)$ and $(\Omega_1,\; \Omega_2,\; \Omega_3)  = (-0.1, \;  -2.76\times 10^{-6},\;  -2.9\times 10^{-5})$.
	The  dotted (blue), dash-dotted (red)  and  solid (green)  lines  denote, respectively, 
the parts of quadrupole, monopole and dipole radiations given in Eq.(\ref{4.1a}). }
	\label{fig5c}
\end{figure}

In Figs. \ref{fig1c}-\ref{fig5c}, we plot out, respectively,  the trajectory of the Broucke R7 3-body system provided in \cite{Website}, the polarization modes $h_{N}$, the response function  $h(t)$, its
Fourier transform $\tilde{h}(f)$ and the radiation powers  $P_{{\cal{A}}}$, $P_{{\cal{B}}}$ and $P_{{\cal{C}}}$,  for $(\vartheta, \varphi;   \theta, \phi, \psi) = (0.6, 5.2; 1.3, 1.2, 1.8)$,
and $(\Omega_1,\; \Omega_2,\; \Omega_3)  = (-0.1, \;  -2.76\times 10^{-6},\;  -2.9\times 10^{-5})$.

\begin{figure}[h!]
	{
		\includegraphics[width=\linewidth]{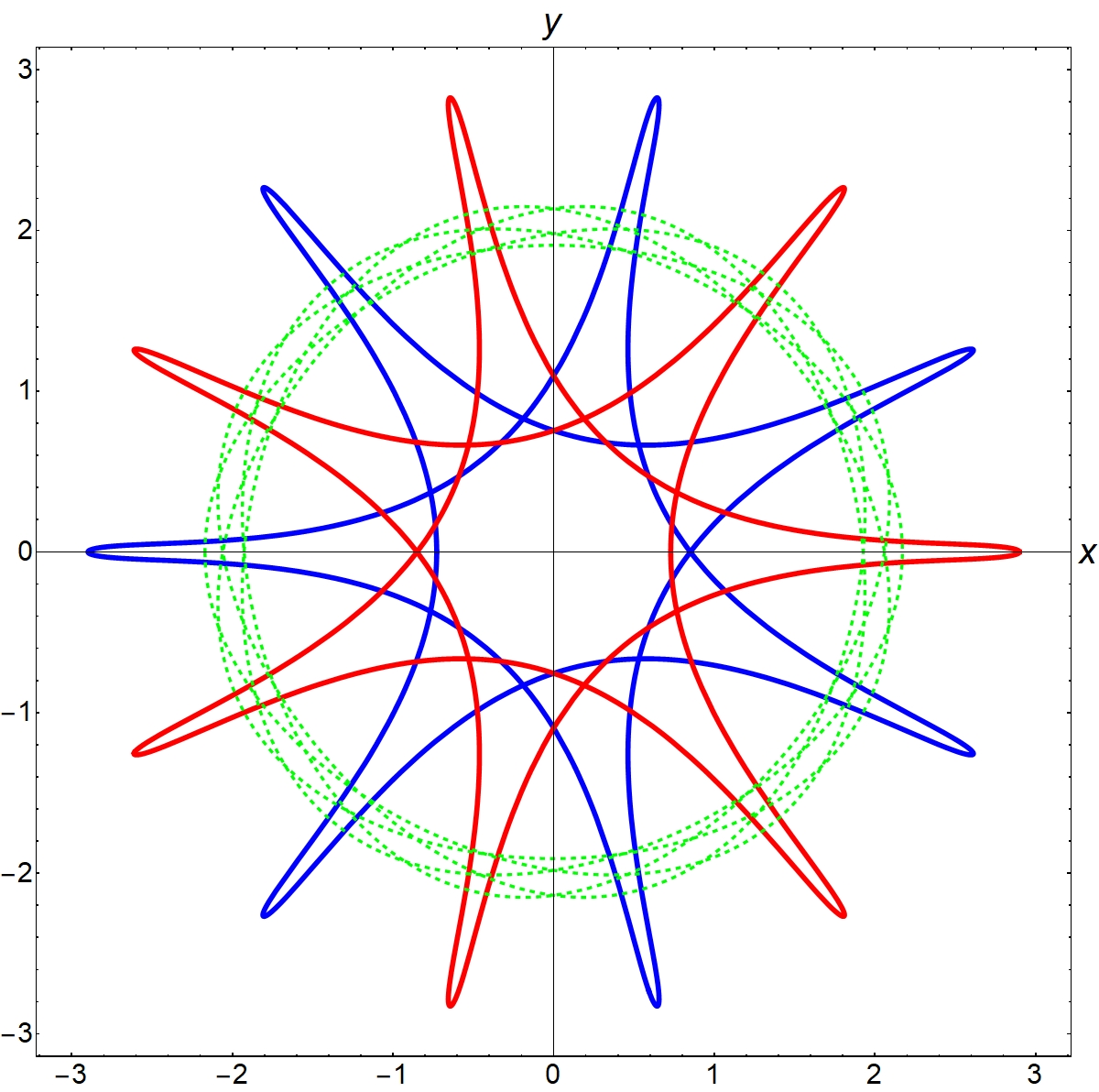}
		
	}
	\caption{Trajectory of the 3-body system for the Broucke A16 figure provided in \cite{Website}. }
	\label{fig1d}
\end{figure}

\begin{figure}[h!]
	{
		\includegraphics[width=\linewidth]{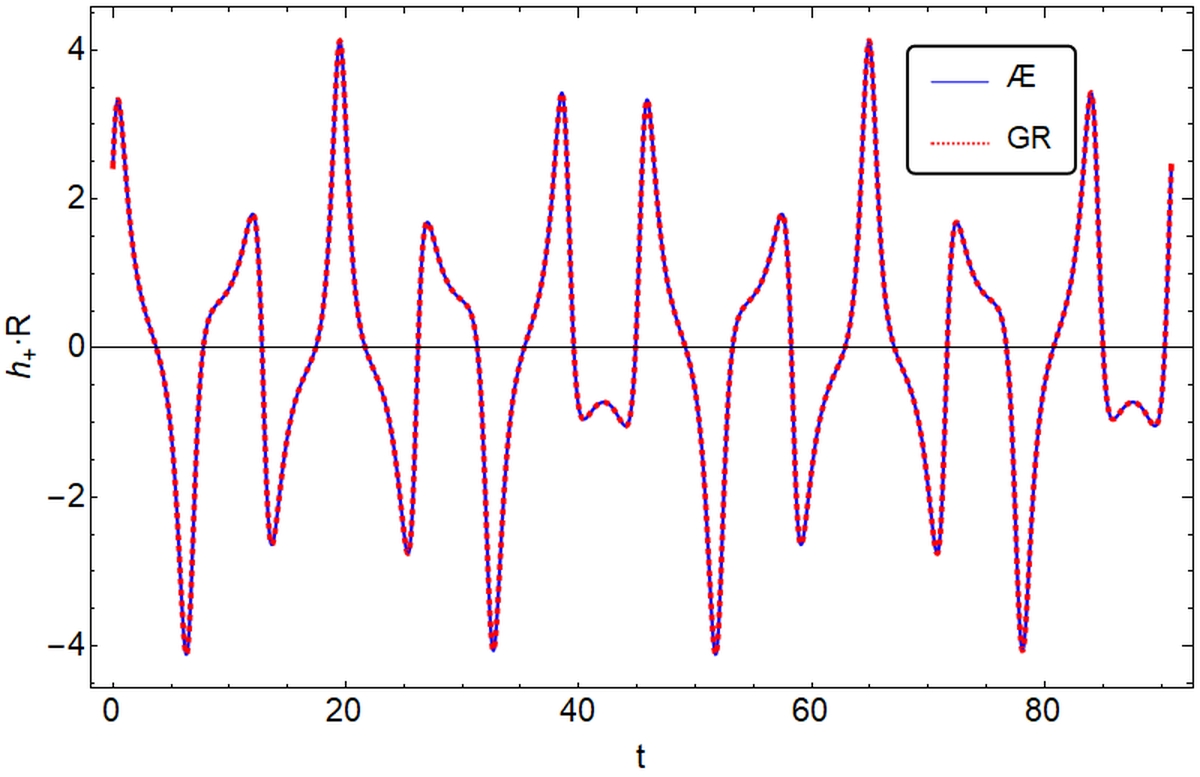}
\includegraphics[width=\linewidth]{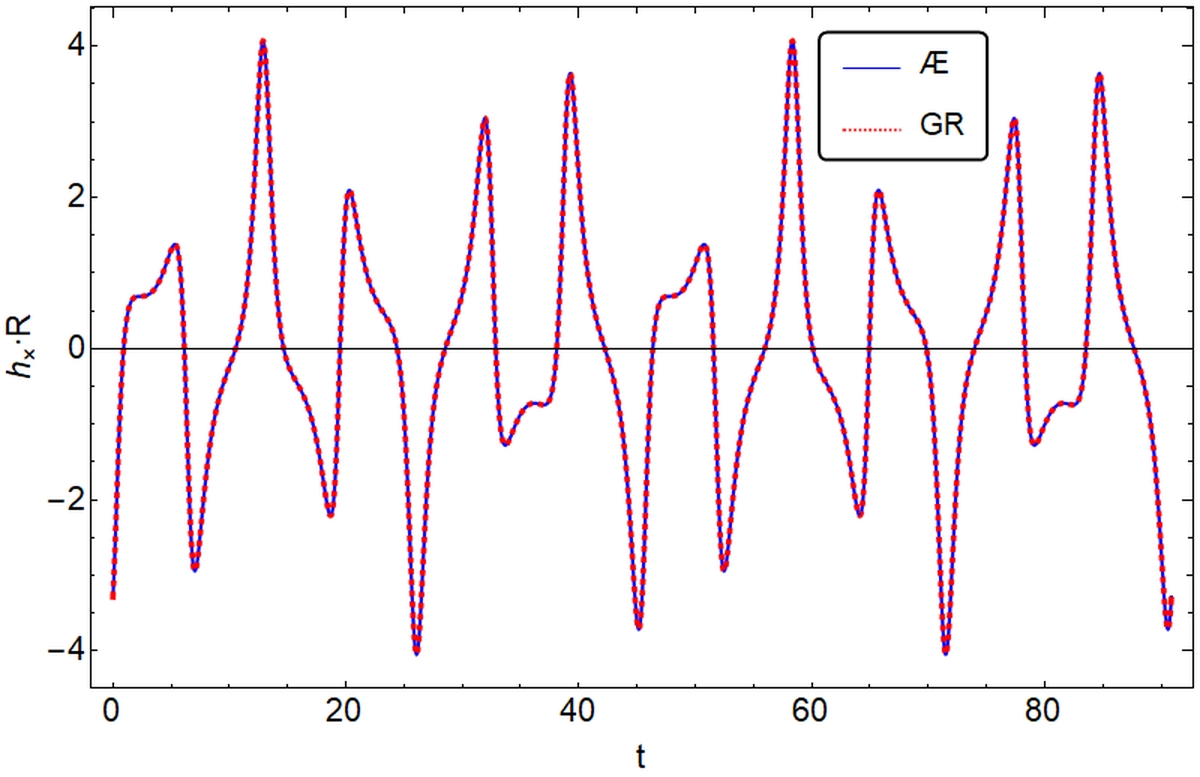}
\includegraphics[width=\linewidth]{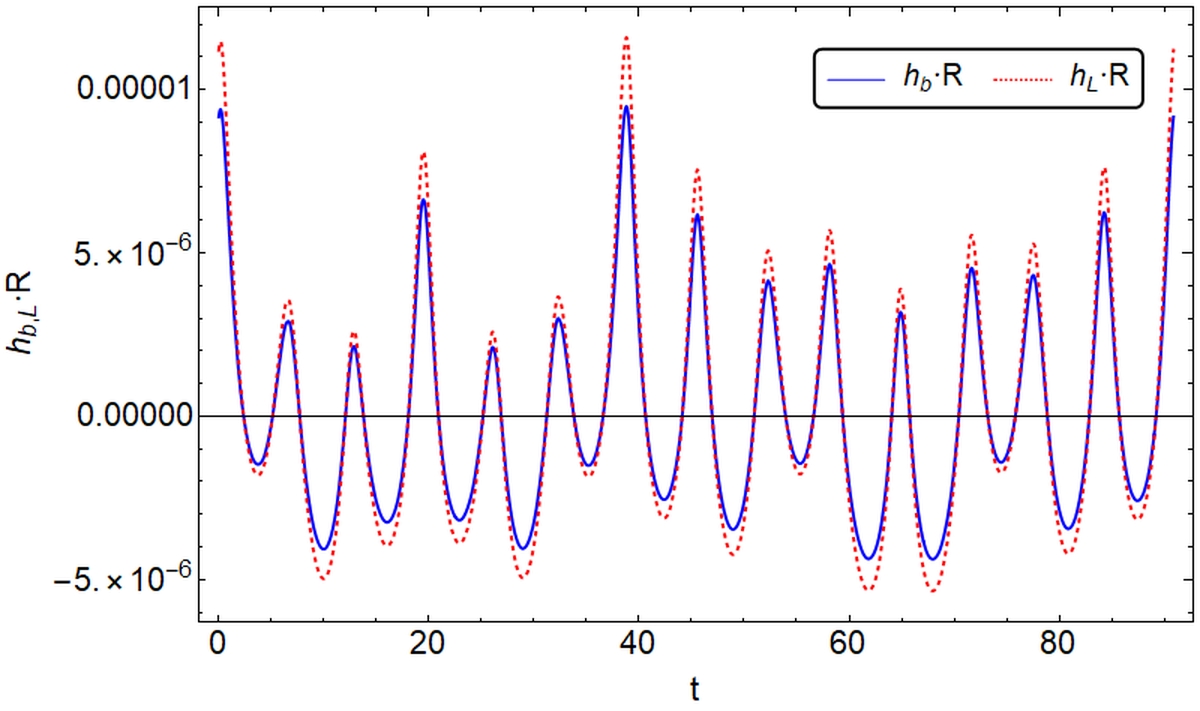}
		
	}
	\caption{The polarization modes $h_{N}$ defined in Eq.(\ref{polarizations}) for the Broucke A16 3-body system with the choice,
	$(\vartheta, \varphi;   \theta, \phi, \psi) = (0.6, 5.2; 1.3, 1.2, 1.8)$ and $(\Omega_1,\; \Omega_2,\; \Omega_3)  = (-0.1, \;  -2.76\times 10^{-6},\;  -2.9\times 10^{-5})$. }
	\label{fig2d}
\end{figure}

\begin{figure}[h!]
	{
		\includegraphics[width=\linewidth]{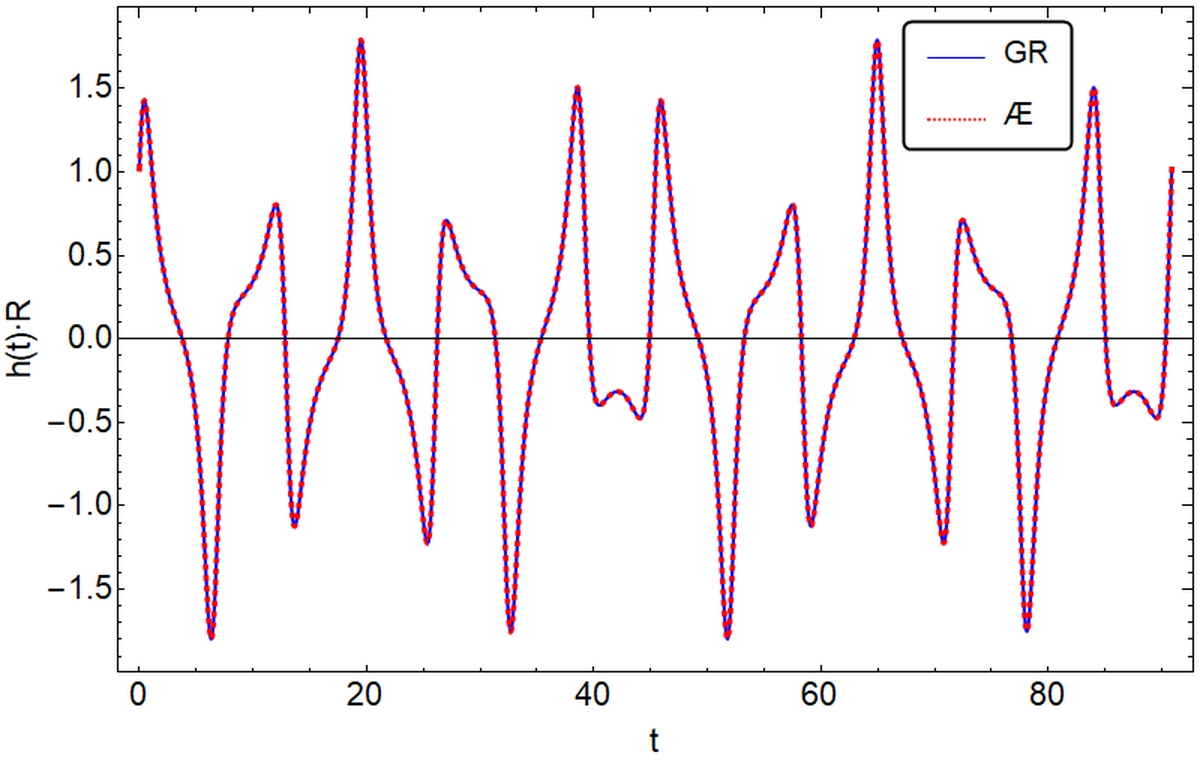}
		
	}
	\caption{ The response function $h(t)$ for the Broucke A16  3-body system with the choice,
	$(\vartheta, \varphi;   \theta, \phi, \psi) = (0.6, 5.2; 1.3, 1.2, 1.8)$ and $(\Omega_1,\; \Omega_2,\; \Omega_3)  = (-0.1, \;  -2.76\times 10^{-6},\;  -2.9\times 10^{-5})$.}
	\label{fig3d}
\end{figure}

\begin{figure}[h!]
	{
		\includegraphics[width=\linewidth]{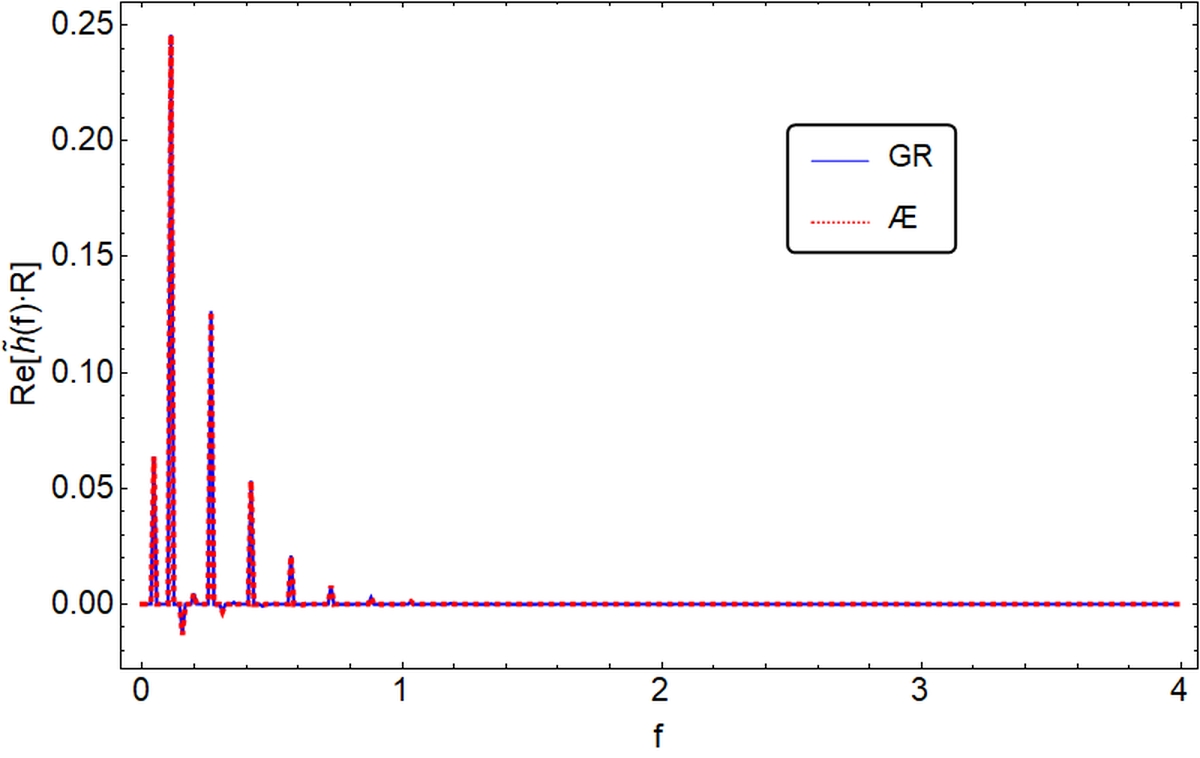}
\includegraphics[width=\linewidth]{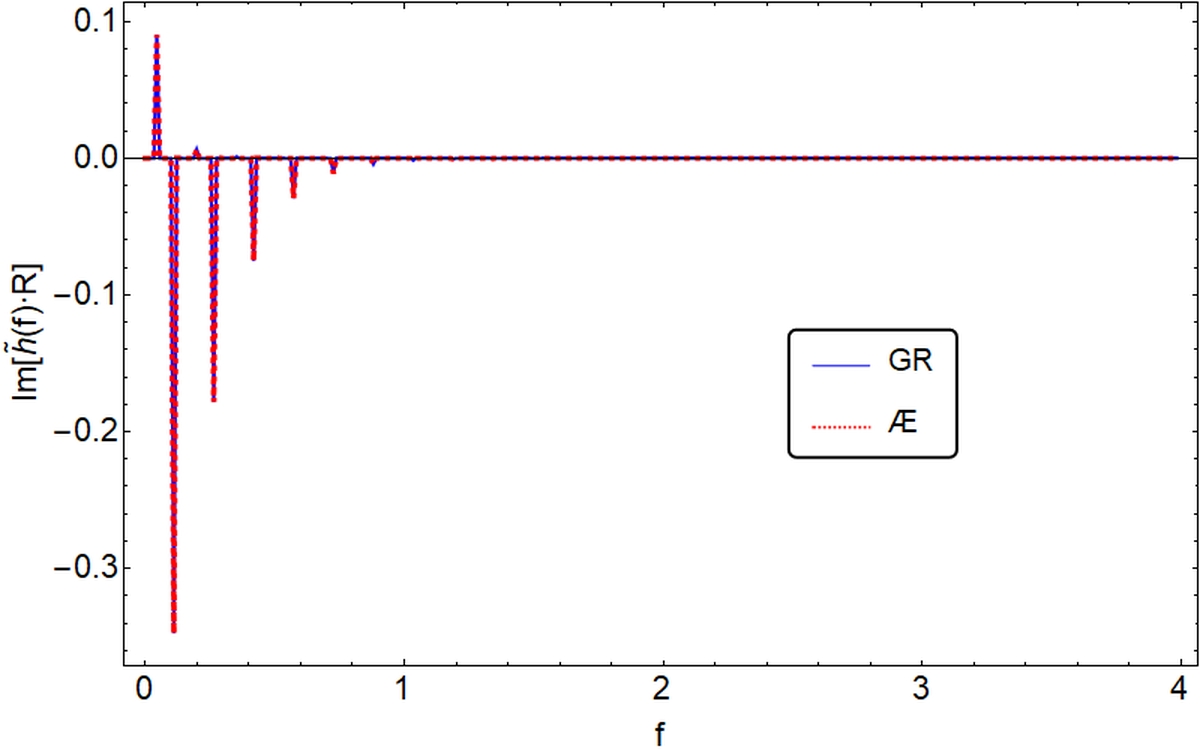}
		
	}
	\caption{The Fourier transform $\tilde{h}(f)$ of the response function  $h(t)$  for the Broucke A16 3-body system with the choice,
	$(\vartheta, \varphi;   \theta, \phi, \psi) = (0.6, 5.2; 1.3, 1.2, 1.8)$ and $(\Omega_1,\; \Omega_2,\; \Omega_3)  = (-0.1, \;  -2.76\times 10^{-6},\;  -2.9\times 10^{-5})$. }
	\label{fig4d}
\end{figure}

\begin{figure}[h!]
	{
		\includegraphics[width=\linewidth]{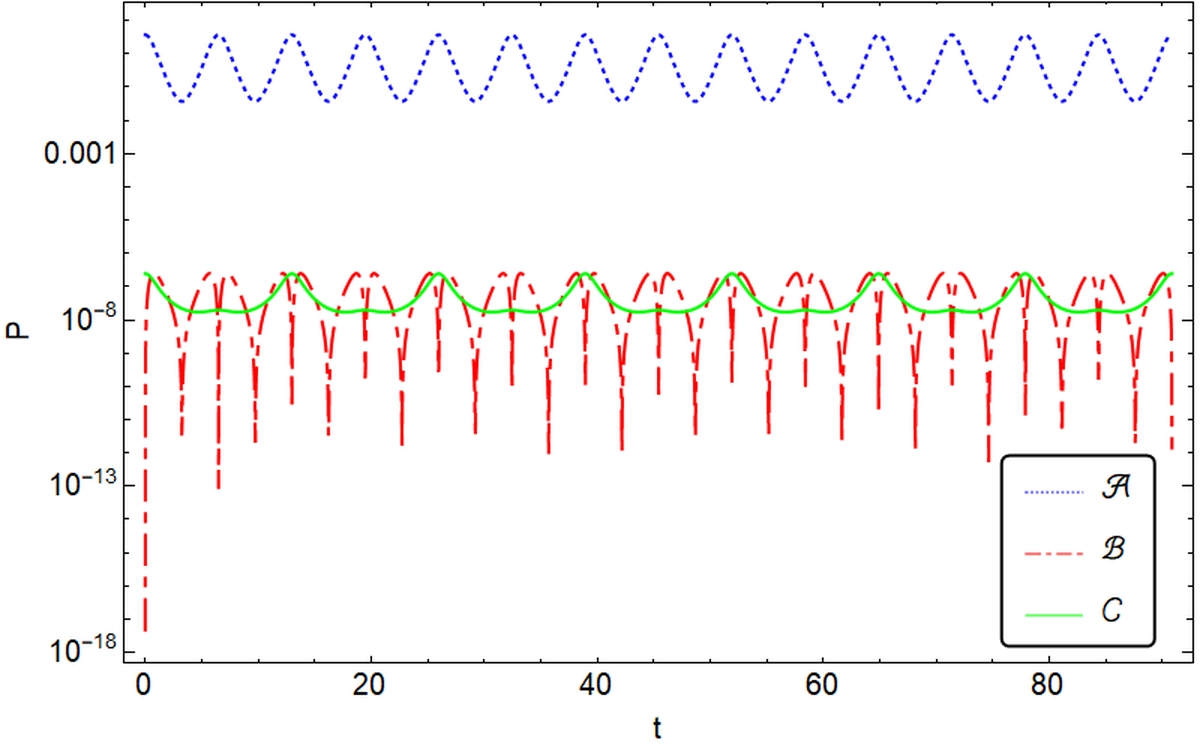}
		
	}
	\caption{The radiation power $P (\equiv - \dot{\cal{E}})$ of the 3-body system of the Broucke A16 figure with the choice,
	$(\vartheta, \varphi;   \theta, \phi, \psi) = (0.6, 5.2; 1.3, 1.2, 1.8)$ and $(\Omega_1,\; \Omega_2,\; \Omega_3)  = (-0.1, \;  -2.76\times 10^{-6},\;  -2.9\times 10^{-5})$.
	The  dotted (blue), dash-dotted (red)  and  solid (green)  lines  denote, respectively, 
the parts of quadrupole, monopole and dipole radiations given in Eq.(\ref{4.1a}).  }
	\label{fig5d}
\end{figure}

In Fig. \ref{fig1d}  we plot out   the trajectory of the Broucke A16   3-body system provided in \cite{Website},  while in  Fig. \ref{fig2d}-\ref{fig5d} we plot out the corresponding physical quantities for the
same choice of the five angular parameters as selected in the case for the Broucke R7 3-body system   in both GR and $\ae$-theory with $(\Omega_1,\; \Omega_2,\; \Omega_3)  = (-0.1, \;  -2.76\times 10^{-6},\;  -2.9\times 10^{-5})$.

From these figures we can see clearly that the GW forms and radiation powers not only depend on the relative positions, orientations between the source and detector, but also depend on the 
configurations of the orbits of the 3-body system. In addition, they also depend on  their binding energies of the three compact bodies.


\section{Conclusions}
 \renewcommand{\theequation}{6.\arabic{equation}} \setcounter{equation}{0}
 
 Three-{body} systems have been attracting more and more attention recently \cite{Naoz16}, specially after the detections of GWs from binary systems \cite{GW150914,GW151226,GW170104,GW170608,GW170814,GW170817},
 as they are common in our Universe \cite{FC17}, and can be ideal sources for periodic GWs.  In particular, we are very much interested in the cases in which the orbits 
 of two   bodies pass each other very closely and yet avoid collisions, so they can produce intense periodic gravitational waves and provide natural sources for the future detections of GWs.
 
 In this paper, we have studied the lowest PN order of three-body problems in the framework of Einstein-aether theory \cite{Jacobson}, a theory that violates locally the Lorentz symmetry and yet passes all the theoretical and observational 
 tests carried out so far \cite{OMW18}. Although these tests were mainly in the weak field limits, strong-field effects of binary neutron stars systems have been also investigated \cite{Foster07,Yagi14,HYY15,SY18}. In particular, the accelerations and
  ``strong-field" Nordtvedt parameter were calculated recently for a triple system to the quasi-Newtonian order \cite{Will18}. 
  
  In this paper, we have first shown that the contributions of the presence of the aether field to the quadruple part of the energy loss rate of a binary system  is the order of ${\cal{O}}\left(c_{14}\right) \lesssim {\cal{O}}\left(10^{-5}\right)$ lower 
  than that of GR when only the lowest PN order is taken into account [cf. Eq.(\ref{4.1ka})]. Due to the presence of two additional modes, the scalar and vector, in Einstein-aether theory, two additional parts also appear in the energy loss rate of a binary
  system \cite{Foster06},  given respectively by the second and third   terms in Eq.(\ref{4.1a}), representing the monopole and dipole contributions.   In comparison with the quadruple contributions of GR, which is the order of  
  ${\cal{O}}\left(v^2\right)$, the monopole contributions is only of the order of  ${\cal{O}}\left(c_{14}\right){\cal{O}}\left(v^2\right)$,   that is, it is  about ${\cal{O}}\left(c_{14}\right) \lesssim {\cal{O}}\left(10^{-5}\right)$ order  lower than that of GR. 
  Here $v$ is the relative velocity of the two compact objects.  However, the dipole contributions can be much larger than those of monopole. In particular, for a binary system with large differences between their binding energies, the dipole 
  part can be as large as ${\cal{O}}\left(c_{14}\right) {\cal{O}}\left(G_N m/d\right)$, where $m$ denotes the mass of a binary system and $d$ the distance between the two stars. For a realistic neutron star, we have ${\cal{O}}\left(G_N m/d\right)
  \simeq 0.1 \sim 0.3$, so that  ${\cal{O}}\left(c_{14}\right) {\cal{O}}\left(G_N m/d\right) \simeq  10^{-2} {\cal{O}}\left(v^2\right)$. 
  It should be noted that the scalar mode has contributions to all the three parts, quadrupole, dipole and monopole, while the vector mode has contributions only to  the quadrupole and dipole parts, as can be seen clearly from Eqs.(102)-(104) 
  of Ref. \cite{Foster06}. On the other hand, the strong-field contributions are only of the orders of
  \bqn
\lb{6.1}
\delta{\cal{W}}^{\mathrm{NS}}_{\cal{A}}    &\lesssim& {\cal{O}}\left(10^{ {-11}}\right), \quad
\delta{\cal{W}}^{\mathrm{NS}}_{\cal{B}}    \lesssim {\cal{O}}\left(10^{ {-11}}\right),\nb\\
\delta{\cal{W}}^{\mathrm{NS}}_{\cal{C}}    &\lesssim& {\cal{O}}\left(10^{ {-12}}\right), \quad
\delta{\cal{W}}^{\mathrm{NS}}_{\cal{D}}    \lesssim {\cal{O}}\left(10^{ {-11}}\right),
\eqn
for a binary neutron star system, where $\delta{\cal{W}}^{\mathrm{NS}}_{\cal{A}}, \; \delta{\cal{W}}^{\mathrm{NS}}_{\cal{B}}$ and $\delta{\cal{W}}^{\mathrm{NS}}_{\cal{C}}$ represent the contributions of the strong-field effects to
the quadrupole, monopole and dipole parts of Eq.(\ref{4.1a}), while $\delta{\cal{W}}^{\mathrm{NS}}_{\cal{D}}$ denotes a cross term due to the motion of the center-of-mass of the system \cite{Yagi14}. Clearly, these effects are much smaller 
than the ones mentioned above, and are out of the  detectability of the current generation of detectors. 

So, in this paper we have ignored these effects, and simply set the sensitivities $s_{A}$ of the compact bodies to zero. However, in the development of the general formulas, we have kept $T_{\mu}$ not zero to its first-order of perturbations in
Sec. IV, for our later applications of the formulas. Here $T_{\mu}$ represents the coupling between aether and matter fields [cf. Eq.(\ref{2.5})]. Setting $T_{\mu} = 0$, our results presented in Sec. IV reduce to the ones of Ref. \cite{Foster06}, 
subjected to some corrections of typos.  From the expressions of the six polarization modes of Eq.(\ref{polarizationsB}) we can see   that
the scalar longitudinal  mode $h_L$ is proportional to the scalar  breather mode $h_b$. Therefore, out of these six components, only five of them are independent. 
In addition,   the scalar breather  and the scalar longitudinal   modes are all suppressed by a factor ${\cal{O}}\left(c_{14}\right) \lesssim {\cal{O}}\left(10^{-5}\right)$
with respect to the transverse-traceless modes $h_{+}$ and $h_{\times}$, while the vectorial modes $h_{X}$ and $h_{Y}$ are suppressed by a factor
 ${\cal{O}}\left(c_{13}\right) \lesssim {\cal{O}}\left(10^{-15}\right)$. These conclusions should be also valid for general cases, and consistent with the analysis of triple systems presented  in Section V.  

Applying the general formulas developed in Sec. IV to a triple system, in Sec. V we have studied the GW forms, response function, its  {Fourier transform}, and the energy loss rates
 due to each part of the GW radiations given in Eq.(\ref{4.1a}) for three different kinds of
periodic orbits of three-body problems: one is the Simo's figure-eight configuration \cite{Simo02}, given by Fig. \ref{fig1ab}, and the other two are, respectively,  the Broucke R7 and A16 configurations provided in \cite{Website} and illustrated by
Figs. \ref{fig1c} and \ref{fig1d} in the current paper. In the case of the Simo's figure-eight configuration, we have studied the effects of the relative  orientations  between the source and  detector, as well as the effects of binding energies of
the three compact bodies. Through this case, we have shown explicitly that the  GW form, response function and its Fourier transform all depend on the relative  orientations  and binding energies, as they are expected from Eq.(\ref{4.1a}) 
 [cf. Figs. \ref{fig2aa1} and \ref{fig5d}]. In the cases of  the Broucke R7 and A16 configurations, the five angles are chosen
as the same as in the second case of the Simo's figure-eight configuration, and the corresponding GW form, response function, its Fourier transform and powers of radiation are given, respectively, by Figs. \ref{fig2c} - \ref{fig5c} and 
Figs. \ref{fig2d} - \ref{fig5d}. From these figures we find that all these physical quantities  are different. Therefore,   {\em the GW form, response function, its Fourier transform  of a triple
system depend not only on their configuration of orbits, but also on their orientation with respect to the detector and binding energies of the three compact bodies}.

\section*{Acknowledgments}

We would like to thank  V. Dmitrasinovic,  A. Hudomal, M. Suvakov, and L. Shao for providing us their numerical codes and valuable discussions and comments. 
This work is supported in part by the National Natural Science Foundation of China (NNSFC) with the grant numbers:  Nos. 11603020, 11633001, 11173021, 11322324, 11653002, 11421303, 11375153, 11675145, 
11675143,  11105120, 11805166, 11835009, 11773028,  11690022, 11375247, 11435006, 11575109, and 11647601.


\end{document}